\documentclass[12pt]{article}

\usepackage[T1]{fontenc} 
\usepackage{lmodern}
\usepackage{microtype} 

\usepackage[left=2.5cm,right=2.5cm,top=32mm,columnsep=20pt]{geometry} 
\usepackage{fancyhdr} 
\usepackage{abstract}

\usepackage{titlesec} 
\titleformat{\section}[block]{\large\bfseries\centering}{\thesection}{1em}{} 
\titleformat{\subsection}[block]{\bfseries}{\thesubsection}{1em}{} 

\usepackage{color}
\definecolor{dark-gray}{gray}{0.20}
\definecolor{gray}{gray}{0.30}
\definecolor{light-gray}{gray}{0.80}
\definecolor{dark-red}{rgb}{0.7,0,0}
\definecolor{dark-green}{rgb}{0.1,0.4,0}
\definecolor{dark-blue}{rgb}{0.3,0.3,0.7}
\definecolor{light-blue}{rgb}{0.8,0.8,1}

\usepackage{cite}    
\usepackage[pdfencoding=auto,pagebackref]{hyperref}
\usepackage{hyperref}
\hypersetup{
	colorlinks=true,
	linkcolor=dark-blue,
	citecolor=dark-red,
	urlcolor=dark-green,
	linktoc=page
}

\usepackage{enumerate}
\usepackage{graphicx}
\usepackage[percent]{overpic}

\usepackage{booktabs}
\usepackage{bbm}
\usepackage[margin=15pt,small]{caption}

\usepackage{tikz}
\usepackage{tikz-3dplot}

\usepackage{amsmath,amssymb,slashed,amsthm}

\numberwithin{equation}{section}
\usepackage{slashed}
\usepackage{mathabx}

\usepackage{subcaption} 

\newcommand{\rmi}{i}

\newcommand{\rmd}{d}

\renewcommand{\Re}{\text{Re}}
\renewcommand{\Im}{\text{Im}}

\newcommand{\Tr}{\text{Tr}~}

\newcommand{\SU}{\mathop{\rm SU}}

\newcommand{\SL}{\mathop{\rm SL}}

\newcommand{\OSp}{\mathop{\rm {}OSp}}

\newcommand{\roberto}[1]{\textbf{\textcolor{brown}{[RV:~#1]}}}

\usepackage{braket}
\usepackage{tikz-feynman}\tikzfeynmanset{compat=1.0.0}
\usepackage{cancel}
\usepackage{clipboard}
\usepackage{simplewick}
\usepackage{mathrsfs}

\renewcommand{\a}{\alpha}
\renewcommand{\b}{\beta}
\newcommand{\g}{\gamma}
\renewcommand{\d}{\delta}
\newcommand{\ep}{\epsilon}
\newcommand{\ve}{\varepsilon}

\renewcommand{\l}{\lambda}

\newcommand{\s}{\sigma}
\renewcommand{\t}{\theta}

\newcommand{\z}{\zeta}
\renewcommand{\o}{\omega}

\newcommand{\G}{\Gamma}   

\newcommand{\Y}{\Upsilon}

\newcommand{\da}{{\dot{\alpha}}}
\newcommand{\db}{{\dot{\beta}}}

\newcommand{\bep}{{\bar{\epsilon}}}
\newcommand{\bphi}{{\bar{\phi}}}
\newcommand{\bl}{{\bar{\lambda}}}
\newcommand{\bs}{{\bar{\sigma}}}

\newcommand{\btau}{{\bar{\tau}}}
\newcommand{\bpsi}{{\bar{\psi}}}

\newcommand{\cC}{{\mathcal{C}}}

\newcommand{\cL}{{\mathcal{L}}}
\newcommand{\cN}{{\mathcal{N}}}

\newcommand{\cR}{{\mathcal{R}}}

\newcommand{\bbZ}{{\mathbb{Z}}}

\newcommand{\pd}{\partial}

\newcommand{\ar}{{\rightarrow}}

\newcommand{\mintt}[2]{\overset{#2}{\underset{#1}{\int}}\!}

\newcommand{\tr}{{\text{tr}}}
\newcommand{\bk}[1]{\braket{#1}}

\newcommand{\eqfig}[2]{\parbox[c]{#2\textwidth}{\includegraphics[width = #2\textwidth]{#1}}}

\renewcommand{\tt}[1]{{\text{#1}}}

\newcommand{\cmt}[1]{}
\newcommand{\q}[1]{``#1''} 

\graphicspath{{./Figures/}}

\title{}

\begin{document}
	
	\begin{titlepage}
		
		\medskip
		\begin{center}
			{\Large \bf BCFT One-point Functions \\[.5em]of Coulomb Branch Operators}
			
			\bigskip
			\bigskip
			\bigskip
			\bigskip
			
			{\bf Davide Bason,${}^{1,2}$ Lorenzo Di Pietro,${}^{1,2}$ Roberto Valandro${}^{1,2}$\\ and Jesse van Muiden${}^{2,3}$   \\ }
			\bigskip
			\bigskip
			\bigskip
			\bigskip
			${}^{1}$Dipartimento di Fisica, Universit\'a di Trieste,
			Strada Costiera 11, I-34151 Trieste, Italy
			\vskip 8mm
			${}^{2}$INFN, Sezione di Trieste, Via Valerio 2, I-34127 Trieste, Italy
			\vskip 8mm
			${}^{3}$SISSA, Via Bonomea 265, I-34136 Trieste, Italy\\
			
			\bigskip
			\texttt{davide.bason@phd.units.it,~lorenzo.dipietro@ts.infn.it,\\roberto.valandro@ts.infn.it,~jvanmuid@sissa.it} \\
		\end{center}
		
		\bigskip
		\bigskip
		
		\begin{abstract} 
			
			\noindent We show that supersymmetry can be used to compute the BCFT one-point function coefficients for chiral primary operators, in 4d $\mathcal{N}=2$ SCFTs with $\frac{1}{2}$-BPS boundary conditions. The main ingredient is the hemisphere partition function, with the boundary condition on the equatorial $S^3$. A supersymmetric Ward identity relates derivatives with respect to the chiral coupling constants to the insertion of the primaries at the pole of the hemisphere. Exact results for the one-point functions can be then obtained in terms of the localization matrix model. We discuss in detail the example of the super Maxwell theory in the bulk, interacting with 3d $\mathcal{N}=2$ SCFTs on the boundary. In particular we derive the action of the SL(2,$\mathbb{Z}$) duality on the one-point functions. 
			
		\end{abstract}
		
		\noindent

	\end{titlepage}
	

\newpage

	\tableofcontents
	
	\newpage
	
	\section{Introduction}
Supersymmetry is a celebrated tool to study quantum field theory (QFT) at strong coupling. Exact results can be often obtained using non-renormalization theorems or supersymmetric localization \cite{Pestun:2016zxk}, and they serve as a theoretical laboratory to test methods and ideas. The set of QFT observables that can be computed exactly using supersymmetry is constantly extended as new techniques are discovered, see e.g. \cite{Binder:2019jwn, Chester:2020dja, Dorigoni:2021bvj, Dorigoni:2021guq} for interesting recent developments.

In this paper we discuss a new example of observables that can be obtained from supersymmetry: the bulk one-point functions in a boundary conformal field theory (BCFT). We recall that BCFTs are defined by placing a conformal field theory in a half space, i.e. restricting one of the coordinates to be positive $y\geq 0$, with a conformal boundary condition at $y=0$, namely a boundary condition that preserves the full conformal group acting on the boundary. BCFTs describe the possible critical behaviors near the boundary of a system that undergoes a second order transition while enclosed in a finite region of space. In this setup, operators inserted at bulk points can have one-point functions
\begin{equation}
O(y,\vec{x}) = \frac{a_O}{y^{\Delta_O}}~,
\end{equation}
that are fixed in terms of the scaling dimensions $\Delta_O$ up to the coefficients $a_O$, which are data of the BCFT. Our goal is to determine these coefficients using supersymmetry. Having a computational tool for this problem is valuable because not much is known about interacting BCFTs in bulk dimension above 2.

More precisely, the object of our study are $\tfrac 12$-BPS boundary conditions for 4d $\mathcal{N}=2$ super-conformal field theories (SCFT). Here ``$\tfrac 12$-BPS'' means that the symmetry preserved is a 3d $\mathcal{N}=2$ super-conformal algebra. In this context we consider one-point functions of a particular class of bulk operators, the chiral primaries \cite{Dolan:2002zh}, that parametrize the Coulomb branch of vacua and are therefore also referred to as ``Coulomb branch operators'' \cite{Gerchkovitz:2016gxx}.  

For SCFTs without boundaries the two-point functions of chiral/anti-chiral primaries were shown in \cite{Gerchkovitz:2016gxx} to be computable by considering the theory on the curved background of $S^4$, and taking derivatives of the partition function with respect to the associated couplings. This involves several ingredients: firstly, a supersymmetric Ward identity that allows to relate correlation functions of integrals of the top component --- i.e. the supersymmetric Lagrangian associated to the chiral multiplet --- to insertions of the chiral/anti-chiral primaries at the north/south poles. Secondly, extracting the flat space result from the one on the sphere requires taking into account the mixing of operators caused by the Weyl rescaling, which can be done by diagonalizing the matrix of two-point functions via a Gram-Schmidt procedure. Thirdly, to perform explicit calculations, one needs to rely on exact results for the $S^4$ partition function, which can be obtained for many SCFTs of interest using supersymmetric localization.

In this paper we will show how this approach can be adapted to the problem at hand to obtain the BCFT one-point functions using the partition function on a hemisphere $HS^4$, with the supersymmetric boundary condition on the equatorial $S^3$. The inclusion of the boundary entails several non-trivial generalizations. To start with, we need to define the relevant Lagrangians on the hemisphere in such a way that they are supersymmetric off-shell, for a given choice of supersymmetric boundary condition, and this requires the addition of appropriate boundary terms. Moving on to the supersymmetric Ward identity, one naturally finds that the integral of the supersymmetric Lagrangian on $HS^4$ gives not only the insertion of the chiral primary at the pole, but also a boundary term integrated on $S^3$. Isolating the local insertion requires a nontrivial cancellation of all the contributions of boundary terms. Once the one-point functions are obtained on the hemisphere, the matrix that diagonalizes the $S^4$ two-point functions needs to be applied in order to recover the flat space BCFT result. 

Localization formulas have been derived on $HS^4$ \cite{Gava:2016oep, Dedushenko:2018tgx} and this allows us to obtain explicit results in terms of matrix integrals in specific models. The localization formulas for the hemisphere partition function are holomorphic functions of the couplings associated to the chiral operators. However, they also depend on real parameters that encode the possible mixings of the $U(1)_R$ superconformal $R$-symmetry with other $U(1)$ symmetries preserved by the boundary condition. This is an important novelty introduced by the boundary, due to the fact that it preserves only the 3d $\mathcal{N}=2$ subalgebra which has an abelian $R$-symmetry. Fixing this ambiguity requires a maximization procedure \cite{Gaiotto:2014gha} which introduces a non-holomorphic dependence in the hemisphere partition function of the BCFT. We explain how one can still obtain one-point functions from derivatives with respect to the holomorphic coupling, by considering a judicious combination of the BCFT partition function and its complex conjugate.

A rich class of boundary conditions that we can study in this way are obtained from coupling $\mathcal{N}=2$ superconformal gauge theories to 3d $\mathcal{N}=2$ SCFTs, using the Neumann bondary condition of the gauge field to gauge the global symmetry of the boundary theory. In this setup the complexified gauge coupling $\tau$ is an exactly marginal bulk coupling and one naturally gets a continuous family of BCFTs. The one-point function coefficients that we compute are therefore functions of $\tau$. Different values of $\tau$ are related to each other by dualities, and the dualities also act on the boundary conditions. 

While the formulas we obtain for the one-point functions are valid in general, the main example that we consider in this paper is that of the $\mathcal{N}=2$ super-Maxwell theory, which is a family of free SCFTs in 4d parametrized by $\tau$, with an action of the SL$(2,\mathbb{Z})$ duality group. This theory is free in the bulk but it still admits interacting boundary conditions of the type described above, on which the duality acts as the 3d $\mathcal{N}=2$ supersymmetric version of Witten's SL$(2,\mathbb{Z})$ action \cite{Witten:2003ya}. The Coulomb branch operators in this example are simply given by the powers $\phi^n$ of the scalar in the vector multiplet. The mixing problem can be completely solved and we find that the flat-space one-point function are obtained from the insertion of the $n$-th Hermite polynomial in the localization integral. This allows us to write explicitly the action of the duality on the BCFT one-point functions. We examine in detail the boundary condition given by two chiral fields of opposite charge, related by an $S$ transformation to the boundary condition defined by the XYZ model, and the boundary condition with a single chiral field, which is invariant under the $\mathbb{Z}_3$ subgroup generated by an $ST^{-1}$ transformation. We prove these BCFT dualities at the level of the hemisphere partition function, making use of known integral identities associated to dualities in 3d $\mathcal{N}=2$ theories. Moreover, we compute many orders in perturbation theory by expanding the localization formulas, both for the partition function and for the one-point functions, and we test the validity of the extrapolation of these perturbative results by checking their agreement with the predictions of the dualities. The hemisphere partition function and the action of dualities in these examples were also studied in \cite{KumarGupta:2019nay, Gupta:2020eev}.

The rest of the paper is organized as follows: in Section \ref{Sec: N=2 gauge theories on HS4} we describe how to define theories on the hemisphere, deriving the boundary terms needed to preserve supersymmetry, and we review the localization results on $HS^4$; in Section \ref{Sec: Ward identities on HS4} we use a supersymmetric Ward identity to obtain the general formula for one-point functions of chiral primaries in terms of the hemisphere partition function, we offer a second derivation of this formula based on localization, and we comment on the geometric interpretation of the hemisphere partition function viewed as a function on the conformal manifold; in Section \ref{Sec:Dualities} we specify to the Abelian gauge theory in the bulk and we derive the action of SL$(2,\mathbb{Z})$ on the partition function and on the one-point functions; finally in Section \ref{Sec:Examples} we study in detail the specific boundary conditions mentioned above for the Abelian theory, both exactly and using perturbative expansions. Many technical results are relegated to the appendices.

%
\section{$\mathcal{N}=2$ gauge theories on $HS^4$}\label{Sec: N=2 gauge theories on HS4}
In this section we discuss the action and supersymmetric boundary conditions for $\mathcal N=2$ gauge theories on $HS^4$.  We then explain how to relate correlation functions of chiral primary operators on $HS^4$ to those on $H\mathbb{R}^4$ when conformal symmetry is also preserved. Finally, we summarize some localization results on $HS^4$ that we will use throughout this paper. Gauge theories preserving $\mathcal{N}=2$ supersymmetry on a 4d hemisphere have been studied in \cite{Gava:2016oep, Dedushenko:2018tgx}. 
\begin{table}[h]
	\centering
	\begin{tabular}{cccc}
		\hline
		{Field}    & {Description}    & {$\text{U}(1)_{R}$ charge} & {$\text{SU}(2)_{R}$ repr.} \\ 
		$g_{\mu\nu}$       & metric                  & $0$                     & $\mathbf{1}$                  \\ 
		$\mathcal A_\mu$   & $\mathfrak{u}(1)_R$ gauge field   & $0$                     & $\mathbf{1}$                  \\ 
		$\mathcal V_\mu$   & $\mathfrak{su}(2)_R$ gauge field  & $0$                     & $\mathbf{3}$                       \\ 
		$T_{\mu\nu}$       & anti-self-dual tensor   & $+2$                    & $\mathbf{1}$                  \\ 
		$\bar{T}_{\mu\nu}$ & self-dual tensor        & $-2$                    & $\mathbf{1}$                  \\ 
		$\tilde{M}$        & scalar                  & $0$                     & $\mathbf{1}$                  \\  \hline
	\end{tabular}
	\caption{Bosonic field content of $\mathcal N=2$ supergravity multiplet.}\label{Tab: weyl multiplet}
\end{table}
\subsection{The action and boundary conditions on the hemisphere}
In order to put a $\mathcal{N}=2$ field theory on the sphere the starting point is the coupling to a background $\mathcal N=2$ Weyl multiplet, whose component fields are spelled out in Table \ref{Tab: weyl multiplet}. We use $\mathscr{R}$ to denote the scalar curvature of the background. The background values on $S^4$ with the round metric of radius $1$ are
\begin{equation}\label{Eq: background sugra values}
	\tilde{M} = 0\:,\quad \mathcal A_\mu = 0 \:,\quad \mathcal V_\mu = 0\:,\quad  T_{\mu\nu} =0 \:,\quad  \bar T_{\mu\nu} = 0\:,\quad g_{\mu\nu} = g_{\mu\nu}^{S^4} \quad 
	\text{with}\quad\mathscr{R} = 12~,
\end{equation}
namely only the metric is activated. The Killing spinors on $S^4$ satisfy the equation
\begin{equation}\label{eksemain}
		D_\mu \epsilon_A=-\frac{i}{2}(\tau_3)_A^{\ B}\sigma_\mu\bar{\epsilon}_B~,\ \ \ \ \ D_\mu \bar{\epsilon}_A=-\frac{i}{2}(\tau_3)_A^{\ B}\bar{\sigma}_\mu\epsilon_B~,
	\end{equation}
where the conventions for the $\sigma$ matrices and the $\text{SU}(2)_R$ matrices $\tau$ are given in Appendix~\ref{app:conv}. The SUSY variations can be obtained from the supergravity ones by plugging in the Killing spinors that solve these equations, see Section \ref{s4dmultiplets}. For a massive theory the resulting supersymmetry algebra on $S^4$ is that of $\text{OSp}(2|4)$. 

To define the theory on the hemisphere $H S^4$ we simply restrict the $S^4$ Lagrangian but we further need to supplement the theory with boundary conditions on the equatorial $S^3$. At most half of the supersymmetry can be preserved, namely a 3d $\mathcal{N}=2$ subgroup which on $S^3$ gives $\text{OSp}(2|2)\times \text{SU}(2)$. The Killing spinor of this subgroup are obtained from the 4d ones imposing the boundary condition
\begin{align}\label{eq:3dcondKS}
\bep_A=i(\tau_3)_A^{\ B}\bs^\perp\ep_B\,.
\end{align}
Here the superscript $\perp$ refers to the following choice for the vierbein $e^a_\mu$ ($a=1,\dots,4$ being the flat index and $\mu$ the spacetime index): $e^4_\mu  \equiv e^\perp_\mu$ is chosen to be the normal $n_\mu$ to the boundary, and $e^{a'}_\mu$ with $a'=1,2,3$ are a completion to an orthonormal basis. The condition \eqref{eq:3dcondKS} leaves two independent Killing spinors, namely $\epsilon_{A=1,2}$ that generate supersymmetry along $S^3$. Indeed defining
\begin{align}\label{e3dmapgeneralmain}
\begin{split}
\ep_{1\a}=\frac{1}{\sqrt{2}}\zeta_\alpha\,,~~~~\ep_{2\a}=\frac{1}{\sqrt{2}}\tilde{\z}_\alpha\,,
\end{split}
\end{align}
we can see that these spinors satisfy the Killing spinor equations on $S^3$ \cite{Willett:2016adv}
\begin{align}
D_i\z=-\frac{i}{2}\g_i\z\,,~~~~D_i\tilde{\z}=-\frac{i}{2}\g_i\tilde{\z}~,
\end{align} 
where we defined
\begin{equation}
(\g^i)_\a^{\ \b}=i(\s^i\bs^\perp)_\a^{\ \b}=-i(\s^\perp\bs^i)^\b_{\ \a}~,
\end{equation}
and $i=1,2,3$ is the 3d spacetime index. Rewriting the supersymmetric variations of the bulk 4d $\mathcal{N}=2$ supersymmetry in terms of $(\zeta,\tilde{\zeta})$ allows to reorganize each 4d $\mathcal{N}=2$ multiplet into 3d $\mathcal{N}=2$ multiplets. We show the associated variations in Appendix~\ref{s3d}.

On the hemisphere we need to impose supersymmetric boundary conditions for all the fields in order to preserve the 3d $\mathcal{N}=2$ subalgebra. Here by supersymmetric boundary conditions we mean that:
\begin{itemize}
\item[1)]{they make the action stationary on-shell;}
\item[2)]{they are compatible with the structure of the SUSY multiplets, i.e. the boundary conditions on all fields in the same multiplet can be obtained from each other by a SUSY variation in the 3d $\mathcal{N}=2$ subalgebra preserved by the boundary.}
\end{itemize} 
In order for 1) and 2) to happen at the same time we also need to add appropriate boundary terms to the bulk action. To make  the property 2) manifest it is convenient to rearrange the four-dimensional fields into three-dimensional $\mathcal N=2$ multiplets, see Appendix~\ref{s4dto3d} for the details. Provided condition 1) is satisfied, condition 2) can be phrased equivalently as the requirement that the SUSY variation of the action vanishes on the space with boundary, or that the normal component of the supercurrent vanishes at the boundary.  

Let us now review the Lagrangians and some examples of supersymmetric boundary conditions. 

\paragraph{Vector multiplet}

The $\mathcal N=2$ vector multiplet consists of a vector, two complex scalars, two gaugini, and a triplet of auxiliary scalars:
\begin{equation}
	\mathcal N=2 \quad \text{vector multiplet:} \quad A_\mu,\, \phi,\, \bar\phi,\, \lambda_A,\, D_{AB}\,,
\end{equation}
where $\mu,\nu,\ldots$ are spacetime indices,  $A,B,\ldots$ are $\text{SU}(2)_R$ indices, and all the fields are valued in the Lie algebra $\mathfrak{g}$ of the gauge group $G$. 
Following the conventions of \cite{Hama:2012bg} we write down the vector multiplet action on a generic curved background 
\begin{equation}\label{eactiononS4}
	\begin{aligned}
		S&=\frac{\Im\:\tau}{4\pi}\int\rmd^4 x \sqrt{g}\,\mathcal L_{\text{YM}} + i\frac{\Re\: \tau}{8\pi}  \int\rmd^4 x\sqrt{g} \,\mathcal L_{\theta}~,
	\end{aligned}
\end{equation}
where
\begin{equation}\label{eYMCS}
	\begin{aligned}
		\mathcal L_{\text{YM}} &=~ \text{Tr}\Big[\frac{1}{2}F^{\mu\nu}F_{\mu\nu}-4D_\mu\bar{\phi}D^\mu\phi+2\big(\tilde{M}-\frac{\mathscr{R}}{3}\big)\bar{\phi}\phi+i\bar{\lambda}^A\bar{\slashed{D}}\lambda_A-i\lambda^A\slashed{D}\bar{\lambda}_A\\
		&\qquad -\frac{1}{2}D^{AB}D_{AB} +4[\phi,\bar{\phi}]^2-2\lambda^A[\bar{\phi},\lambda_A]+2\bar{\lambda}^A[\phi,\bar{\lambda}_A]\\
		&\qquad +16F_{\mu\nu}(\bar{\phi}T^{\mu\nu}+\phi\bar{T}^{\mu\nu})+64(\bar{\phi}^2T^{\mu\nu}T_{\mu\nu}+\phi^2\bar{T}^{\mu\nu}\bar{T}_{\mu\nu})\Big]~,\\
		\mathcal L_{\theta} &= \text{Tr}\Big[\frac12 \varepsilon^{\mu\nu\rho\sigma} F_{\mu\nu} F_{\rho\sigma} \Big]~.
	\end{aligned}
\end{equation}
Here  $\tau$ is the complexified gauge coupling 
\begin{equation}\label{etaug}
	\tau=\frac{\t}{2\pi}+\frac{4\pi}{g^2}\,.
\end{equation}
To specify the $S^4$ background we simply plug \eqref{Eq: background sugra values} into this Lagrangian.

The vector multiplet can be organized in a couple of 3d $\mathcal{N}=2$ linear multiplets as follows (see Appendix~\ref{App: Supersymmetry transformations} for more details on these multiplets)\footnote{Note that in order to identify the variations of $\mathcal J_{\text{N}}$ with those of a 3d linear multiplet we need to use that the components $D_{11}$ and $D_{22}$ of $D_{AB}$ vanish at the boundary. This is true in general, even in in the presence of interactions, due to the fact that the boundary condition preserves the Cartan of $SU(2)_R$ symmetry under which these components are charged.}
\begin{equation}\label{ecurrendDN}
	\begin{aligned}
		\mathcal J_{\text{D}} =&\, \Big\{2\phi_1,-i \l_2^+,i\l_1^-,-\frac{i}{2}\varepsilon_{ijk}F^{jk},-2\big(D_\perp\phi_2-i\phi_1+\frac{i}{2} D_{12}\big) \Big\}\,,\\
		\mathcal J_{\text{N}} =&\, \Big\{2\phi_2,\l_2^-,-\l_1^+,F_{\perp i},2(D_\perp\phi_1+ i\phi_2)\Big\}\,,
	\end{aligned}
\end{equation}
where we have defined
\begin{equation}
	\phi_1 = \frac{\phi+\bar\phi}{2i}\:,\quad \phi_2 = \frac{\phi-\bar\phi}{2}\,,\quad \lambda_A^{\pm} = \frac{1}{\sqrt{2}} \left( \lambda_A \pm i \sigma_4 \bar\lambda_A \right).
\end{equation}
The simplest supersymmetric boundary conditions we can impose are the Dirichlet (D) and Neumann (N) boundary conditions
\begin{equation}\label{ebc}
	\text{D}\,: \quad \mathcal J_{\text{D}} = 0\,,\qquad \text{and} \qquad \text{N}\,: \quad  \mathcal J_{\text{N}} + \gamma\mathcal{J}_{\text{D}} = 0\,,
\end{equation}
where we defined
\begin{equation}\label{Eq: definition of gamma}
	\gamma = \frac{\Re \, \tau}{\Im\,\tau}\,.
\end{equation}
These boundary conditions are the supersymmetrized versions of the Dirichlet and Neumann boundary conditions of the Maxwell theory
\begin{equation}\label{Eq: bndry conditions for fluxes}
	\text{D}\,: \quad F_{ij} = 0\,,\qquad \text{and}\qquad \text{N}\,: \quad F_{\perp i} - \frac{i}{2}\gamma \varepsilon_{ijk} F^{kl}=0\,.
\end{equation}
The additional terms needed to preserve 3d $\mathcal N=2$ supersymmetry are\footnote{\label{forientation}Our conventions for the orientation are such that $\int_{HS^4}\partial\cdot V=\int_{S^3}V^\perp$.}
\begin{equation}\label{eactiononHS4}
	S_{\partial} =  \frac{\Im\:\tau}{4\pi} \int_{S^3} \rmd^3 x \sqrt{h}\, \mathcal{L}_{\text{YM}}^{\partial} + i\frac{\Re\: \tau}{8\pi}  \int_{S^3}\rmd^3 x\sqrt{h}\,\mathcal{L}_{\theta}^{\partial}~,
\end{equation}
where 
\begin{equation}\label{Eq: explicit boundary terms}
	\begin{aligned}
		&\mathcal L_{\text{YM}}^\partial = \text{Tr}\Big[-8 \phi_2 \left( D_\perp \phi_2 + \tfrac{i}{2} D_{12}\right) + \lambda_{1} \lambda_2 + \bar\lambda_1 \bar\lambda_2\Big]\:,\\
		&\mathcal L_{\theta}^{\partial} = 2\text{Tr}\Big[ 8i \phi_1 \left(D_\perp \phi_2+\tfrac{i}{2} D_{12}\right) - \lambda_1 \lambda_2 + \bar\lambda_1 \bar\lambda_2 - i \lambda_1 \sigma^\perp \bar\lambda_2 + i\lambda_2\sigma^\perp\bar{\lambda}_1\Big]\:,
	\end{aligned}
\end{equation}
and $h$ denotes the induced metric on the equatorial $S^3$, i.e. the round metric with radius 1. With this boundary action both N and D are supersymmetric boundary conditions according to the definition given above.

Imposing the N boundary condition we have the freedom to perform gauge transformations on the boundary, while the D boundary condition fixes it and the gauge group $G$ becomes a global symmetry on the boundary, with associated current $\mathcal{J}_N$.\footnote{The operator $\mathcal{J}_D$ that survives in the case of the Neumann boundary condition is not gauge invariant and so it does not give rise to a global symmetry, except in the abelian case or more generally if the group has $\text{U}(1)$ factors so that $\tr[\mathcal{J}_D]\neq 0$.\label{foot:JD}} It is also possible to impose  D boundary conditions on some of the gauge components of the gauge field and N on the others in order to preserve only a subgroup of the gauge symmetry at the boundary.

A rich class of boundary conditions for 4d $\mathcal{N}=2$ gauge theories preserving 3d $\mathcal{N}=2$ can be obtained from deformations of the N and D boundary conditions. These deformations are defined by adding local degrees of freedom on the boundary and coupling them with the boundary modes of the bulk fields. In particular, a natural deformation of the N boundary condition consists in using the boundary mode of the bulk gauge field to gauge a global symmetry of additional local boundary degrees of freedom. In the abelian case some deformation is necessary in order to introduce interactions, because the bulk theory is free and the N and D boundary conditions are linear constraints that keep the theory gaussian. 

\paragraph{Hypermultiplet}\label{parhypermultiplet}

A theory with $n$ hypermultiplets is described by $q_{IA}$ scalars, carrying both a fundamental $SU(2)_{\mathcal{R}}$ index $A$ and a fundamental $USp(2n)$ index $I$, and $\psi_I,\, \bar{\psi}_I$ fermions. The index is lowered with the symplectic matrix $\Omega_{IJ}$ and raised with $\Omega^{IJ}$ (defined such that $\Omega^{IJ}\Omega_{JK}=\delta^I_K$). An off-shell formalism for the full 4d $\mathcal{N}=2$ algebra requires an infinite number of auxiliary fields. On the other hand, an off-shell formalism for a 3d $\mathcal{N}=2$ subalgebra, like the one preserved by the boundary, only requires to add auxiliary scalars $F_{IA}$.

The action for the hypermultiplets coupled to the background Weyl multiplet is \cite{Hosomichi:2016flq}
\begin{align}
\begin{split}
&S_{\text{hyp}}=\int\rmd^4 x \sqrt{g}\left[\frac{1}{2}D_\mu q^{IA} D^\mu q_{IA}-q^{IA}\{\phi,\bphi\}_I^{\ J}q_{JA}+\frac{i}{2} q^{IA} (D_{AB})_I^{\ J} q_J^B\right.\\
&+\frac{1}{8}(\tilde{M}+\frac{2}{3}\mathscr{R})q^{IA} q_{IA} -\frac{i}{2}\bpsi^I \bs^\mu D_\mu\psi_I -\frac{1}{2}\psi^I(\phi)_I^{\ J}\psi_J+\frac{1}{2}\bpsi^I(\bphi)_I^{\ J}\bpsi_J+\frac{i}{2}\psi^I \s^{\mu\nu}T_{\mu\nu}\psi_I\\
&\left.-\frac{i}{2}\bpsi^I\bs^{\mu\nu}\bar{T}_{\mu\nu}\bpsi_I -q^{IA}(\l_A)_I^{\ J}\psi_J+\bpsi^I(\bl_A)_I^{\ J} q^A_J-\frac{1}{2}F^{IA}F_{IA} \right]~.
\end{split}
\end{align}
To get the theory on $S^4$ we simply plug \eqref{Eq: background sugra values} into this action. As for the vector multiplet, also in this case we need to specify boundary conditions and add a boundary action to preserve 3d $\mathcal{N}=2$ on $HS^4$. 

A single 4d hypermultiplet decomposes in two 3d $\cN=2$ chiral multiplets (see Section \ref{s4dto3d}) 
\begin{align}
\begin{split}\label{eq:bchyp}
{\Phi}_{11}&=\Big\{q_{11},\ -\frac{i}{2}(\psi_1+\rmi\s_\perp\bpsi_1),\ D_\perp q_{12}+2\phi_2 q_{12}-\rmi F_{11}\Big\}\equiv {\Phi}_{22}^\dagger~,\\
{\Phi}_{12}&=\Big\{q_{12},\ \frac{i}{2}(\psi_1-\rmi\s_\perp\bpsi_1),\ -D_\perp q_{11}+2 \phi_2 q_{11}-\rmi F_{12}\Big\}\equiv - {\Phi}_{21}^\dagger~.
\end{split}
\end{align}
The simplest boundary conditions for a single 4d hypermultiplet, i.e. $n=1$, sets to zero either ${\Phi}_{11}$ or ${\Phi}_{12}$. We will refer to them as boundary conditions $1$ and $2$, respectively. These conditions are easily generalised to the case of $n$ hypermultiplets. Namely, a similar binary choice is obtained by picking $n$ components labeled by $i$ among the $2n$ components labeled by $I$, and setting to zero one between
${\Phi}_{i 1} = (\Omega^{iJ}{\Phi}_{J2})^\dagger$ or ${\Phi}_{i 2}= -(\Omega^{iJ}{\Phi}_{J1})^\dagger$. A subgroup of $\text{USp}(2n)$ could be gauged, in which case the commutant in $\text{USp}(2n)$ gives a flavour symmetry. The representations of the hypermultiplets under both the gauge and the flavor symmetry are consistent with the reality condition \eqref{ehyperchiral3dmap}. The flavour symmetry, or even the gauge symmetry if the corresponding gauge fields have Dirichlet conditions, might be broken by the boundary conditions above. The boundary term that needs to be added is
\begin{align}
S_{\text{hyp},\partial}=\int d^3x\sqrt{h}\bigg[-\frac{1}{8}\psi^I\psi_I+\frac{1}{8}\bar{\psi}^I\bar{\psi}_I+\frac{i}{4}\psi^I\sigma^4\bar{\psi}_I-2q_{A=1}^I(\phi_2)_I^{\ J}q_{J,A=2}\bigg]\,.
\end{align}
Like for the vector multiplet, also for the hypers the simple boundary conditions that we just mentioned can be deformed by interactions. For instance, if we perturb the boundary condition 1 by a boundary superpotential $W_\tt{3d}(q_{12},\hdots)$, where the dots refer to possible local 3d degrees of freedom, then the modified boundary condition is $q_{11}\propto \frac{\pd W_\tt{3d}}{\pd q_{12}}$. 
\subsection{Chiral primary operators and mixing}\label{SSec: chiral ring}
Let us now review some properties of the chiral primary operators in four-dimensional $\mathcal N=2$ conformal field theories.  These operators are annihilated by all conformal supercharges ($S_\alpha^A$, $\bar S_{\dot{\alpha}}^A$), and additionally by all supercharges of negative helicity
\begin{equation}
	[\bar{Q}_{\dot{\alpha}}^A , A] = 0\,.
\end{equation}
Similarly, their complex conjugate are annihilated by all the supercharges with positive helicity $[Q_{\alpha}^A , \bar{A}] = 0$ and are called anti-chiral. The superconformal algebra relations imply that chiral operators are scalars with respect to the $\text{SU}(2)_R$ symmetry, and we will also take them to be Lorentz scalars. Furthermore, their conformal dimension is fixed in terms of their $\text{U}(1)_R$-charges
\begin{equation}
	\Delta[A] = R/2\,,\quad \text{and}\quad \Delta[\bar{A}] = -R/2\,.
\end{equation}
Chiral primary operators sit in an $\mathcal N=2$ chiral multiplet whose content is provided in Table~\ref{Tab: N=2 chiral with w=2}. In the case that $\Delta = 2$ the top component, denoted by $C$, is exactly marginal.

The Operator Product Expansion (OPE) of chiral primaries is non-singular, and therefore these operators form a ring. Take $A_I$ to be a set of superconformal chiral primaries, then we have, up to $\bar Q$-exact terms
\begin{equation}\label{enonsingularOPE}
\underset{x\to 0}{\lim}	A_I(x) A_J(0) = C^K_{IJ} A_K(0) \,.
\end{equation}
These operators generate the Coulomb branch of the theory, and for this reason they are also referred to as Coulomb branch operators. In the subsequent sections we will study their correlation functions both on the hemisphere $HS^4$ and on the half-space $H\mathbb{R}^4$. To map correlation functions between curved and flat space one has to be careful about  Weyl anomalies. This phenomenon was studied for $\mathcal N=2$ theories on $S^4$ in \cite{Gerchkovitz:2016gxx}, where it was showed that the non-zero curvature invariants generate mixings on $S^4$ between chiral primaries of different scaling dimensions, and it was explained how to undo the mixing using a Gram-Schmidt procedure. This procedure applies also to our setup of $HS^4$, indeed since the curvature invariants that cause the mixing are local they are unaffected by the presence of the boundary. This implies that the same linear combinations of operators that undo the mixing on $S^4$ can be used on $HS^4$ to compute un-mixed bulk correlation functions in $H\mathbb{R}^4$.
\begin{table}
	\centering
	\begin{tabular}{ccccccc}
		\hline
		Field & $A$ & $\Psi$ & $B$ & $G_{ab}^-$ & $\Lambda$ & $C$  \\
		$j$ & $0$ & $1/2$ & $0$ & $1$ & $1/2$ & $0$ \\
		$\Delta$ & $w$ & $w+1/2$ & $w+1$ & $w+1$ & $w+3/2$ & $w+2$ \\
		$\text{U}(1)_R$ & $w$ & $w-1/2$ & $w-1$ & $w-1$ & $w-3/2$ & $w-2$ \\
		$\SU(2)_R$ & $\mathbf{1}$ & $\mathbf{2}$ & $\mathbf{3}$ & $\mathbf{1}$ & $\mathbf{2}$ & $\mathbf{1}$\\
		\hline
	\end{tabular}	
	\caption{Field content of a four-dimensional $\mathcal N=2$ chiral multiplet. The second, third, and fourth rows respectively present the spin $j$, conformal dimensions $\Delta$, $\text{U}(1)_R$-charges, and $\text{SU}(2)_R$ representations of the individual fields.}\label{Tab: N=2 chiral with w=2}
\end{table}
\subsection{Supersymmetric localization on the hemisphere}\label{SSec: supersymmetric localization on HS4}
Let us review some known results about localization on $HS^4$ \cite{Gava:2016oep,Dedushenko:2018tgx} that we will use in the following sections. 
Using the embedding coordinates
\begin{equation}
	\mathbf{z} = (\mathbf{z_\parallel}, z_{\perp})\,,\quad \text{such that}\quad \mathbf{z}\cdot \mathbf z = 1 \,,~ \text{with}~0 \leq z_{\perp} \leq 1~,
\end{equation}
where $\mathbf{z_\parallel}$ are the coordinates parallel to the boundary located at $z_\perp = 0$, the supercharge $Q$ used for localization can be defined as the one implementing the transformation with the following parameters
\begin{align}\label{eq:Qspinor}
		\ep_{\a A}  =\frac{\sqrt{1-z_\perp}}{2}\begin{pmatrix}
			1 & 0\\
			0 & 1
		\end{pmatrix}~,\ \ \ \bep^\da_{\ A}=\frac{i\sqrt{1+z_\perp}}{2}\begin{pmatrix}
			1 & 0\\
			0 & -1
		\end{pmatrix}~,
\end{align}
that are a solution of the Killing spinor equation \eqref{eksemain} and satisfy the boundary condition~\eqref{eq:3dcondKS}. 

In \cite{Gava:2016oep} it was shown that the supersymmetric locus for the $\mathcal N=2$ vector multiplet is the same as on $S^4$, namely
\begin{equation}\label{eq:loclocvec}
	\phi = \bar\phi = -\frac{i}{2} a\,,\quad D^{AB} = - a (\tau_3)^{AB}\,, \quad \text{and} \quad F_{\mu\nu} = 0\,.
\end{equation}
The parameter $a$ takes values in the Lie algebra of the gauge group. Moreover, there can be contributions from instantons localized at the pole of the $HS^4$. At the supersymmetric saddle the path integral reduces significantly, and the final expression depends on the boundary conditions chosen on $HS^4$. 

Let us first consider the case of a Dirichlet boundary condition for the gauge field, deformed by the addition of boundary degrees of freedom in a 3d $\mathcal{N}=2$ theory, in such a way that a subgroup $G$ of the global symmetry of the boundary degrees of freedom is identified with the global symmetry $G$ of the Dirichlet condition (this can be achieved through boundary superpotential couplings). Without loss of generality, we can choose $a$ in the Cartan. In this case the partition function can be written as following product:
\begin{equation}\label{eq:Dpfgen}
	\begin{aligned}
		\mathcal Z^{\text{D}}_{HS^4} ( \tau,a , z_i ) =&\, Z^{\text{D}}_{\text{vector}}(\tau,a) \cdot Z_{\text{hypers}}^{1,2}(a)  \cdot Z_{\text{inst}}(\tau,a) \cdot Z_{3\rmd}(a,z_i)\,.
	\end{aligned}
\end{equation}
Let us detail the form of each factor in \eqref{eq:Dpfgen}:
\begin{itemize}

\item The contribution of the bulk vector is given by
\begin{equation}\label{eq:Dpf}
	Z^{\text{D}}_{\text{vector}}(\tau,a) = e^{i \pi \tau\, \text{Tr}\,a^2}\prod\limits_{\alpha \in \Delta^+} H(i a \cdot \alpha) \frac{a\cdot \alpha}{\sinh (\pi a \cdot \alpha)}\,,
\end{equation}
where the product is taken over the set of positive roots $\Delta^+$ of the gauge group, and the function $H(x)$ is determined in terms of Barnes $G$-functions
\begin{equation}
	H(x) = G(1+x) G(1-x)\,.
\end{equation}
Note that the classical action contributes a term holomorphic in $\tau$, namely $e^{i \pi \tau\, \text{Tr}\,a^2}$, as opposed to the contribution on the full $S^4$ which only contains the imaginary part of $\tau$. This is due to the contribution of the boundary term \eqref{Eq: explicit boundary terms}, and ultimately to the fact that the parameter $\theta$ is visible even classically on a space with a boundary. 

\item The hypermultiplet contribution depends on the choice of boundary condition  $1$ or $2$. Fixing $\Phi_{i1}$ and $\Phi_{i2}$ to transform under the representation $R$ of the gauge symmetry with weights $w$, the result is
\begin{align}
\begin{split}
Z_\tt{hypers}^1=\underset{w\in R}{\prod}\frac{1}{H^{\frac{1}{2}}(i a\cdot w)} e^{-\frac{1}{2}\ell(i a\cdot w)}\,,~Z_\tt{hypers}^2=\underset{w\in R}{\prod}\frac{1}{H^{\frac{1}{2}}(i a\cdot w)} e^{\frac{1}{2}\ell(i a\cdot w)}~,
\end{split}
\end{align}
with $\ell$ being the following function 
\begin{align}\label{eq:ellFnct}
\ell(z)=-z\log(1-e^{2\pi i z})+\frac{i}{2}\big(\pi z^2+\frac{1}{\pi}\tt{Li}_2(e^{2\pi i z})\big)-\frac{i\pi}{12}\,.
\end{align}

\item $Z_{\text{inst}}(\tau,a)$ is Nekrasov's instanton partition function \cite{Nekrasov:2002qd} with Omega deformation parameters $\epsilon_{1,2}$ identified with the inverse radius of the sphere. This function receives contributions from instanton configurations localized at the north pole. On $S^4$ there is also a complex conjugate contribution coming from the south pole, and as a result the squared modulus of $Z_{\text{inst}}(\tau,a)$ appears in the $S^4$ partition function. On $HS^4$ instead only the holomorphic function of $\tau$ appears. Both in the contribution of the classical action and in the instanton contribution we see that restricting to the hemisphere makes the partition function holomorphic in $\tau$. However this holomorphicity is broken in the SCFT boundary partition function for reasons we discuss in the next section.

\item The factor $Z_{3\rmd}(a,z_\kappa)$ denotes the $S^3$ partition function of the boundary degrees of freedom. For Lagrangian theories it can also be computed using localization \cite{Jafferis:2010un}. The partition function depends on the fugacity $a$ for the global $G$ symmetry. It moreover depends on the fugacities $u_i$ for the remaining continuous global symmetries of the 3d theory.\footnote{A fugacity is a vev for the scalar component of a background vector multiplet associated to a global U(1) symmetry.}
The $\mathcal{N}=2$ 3d partition function depends finally on the coefficients $\Delta_\kappa$ that defines the IR $R$-symmetry: in fact, the 3d $R$-charge is a linear combination of the $\text{U}(1)$ symmetries:
\begin{equation}
R = \sum_{\kappa} \Delta_\kappa R_\kappa\,,
\end{equation}
where $R_0$ is a reference $R$-charge and $\Delta_0=1$. 
In \cite{Jafferis:2010un} it was found that the sphere 3d partition functions depend on $u_\kappa$ and $\Delta_\kappa$ only through the complex combination $z_\kappa=\Delta_\kappa - i u_\kappa$ and that $Z_{3d}$ is actually holomorphic in $z_\kappa$. In principle one should also consider the mixing coefficient $\Delta_\kappa$ related to the $G$ symmetry. If however $G$ does not contain abelian factors then it cannot mix with the 3d $R$-symmetry under the assumption that the boundary condition preserves the entire $G$.

\end{itemize}

Next, we consider the Neumann boundary condition. Also in this case we consider a possible deformation due to the presence of local 3d degrees of freedom. This 3d theory is assumed to have a $G$ global symmetry and the bulk-boundary interaction is defined by gauging the $G$ symmetry with the boundary mode of the bulk gauge field. Since the latter is still fluctuating on the boundary, the localization formula in this case takes the form of an integral over the Cartan-valued matrix $a$\footnote{More generally, also global symmetries acting on the hypermultiplets might contain $U(1)$ factors that are preserved by the boundary condition and can mix with the $R$ symmetry. In that case we need to introduce appropriate $z_\kappa$'s also in the hypermultiplet partition function.}
\begin{align}
\mathcal Z^{\text{N}}_{HS^4} ( \tau,z_\kappa )=\int \rmd a\, D(a)\,Z^{\text{N}}_{\text{vector}}(\tau,a) \cdot Z_{\text{hypers}}^{1,2}(\tau,a) \cdot Z_{\text{inst}}(\tau,a) \cdot Z_{3\rmd}(a,z_\kappa) \, ,
\end{align}
where the only new ingredients compared to the Dirichlet case are the Vandermonde determinant $D(a)=\prod\limits_{\alpha \in \Delta^+}(\alpha\cdot a)^2$ and the contribution of the vector multiplet, namely
\begin{align}
Z^{\text{N}}_{\text{vector}}(\tau,a)= e^{i \pi \tau\, \text{Tr}\,a^2}\prod\limits_{\alpha \in \Delta^+} H(i a \cdot \alpha) \frac{\sinh (\pi a \cdot \alpha)}{a\cdot \alpha}\,.
\end{align}

\paragraph{Abelian bulk theory} In the special case of the abelian theory, the product over roots is absent and one gets $Z^{\text{D}}_{\text{vector, abelian}}(\tau,a) =Z^{\text{N}}_{\text{vector, abelian}}(\tau,a)= e^{i \pi \tau a^2}$, where $a$ is a real number. Let us focus on the case of a Neumann boundary condition, with boundary degrees of freedom that form a certain 3d $\mathcal{N}=2$ CFT with a $\text{U}(1)$ symmetry, coupled to the bulk by gauging the $\text{U}(1)$ symmetry with the boundary components of the bulk gauge field. The resulting modified Neumann boundary condition is
\begin{equation}
\frac{1}{g^2}(\mathcal{J}_\text{N} + \gamma \mathcal{J}_\text{D}) = \mathcal{J}_\text{CFT}~,
\end{equation}
where $\mathcal{J}_\text{CFT}$ denotes the linear multiplet of the $\text{U}(1)$ symmetry that is gauged by the bulk gauge field. We can think of this boundary condition as setting to zero a linear combination of the two global symmetry with currents $\mathcal{J}_\text{N} + \gamma \mathcal{J}_\text{D} $ and $\mathcal{J}_\text{CFT}$, while an independent linear combination survives as a U(1) global symmetry of the boundary condition. The partition function of the 3d theory is denoted as $Z_{3\rmd}(a,z_\kappa)$, where $a$ is the fugacity for the U(1) symmetry coupled to the bulk gauge field, and the $z_\kappa$ denote fugacities for additional U(1) symmetries. In addition, we can turn on a fugacity $a'$ for the global symmetry with current $\mathcal{J}_D$ (see foonote \ref{foot:JD}), and a fugacity $x'$ for the surviving combination of $\mathcal{J}_\text{N}$ and $\mathcal{J}_\text{CFT}$, which can take for definiteness to be simply $\mathcal{J}_\text{CFT}$. The resulting partition function is \cite{Gaiotto:2014gha}
\begin{align}\label{eq:Npf}
\begin{split}
\mathcal Z^{\text{N}}_{HS^4,\text{abelian}}( \tau, a' , x',z_\kappa) &  = \int \rmd a \,e^{-2\pi i a a'}\, e^{i \pi \tau a^2}\, Z_{3\rmd}(a - x',z_\kappa)  ~.
\end{split}
\end{align}
The two symmetries with fugacities $a'$ and $x'$ descend from magnetic and electric one-form symmetries of the bulk theory, and therefore can be called magnetic and electric U(1) symmetries. We will see later that they are indeed exchanged by electric-magnetic duality. Note that the instanton contribution in the abelian case is $a$-independent \cite{Pestun:2007rz}, and  it only contributes an overall factor holomorphic in the $\tau$ that can be reabsorbed by a counterterm, see \eqref{eonlyholomorphic}. As a result it does not contribute to physical quantities and it can be omitted.

Also the magnetic and the electric U(1)'s, like any other U(1) symmetry of the boundary condition, can in principle mix with the 3d R-symmetry. The mixing parameters, which we will denote as $\tilde{\Delta}$ and $\hat{\Delta}$, respectively, appear together with $a'$ and $x'$ in a holomorphic combination in the partition function. The expression of $\mathcal Z^{\text{N}}_{HS^4,\text{abelian}}$ where $\tilde{\Delta}$ and $\hat{\Delta}$ explicitly appear is then
\begin{align}\label{eq:NpfDeltatilde}
\begin{split}
\mathcal Z^{\text{N}}_{HS^4,\text{abelian}}( \tau, a' - i \tilde{\Delta} , x' - i \hat{\Delta},z_\kappa) 
& = \int \rmd a \,e^{-2\pi i a a'}\,e^{-2\pi  a \tilde{\Delta}}\, e^{i \pi \tau a^2}\, Z_{3\rmd}(a - x' + i\hat{\Delta},z_\kappa)  ~.
\end{split}
\end{align}

\subsection{$F_\partial$-maximization}\label{Sec:Fmaximiz}
So far we have given formulas for general $\mathcal{N}=2$ bulk gauge theories, but now let us further suppose that the bulk theory is conformal, and the gauge coupling $\tau$ is an exactly marginal parameter. There are many examples in this class, to mention a few: the abelian theory, $\SU(N)$ SQCD with $2N$ flavors, $\mathcal{N}=4$ SYM and its orbifold. 

When the bulk theory is conformal the symmetry on $S^4$ is $\SL(1|2,\mathbb{H})$. In this case we can require that the boundary condition on the $S^3$ preserves the maximal possible subgroup, namely $\OSp(2|2,2)$, making it a $\tfrac{1}{2}$-BPS superconformal boundary condition. 
This will happen only when the $\text{U}(1)_R$ symmetry of the 3d $\mathcal{N}=2$ algebra preserved by the boundary is the superconformal $R$-symmetry. As we have said above, different values of the mixing parameters $\Delta_\kappa$ correspond to different choices of the 3d $\text{U}(1)_R$ symmetry. 
A criterion to determine the superconformal values of $\Delta_\kappa$ was given in \cite{Gaiotto:2014gha}: they are the values that maximize the following boundary free energy
 \begin{equation}\label{Eq: boundary F}
	F_\partial = - \frac12 \log \frac{\mathcal{Z}_{HS^4} \bar{\mathcal{Z}}_{HS^4}}{\mathcal{Z}_{S^4}} ~.
\end{equation}
Here $\bar{ \mathcal{Z}}_{HS^4}$ is simply the complex conjugate of the hemisphere partition function, which  by reflection positivity can  equivalently be thought of as the partition function on the orientation-reversed hemisphere, we will come back to this point in Section \ref{SSec: Gaiottos free energy}. On the other side, $\mathcal{Z}_{S^4}$ is the partition function of the bulk theory on the full sphere. The modulus squared ensures that $F_\partial$ is a real function, and the normalization by $\mathcal{Z}_{S^4}$ ensures that contributions from bulk Weyl anomalies cancel. $F_\partial$ is sensitive only to the boundary modes of the bulk fields and possibly their interaction with boundary-localized degrees of freedom, so it is analogous to the quantity $F = -\log \mathcal{Z}_{S^3}$  defined for local 3d theories.  

The function $F_\partial$ evaluated at the maximum is an interesting observable of the boundary conformal field theory (BCFT), which can be defined for any BCFT as the quantity~\eqref{Eq: boundary F} in which the coupling of the BCFT to the curved space are fixed by Weyl rescaling. We use the subscript BCFT to distinguish partition functions before and after specifying $\Delta$ to the conformal value, if an ambiguity might arise.
 The procedure of maximization in the supersymmetric context is then a way to fix the curvature couplings compatibly with conformal symmetry. The latter are indeed related to the choice of $R$-symmetry because the background gauge field for the $R$-symmetry is in the same multiplet as the metric. An important property of $F_\partial$ for BCFT's is that it is monotonic under boundary RG flows, as conjectured in \cite{Gaiotto:2014gha} and proved in \cite{Casini:2023kyj} using entropic arguments. 

When we apply $F_\partial$-maximization to our setup, the resulting value of $\Delta_\kappa$ is a real function of the marginal coupling, that we write as $\Delta_\kappa^{\text{max}}(\tau,\bar{\tau})$. This has a very important consequence: unlike what one might have thought from the expressions \eqref{eq:Dpf}-\eqref{eq:Npf}, the hemisphere partition function of the BCFT is {\it not} a holomorphic function of the complexified gauge coupling $\tau$, namely
\begin{equation}\label{eq:BCFTZ}
\mathcal Z_{HS^4, \, \text{BCFT}} ( \tau,\bar{\tau})=\mathcal Z_{HS^4} ( \tau)\vert_{\Delta_\kappa=\Delta_\kappa^{\text{max}}(\tau,\bar{\tau})} ~.
\end{equation}
The additional $(\tau,\bar{\tau})$ dependence due to $\Delta_\kappa^{\text{max}}(\tau,\bar{\tau})$ will play an important role in the following when we will use the BCFT partition function to extract BCFT one-point functions in $H\mathbb{R}^4$.

\section{One-point functions from $\mathcal Z_{H S^4}$}\label{Sec: Ward identities on HS4}
In this section we derive the relation between the derivatives of the hemisphere partition function and the one-point function of the Coulomb branch operators, which we can then use to obtain the BCFT one-point functions in flat space. We show two derivations, one that is an adaptation of the approach used in \cite{Gerchkovitz:2016gxx} to compute two-point functions $\langle A \, \bar{A}\,\rangle$, and a second one based purely on localization.

\subsection{SUSY Identities for the derivatives of $\mathcal{Z}_{HS^4}$}\label{ssusyidentities}
The Coulomb branch operator is the bottom component of a $\mathcal{N}=2$ chiral multiplet. We consider a deformation of the BCFT by the bulk coupling associated to the top component of this multiplet. The
$\OSp(2|2)\times \SU(2)$ preserving deformation on $HS^4$ is		
\begin{equation}\label{Eq: deforming with C}
	S_{\text{CFT}} \rightarrow S_\text{CFT} + \int_{H S^4} \rmd^4 x \sqrt{g}\,(\lambda\, \mathcal C + \bar{\lambda}\, \bar{\mathcal C})+ \int_{S^3} \rmd^3 x \sqrt{h}\,(\lambda\, \mathcal C_\partial + \bar{\lambda}\, \bar{\mathcal C}_\partial)\,.
\end{equation} 
The bulk action is the integral of $\OSp(2|4)$ invariant Lagrangian densities that can be expressed in terms of the components of the chiral multiplet, showed in Table \ref{Tab: N=2 chiral with w=2}, as (see~\cite{Gerchkovitz:2016gxx}) 
\begin{align}\label{Eq: definition of cal C}
\begin{split}
		\mathcal C & = C + i ( w - 2 ) (\tau_3)^{AB} B_{AB} + 2 (w - 2)( w - 3) A\,,\\
		\bar{\mathcal C} & = \bar{C} + i ( w - 2 ) (\tau_3)^{AB} \bar{B}_{AB} + 2 (w - 2)( w - 3) \bar{A}\,.
\end{split}
\end{align}
Here the matrices $\tau_{1,2,3}$ are the generators of the $SU(2)_R$ symmetry, see Appendix~\ref{app:conv}, and $\lambda$ is a complex coupling of mass dimension $2-w$. The fact that \eqref{Eq: definition of cal C} are supersymmetric Lagrangians on $S^4$ means that their supersymmetry variation is a total derivative. Indeed using the variations \eqref{eq:Chivar}-\eqref{eq:AntiChivar}, specified to the sphere via \eqref{Eq: background sugra values} and \eqref{eksemain}, we obtain
\begin{align}\label{edeltaC}
\begin{split}
\d\cC&=-D_\mu\big(2i \bep^A\s^\mu\Lambda_A-2(\o-2)(\tau_3)^{AB}\bep_{(A}\bs^\mu\Psi_{B)}\big)\,,\\
\d\bar{\cC}&=-D_\mu\big(2i\ep^A\s^\mu\bar{\Lambda}_A+2(\o-2)(\tau_3)^{AB}\ep_{(A}\s^\mu\bar{\Psi}_{B)}\big)\,.
\end{split}
\end{align}
As a result the supersymmetric variation of the bulk action on $HS^4$ gives a boundary term on $S^3$. Using the same variations, supplemented by the boundary condition \eqref{eq:3dcondKS} on the Killing spinors, one can check that the boundary term is canceled by the variation of the following boundary Lagrangian
\begin{align}
\begin{split}
\mathcal{C}_\partial & = 2 D_\perp A -i (\tau_3)^{AB} B_{AB} - 4 (w-2)A~,\\
\bar{\mathcal{C}}_\partial & = 2 D_\perp \bar{A} +i (\tau_3)^{AB} \bar{B}_{AB} + 4 (w-2)\bar{A}~.\\
\end{split}
\end{align}
 
The case $w=2$ is special because the deformation is marginal and preserves superconformal symmetry. The correlation functions of the marginal operators compute interesting geometric quantities on the conformal manifold \cite{Gomis:2014woa, Gerchkovitz:2014gta}, and this also applies to the boundary one-point function as we discuss in Section \ref{SSec: CFT distances}. The gauge coupling in conformal gauge theories is an example of such an exactly marginal coupling, and it is obtained from the chiral and antichiral multiplets with bottom components $A = \Tr[\phi^2]$ and $\bar{A} = \Tr[\bar{\phi}^2]$, respectively, and with the identification of couplings 
\begin{align}\label{elambdatotau}
\lambda =  i \frac{\tau}{8 \pi}\,.
\end{align}
One can check that making these substitutions in \eqref{Eq: deforming with C} reproduces the bulk action \eqref{eactiononS4} and also the boundary term \eqref{eactiononHS4}.\footnote{In order to match precisely with \eqref{eactiononS4} and \eqref{eactiononHS4} some integrations by parts are needed.} For this reason in the following we will use the letter $\lambda$ when referring to an arbitrary chiral deformation, and $\tau$ in the specific case of an exactly marginal one. 

Taking derivatives of the partition function with respect to the coupling $\lambda$ gives integrated one-point functions 
\begin{align}
\begin{split}\label{Eq: one-point functions on HS4}
\partial_\lambda(- \log \mathcal{Z}_{HS^4} )& =  \int_{HS^4} \rmd^4 x \sqrt{g} \,\langle \mathcal C \rangle + \int_{S^3} \rmd^3 x \sqrt{h}\, \langle \mathcal C_\partial \rangle~, \\
\partial_{\bar{\lambda}}(- \log \mathcal{Z}_{HS^4} )& = \int_{HS^4} \rmd^4 x \sqrt{g}\,  \langle \bar{\mathcal C} \,\rangle + \int_{S^3} \rmd^3 x \sqrt{h}\, \langle \bar{\mathcal C}_\partial \rangle~.
\end{split}	
\end{align}
Let us focus first on the bulk integral term.
The chiral deformations given in \eqref{Eq: definition of cal C} can be rewritten as a total derivative, up to a $Q$-exact contribution\footnote{Our expression for $\mathcal{C}$ as a total derivative plus a $Q$-exact term, while analogous to the one presented in \cite{Gomis:2014woa}, does not agree in details with it. The results in \cite{Gomis:2014woa} are not affected because the total derivative term that is singular at the pole is identical to ours. However the contribution on the other boundary $S^3$ is affected by this difference.}
\begin{equation}\label{Eq: rewriting C in terms of A}
\begin{split}
		\mathcal C =& D_\mu \bigg[2 B_{AB}\frac{\epsilon^A\sigma^\mu \bar{\epsilon}^B}{\epsilon^A \epsilon_A}+2i\frac{\tau_{3,AB}\epsilon^A\sigma^\mu \bar{\epsilon}^B}{\epsilon^A\epsilon_A}\bigg((2w -3)-\frac{\bar{\epsilon}_A\bar{\epsilon}^A}{\epsilon^A\epsilon_A}\bigg)A-2\frac{\bar{\epsilon}_A\bar{\epsilon}^A}{\epsilon^A\epsilon_A} D^\mu A\bigg]\\
		&+\delta \bigg(-\frac{\epsilon^A\Lambda_A}{\epsilon^A\epsilon_A}-i(w-1)\frac{\tau_{3,AB}\epsilon^A\Psi^B}{\epsilon^A\epsilon_A}+\frac{D_\mu\Psi^A\sigma^\mu\bar{\epsilon}_A}{\epsilon^A\epsilon_A}\bigg)\,.
\end{split}
\end{equation}
Since the boundary condition preserves $Q$, the $Q$-exact term drops from the one-point function, and we are only left with the total derivative.
The argument of the total derivative however is singular at the north pole $z_\perp = 1$ and following \cite{Gomis:2014woa} we regulate it by removing a small ball of radius $\delta$ around the pole and taking $\delta \rightarrow 0$ after performing the integral. As a result the integral of the total derivative equals the sum of two boundary contributions: one at the $S^3$ of radius $\delta$ around the singularity which gives the same result as in \cite{Gomis:2014woa}, namely the insertion of the bottom component $A$ at the north pole $N$, and a new second one at the equatorial boundary $S^3$. We then obtain
\begin{equation}\label{Eq: final result ward}
	\int_{HS^4} \rmd^4 x \sqrt{g}\, \langle \mathcal C \rangle = 32 \pi^2 \langle A(N) \rangle - \int_{S^3} \rmd^3 x \sqrt{h}\, \langle -2 D_\perp A +i (\tau_3)^{AB} B_{AB} + 4 (w-2)A\rangle\,.
\end{equation}
Since bulk one-point functions diverge like a power-law when the operator approaches the boundary, this equation contains divergent contributions both from the integral on the left-hand side and from the restriction of the operators to the boundary on the right-hand side. Both of these singularities can be regulated by e.g. a minimal distance from the boundary. Using the relation between the one-point function of $\mathcal{C}$ and the one-point function of $A$ dictated by supersymmetry, these divergent contributions, as well as the finite terms, must match between the left and the right hand side. We have checked this explicitly in the case of $w=2$. 

The computation of the integrated one-point function of $\bar{\mathcal{C}}$ is analogous, and starts from the expression similar to \eqref{Eq: rewriting C in terms of A}
\begin{align}
\begin{split}
	\bar{\mathcal C} =& D_\mu\bigg[-2\bar{B}_{AB}\frac{\epsilon^A\sigma^\mu\bar{\epsilon}^B}{\bar{\epsilon}_A\bar{\epsilon}^A}-2i\frac{\tau_{3,AB}\epsilon^A\sigma^\mu\bar{\epsilon}^B}{\bar{\epsilon}_A\bar{\epsilon}^A}\bigg((2w-3)-\frac{\epsilon^A\epsilon_A}{\bar{\epsilon}_A\bar{\epsilon}^A}\bigg)\bar{A}-2\frac{\epsilon^A\epsilon_A}{\bar{\epsilon}_A\bar{\epsilon}^A}D^\mu\bar{A}\bigg]\\
	&+\delta\bigg(\frac{\bar{\epsilon}_A\bar{\Lambda}^A}{\bar{\epsilon}_A\bar{\epsilon}^A}+i(w-1)\frac{\tau_{3,AB}\bar{\epsilon}^A\bar{\Psi}^B}{\bar{\epsilon}_A\bar{\epsilon}^A}+\frac{D_\mu\bar{\Psi}_A\bar{\sigma}^\mu\epsilon^A}{\bar{\epsilon}_A\bar{\epsilon}^A}\bigg)\,.
\end{split}
\end{align}
The important difference with $\bar{\mathcal{C}}$ instead of $\mathcal{C}$ is that now the argument of the total derivative is not singular anywhere in northern hemisphere $z_\perp \geq 0$. The singularity on $S^4$ would be at the south pole $z_\perp=-1$. As a result in this case, after dropping the $Q$-exact term, we only receive a contribution from the equatorial boundary $S^3$. We then obtain
\begin{equation}
	\int_{HS^4} \rmd ^4 x \sqrt{g}\,\langle \bar{\mathcal C} \rangle = \int_{S^3} \rmd^3 x \sqrt{h}\, \langle -2 D_\perp \bar{A} -i (\tau_3)^{AB} \bar{B}_{AB} - 4 (w-2)\bar{A}\rangle\,.
\end{equation}
Plugging this relation back into \eqref{Eq: one-point functions on HS4} we finally obtain
\begin{align}\label{Eq: final 1-pt functions}
\begin{split}
		\partial_\lambda (-\log \mathcal{Z}_{HS^4} )&=  32 \pi^2 \langle A(N) \rangle ~,\\
		\partial_{\bar\lambda} (- \log \mathcal{Z}_{HS^4} )&= 0~.
\end{split}
\end{align}
Just like in the case of $S^4$ two-point functions of \cite{Gerchkovitz:2014gta}, the above results can be generalized to include the additional insertion of generic $Q$-invariant operators.

\subsection{Flat space one-point functions from SUSY identities}\label{SSec: Gaiottos free energy}

In this section we show how to combine the two Ward Identities \eqref{Eq: final 1-pt functions} in such a way to obtain the flat space one-point function from the BCFT partition function \eqref{eq:BCFTZ}. 

Using (\ref{Eq: final 1-pt functions}) we obtain
\begin{align}
\partial_\lambda (-\log \mathcal{Z}_{HS^4} )=32\pi^2\bk{A(N)}-\partial_\lambda (-\log \bar{\mathcal{Z}}_{HS^4} )\,.
\end{align} 
Defining
\begin{equation}
f_\partial = -\frac 12 \log(\mathcal{Z}_{HS^4,\text{BCFT}} \bar{\mathcal{Z}}_{HS^4,\text{BCFT}})~,
\end{equation}
we obtain
\begin{equation}\label{eq:1ptSUSYid}
\partial_\lambda f_\partial\vert_{\tau,\bar{\tau}} = 16 \pi^2 \langle A( N ) \rangle_{HS^4}~,
\end{equation}
where by the evaluation in $\tau,\bar{\tau}$ we mean that all the dimensionful couplings with $w\neq 2$ are turned off while the marginal deformations $\tau,\bar{\tau}$ corresponding to $w=2$ can take an arbitrary value, parametrizing a point on the conformal manifold. One might be worried about the special case in which we are computing the one-point function for a chiral with $w=2$, and $\lambda =i \frac{\tau}{8\pi}$ is one of the marginal couplings, because $\mathcal{Z}_{HS^4,\text{BCFT}}$ has additional dependence on $\tau$ through $\Delta^{\text{max}}_\kappa(\tau,\bar{\tau})$. On the other hand we can write this additional dependence as 
\begin{equation}\label{eq:maxcondDeltacanc}
\frac{1}{2\pi i} \frac{\partial \Delta_\kappa}{\partial \tau}  \partial_{\Delta_\kappa} f_{\partial}(\tau,\bar{\tau})\vert_{\Delta_\kappa = \Delta_\kappa^{\text{max}}}=\frac{1}{2\pi i} \frac{\partial \Delta_\kappa}{\partial \tau}  \partial_{\Delta_\kappa} F_{\partial}(\tau,\bar{\tau})\vert_{\Delta_\kappa = \Delta_\kappa^{\text{max}}}~,
\end{equation}
and this vanishes thanks to the maximization condition. Eq. (\ref{eq:1ptSUSYid}) allows us to compute the BCFT one-point functions on $HS^4$, and we added a subscript to stress that this is the result on the curved background. 

This is very close to our goal of obtaining the BCFT one-point functions in flat space, the only remaining task is to take care of the possible mixing induced by the curvature as we reviewed in Section \ref{SSec: chiral ring}. As we mentioned there, this can be done with the standard Gram-Schmidt procedure that is also used for correlation functions without a boundary. After subtracting the mixing, a Weyl transformation shows that the quantity we are left with is the coefficient 
\begin{equation}
a_A = y^{w} \langle A(y,x^i)\rangle_{H \mathbb{R}^4}~,
\end{equation} 
of the one-point function in the flat-space BCFT. Here $y\geq 0$ is the distance from the boundary at $y = 0$, and $x^{i=1,2,3}$ are the coordinates parallel to the boundary.

We can be more explicit in the special case of $w=2$. In this case the only possible mixing is between $A$ and the identity operator, and this can be undone simply by subtracting the one-point function of $A$ on $S^4$. Using (\ref{elambdatotau}) we obtain
\begin{equation}\label{eq:fromFpart}
w=2:~~ a_A(\tau,\bar{\tau}) = \frac{1}{2 \pi i} \partial_{\tau}\left( -\frac 12 \log\left(\frac{\mathcal{Z}_{HS^4,\text{BCFT}} \bar{\mathcal{Z}}_{HS^4,\text{BCFT}}}{\mathcal{Z}_{S^4}}\right)\right) = \frac{1}{2\pi i} \partial_{\tau} F_{\partial}(\tau,\bar{\tau})~.
\end{equation}
Therefore we see that, for a Coulomb branch operator residing in the same multiplet of an exactly marginal operator, the coefficient of the one-point function is directly related to the quantity $F_\partial$ that we can assign to the BCFT.

Another advantage of expressing the one-point function in terms of $F_\partial$ is that it makes the scheme-independence manifest. The scheme dependence of $\mathcal{Z}_{S^4}$ was analyzed in \cite{Gomis:2014woa}, where the addition of certain chiral density built out of curvature invariants was found to affect the coupling dependence as
\begin{equation}
\mathcal{Z}_{S^4}(\tau,\bar{\tau}) \to e^{f(\tau) + \bar{f}(\bar{\tau})}\mathcal{Z}_{S^4}(\tau,\bar{\tau})
\end{equation}
where $f(\tau)$ is an arbitrary holomorphic function. This matches with the interpretation of $\mathcal{Z}_{S^4}$ in terms of the K\"ahler potential on the conformal manifold, the ambiguity represents a K\"ahler transformation. It can be easily checked that evaluating the same chiral density on $HS^4$, with the appropriate boundary term as discussed above, gives the following transformation 
\begin{equation}\label{eonlyholomorphic}
\mathcal{Z}_{HS^4}(\tau,\bar{\tau}) \to e^{f(\tau)}\mathcal{Z}_{HS^4}(\tau,\bar{\tau})~.
\end{equation}
As a result this ambiguity cancels in the combination $F_\partial$.

\subsection{Multiple derivatives}\label{sec:mulder}

At least heuristically, we would expect that by taking multiple derivatives of the partition function we compute one-point functions of products of Coulomb branch operators (a concept that makes sense thanks to the non-singular OPE). Indeed by applying additional holomorphic derivatives to \eqref{Eq: final 1-pt functions} and using the non-singular OPE we obtain
\begin{align}
\begin{split}
		\partial_\lambda^n (-\log \mathcal{Z}_{HS^4} )&=  (32 \pi^2)^n \langle A^n(N) \rangle ~,
\end{split}
\end{align}
for any $n\geq 1$. However there is an important difference between $n=1$ and $n>1$. In the first case, we showed in the previous section how to express the result directly in terms of the BCFT partition function, evaluated at $\Delta_\kappa = \Delta_\kappa^\text{max}(\tau,\bar{\tau})$. For $n>1$ derivatives, instead, there is no analogous simple manipulation that allows to cancel the additional dependence on the marginal parameter due to the maximization, and re-express this result just in terms of derivatives of the BCFT partition function. One possible way to proceed is to evaluate the result at $\Delta_\kappa = \Delta_\kappa^\text{max}(\tau,\bar{\tau})$ after taking the derivative. However the result obtained in this way is not manifestly scheme-independent, unlike the expression for $n=1$ found in the previous section. 

In the next section we will use the explicit localization results, which apply in a specific
scheme and have a completely determined $\tau$-dependence, to compute the one-point functions of products of chiral operators. This will confirm the procedure of taking multiple derivatives of the localized partition function and afterwards plugging in the extremized BCFT values of $\Delta_\kappa$.

\subsection{Flat space one-point functions from localization}\label{sec:oneptLoc}

\paragraph{One-point functions on $HS^4$} While the approach in the last section was just based on symmetry and did not assume any Lagrangian formulation or the existence of a path integral that localizes to a matrix model, it is also natural to derive the expression for the one-point functions using the localization formulas for the theory on $HS^4$. In this approach, one uses that in a $\mathcal{N}=2$ gauge theory the general Coulomb branch operator~$A_{n_1,\dots,n_k;m_1,\dots,m_k} = \prod_{i=1}^k\mathrm{Tr}[\phi^{n_i}]^{m_i}$, labeled by the $k$-tuples of positive integers $n_i$ (all distinct) and $m_i$ (not necessarily distinct), at the localization locus \eqref{eq:loclocvec} reduces to the monomial~$\prod_{i=1}^k\mathrm{Tr}[(-\frac i 2 a)^{n_i}]^{m_i}$ of the integration variable $a$. More precisely, this is true if one ignores the contribution from instantons localized at the north-pole. Therefore, at least up to non-perturbative corrections, the one point function is directly computed by inserting the monomial in the matrix model. Let us now discuss how to connect this localization result to the previous approach based on taking derivatives of the partition function. This will also allow us to suggest how to include the non-perturbative corrections. Deforming the theory by couplings
\begin{equation}\label{edeformation}
\hspace{-0.09375cm} S_{\text{CFT}} \rightarrow S_\text{CFT} + \sum_{i=1}^k\left[\int_{H S^4} \rmd^4 x \sqrt{g}\,(\lambda_{n_i}\, \mathcal C_{n_i} + \bar{\lambda}_{n_i}\, \bar{\mathcal C}_{n_i})+ \int_{S^3} \rmd^3 x \sqrt{h}\,(\lambda_{n_i}\, \mathcal C_{n_i\,\partial} + \bar{\lambda}_{n_i}\, \bar{\mathcal C}_{n_i\,\partial})\right]\,,  
\end{equation} 
where $\mathcal C_{n_i}$ is the top component of the chiral multiplet with $\mathrm{Tr}[\phi^{n_i}]$ as bottom component, and $\mathcal C_{n_i\,\partial}$ is the associated boundary term, we have that the contribution of the classical action on $HS^4$ is given by\footnote{\label{freflpos}In order to compute \eqref{edeformation} on the supersymmetric locus we use that the integrals of $\mathcal{C}$ and $\bar{\mathcal{C}}$ evaluate to the insertions of $A$ and $\bar{A}$ at the pole, plus a boundary term that combines with the boundary action to give the total boundary contribution in \eqref{Eq: final 1-pt functions}. To evaluate the boundary contributions we use that 
\begin{align}
\begin{split}
&A_n=\text{Tr}[\phi^{n}]~,~~\bar{A}_n=\text{Tr}[(-\bar{\phi})^n]~,\\
&B_{AB\,n}= n \,\text{Tr}\big[D_{AB}\phi^{n-1}+\frac{i}{2}(n-1)\phi^{n-2}\lambda_A\lambda_B\big]\,,\\
&\bar{B}^{AB}_n= (-1)^n n\,\text{Tr}\big[D^{AB}\bar{\phi}^{n-1}-\frac{i}{2}(n-1)\bar{\phi}^{n-2}\bar{\lambda}^A\bar{\lambda}^B\big]\,.
\end{split}
\end{align}
The signs are due to the fact that in our conventions $\big(\phi(\mathbf{z_\parallel},z_\perp)\big)^*=-\bar{\phi}(\mathbf{z_\parallel},-z_\perp)$.}
\begin{equation}\label{eq:Zclassgen}
Z_{\text{class}}(\lambda,a) = e^{-32\pi^2 \sum_{i=1}^k \lambda_{n_i} \mathrm{Tr}[(-\frac i 2 a)^{n_i}]}~.
\end{equation}
We are using the entry $\lambda$ to indicate the dependence on the full set of couplings $\lambda_{n_i}$. Notice that here, unlike the case of the full $S^4$, the dependence on these couplings is holomorphic. 

 As a result we can trade the insertion of the monomial in the localization integral with derivatives of the classical action (ignoring for the moment instanton contributions, which we denote with subscript ``pert'')
\begin{align}
\begin{split}
&\langle A_{n_1,\dots,n_k;m_1,\dots,m_k}(N) \rangle_{HS^4,\text{pert}} \\ 
& = \frac{1}{\mathcal{Z}^{\text{N}}_{HS^4,\text{BCFT}}} \int\, da\, \prod_{i=1}^k\mathrm{Tr}[(-\tfrac i 2 a)^{n_i}]^{m_i}  Z_{\text{class}} (\tau,a) Z_{\text{1loop}}(a) Z_{\text{3d}}(a,\Delta^{\text{max}}(\tau,\bar{\tau})) \\
& = \frac{1}{\mathcal{Z}^{\text{N}}_{HS^4,\text{BCFT}}} \int\, da\, \Big(\prod_{i=1}^k\left.\left(-\tfrac{1}{32\pi^2}\tfrac{\partial}{\partial \lambda_{n_i}}\right)^{m_i} Z_{\text{class}}(\lambda,a)\Big)\right\vert_{\tau} Z_{\text{1loop}}(a)  Z_{\text{3d}}(a,\Delta^{\text{max}}(\tau,\bar{\tau}))~.
\end{split}
\end{align}
Here again the evaluation at $\tau$ denotes that all the dimensionful couplings are turned off while the marginal one denoted by $\tau$ is allowed to vary on the conformal manifold. This expression also suggests a way to include the non-perturbative corrections, by including the instanton partition function in the derivative\footnote{Furthermore this is also how these operators are inserted in extremal correlators, as explained in \cite{Gerchkovitz:2016gxx}.}, obtaining
 \begin{align}
\begin{split}
&\langle A_{n_1,\dots,n_k;m_1,\dots,m_k}(N)\rangle_{HS^4}   \\
& = \frac{1}{\mathcal{Z}^{\text{N}}_{HS^4, \text{BCFT}}} \int\, da\, \Big(\prod_{i=1}^k\left.\left(-\tfrac{1}{32\pi^2}\tfrac{\partial}{\partial \lambda_{n_i}}\right)^{m_i} Z_{\text{class}}(\lambda,a)Z_{\text{inst}}(\lambda,a)\Big)\right\vert_{\tau} \\ & \hspace{8cm} \times Z_{\text{1loop}}(a)  Z_{\text{3d}}(a,\Delta^{\text{max}}(\tau,\bar{\tau}))~.
\end{split}
\end{align}
We almost obtained a derivative of the full integrand, which would allow us to rewrite the final result as a derivative of the $HS^4$ partition function. The obstruction to do so is the additional dependence on $\tau,\bar{\tau}$ coming from the conformal value of $\Delta$, i.e. the function $\Delta^{\text{max}}$ obtained from the extremization. We can circumvent this problem by writing the derivative as acting on the partition function with generic $\Delta$, and plug in the conformal value only after the evaluation of the derivative. In this way we arrive at
\begin{equation}\label{eq:1ptloca}
\langle A_{n_1,\dots,n_k;m_1,\dots,m_k}(N)\rangle_{HS^4}  = \frac{1}{\mathcal{Z}_{HS^4,\text{BCFT}}^\text{N}} \left.\Big(\prod_{i=1}^k\left(-\tfrac{1}{32\pi^2}\tfrac{\partial}{\partial \lambda_{n_i}}\right)^{m_i}\mathcal{Z}^{\text{N}}_{HS^4}(\lambda,\Delta)\Big)\right\vert_{\tau,\Delta= \Delta^{\text{max}}}~.
\end{equation}

Note that for single-trace operators, when we have a single derivative acting on $\mathcal{Z}$, this formula coincides with \eqref{eq:1ptSUSYid}. This is because the additional dependence due to $\Delta_\kappa^{\text{max}}$, which is potentially dangerous for $w=2$, cancels thanks to the extremization condition, see the discussion around eq. \eqref{eq:maxcondDeltacanc}. Moreover this formula also gives a general result for multi-trace one-point functions.

\paragraph{From $HS^4$ to $H\mathbb{R}^4$} Having derived the one-point functions on $HS^4$, the remaining step is to perform correctly the Weyl rescaling to obtain the one-point functions for the BCFT in flat space $H\mathbb{R}^4$. As we have already reviewed, this amounts to applying a certain matrix to the vector of one-point functions. This matrix is the change of basis determined by the Gram-Schmidt procedure to the matrix of two-point functions $\langle A_I(N) \bar{A}_J(S)\rangle$ on $S^4$, without a boundary. In formulas, using $I,J,K,\dots$ as indices that run over the full set of Coulomb branch operators, we have
\begin{align}
\begin{split}
& U_I^{~L} {U^*}_J^{~M} \langle A_L(N) \bar{A}_M (S)\rangle_{S^4} = C_I \delta_{IJ} = \langle A_L(0) \bar{A}_M (\infty)\rangle_{\mathbb{R}^4}~,\\
& \Rightarrow U_I^{~L} \langle A_L(N) \rangle_{HS^4} =  y^{\Delta_I} \langle A_I (y, x^i) \rangle_{H\mathbb{R}^4} = a_{A_I}~.
\end{split}
\end{align}
We stress that the change of basis $U_I^{~J}$, unlike the vector $\langle A_I(N) \rangle_{HS^4}$, does not depend on the boundary condition, it is uniquely fixed by the data of the theory in $S^4$ and the choice of normalization $C_I$ in flat space.

\paragraph{Abelian bulk theory} We now specialize to the case of the boundary conditions for the free bulk CFT of a $\mathcal{N}=2$ vector multiplet. The $S^4$ partition function is simply a Gaussian integral
\begin{equation}
\mathcal{Z}_{S^4}(\tau,\bar{\tau}) =\int d a e^{-2\pi\,\mathrm{Im}\tau \,a^2} = \frac{1}{\sqrt{2 \Im \tau}}~,
\end{equation}
where $\tau$ is the exactly marginal gauge coupling. The Coulomb branch operators are
\begin{equation}
A_n = \phi^n~, ~~\bar{A}_n = (-\bar{\phi})^n = A_n^\dagger~,
\end{equation} 
i.e. powers of the generator $\phi$, $\bar{\phi}$ which is the scalar field in the vector multiplet (the choice of the \q{$-\bar{\phi}$} is explained in footnote \ref{freflpos}). For even $n$, their correlation functions are obtained taking derivatives w.r.t. to $\tau$ and $\bar{\tau}$. The matrix of two-point functions on the sphere is
\begin{equation}
\langle A_n (N) \bar{A}_m (S)\rangle_{S^4} =\frac{1}{\mathcal{Z}_{S^4}(\tau,\bar{\tau})}\int d a \, \left(-\tfrac i2 a\right)^n\, \left(\tfrac i2 a\right)^m\,e^{-2\pi\,\mathrm{Im}\tau \,a^2} ~.
\end{equation}
The Gram-Schmidt orthogonalization problem in this case is thus solved by finding a basis of polynomials which are orthogonal w.r.t. to the Gaussian measure. Hermite polynomials $H_n(x)$ satisfy this condition, so the change of basis is fixed by requiring 
\begin{equation}
\sum_{m=0}^n U_n^{~m} \left(-\tfrac i2 a\right)^m = \left( \tfrac{-i}{4\sqrt{2\pi \, \Im\,\tau}} \right)^n H_n(\sqrt{2\pi \, \Im\, \tau} a)~.
\end{equation} 
The normalization is such that $U_n^{~n} = 1$. This choice gives the following coefficient for the two-point function in flat space: $C_n = n!(2\pi \mathrm{Im}\tau)^n $. 

In a generic boundary condition given by Neumann deformed by gauging a 3d $\mathcal{N}=2$ SCFT, we therefore arrive at the following formula for the flat-space one-point function of the Coulomb branch operators
\begin{align}\label{Eq: 1-pt functions flat space}
\begin{split}
& a_{A_n} (\tau,\bar{\tau})= y^{n}	\langle A_n(y, x^i) \rangle_{H\mathbb{R}^4}(\tau,\bar{\tau}) \\
& = \frac{1}{\mathcal Z_{HS^4,\text{BCFT}}^\text{N}} \int\limits \rmd a\, \left( \tfrac{-i}{4\sqrt{2\pi \, \Im\,\tau}} \right)^n H_n(\sqrt{2\pi \, \Im\, \tau} a) e^{i  \pi \tau  a^2 } e^{2\pi a \tilde\Delta^{\text{max}}(\tau,\bar{\tau})}   Z_{3d}(a,\Delta^{\text{max}}(\tau,\bar{\tau}))~.
\end{split}
\end{align}
In the following we use interchangeably the notation $a_{A_n}$ and $\langle A_n \rangle_{H\mathbb{R}^4}$ to denote the coefficient of the one-point function in flat space with a boundary.

\subsection{$Z_{HS^4}$ and the geometry of the conformal manifold}\label{SSec: CFT distances}
In this section we make some remarks about the geometric interpretation of the hemisphere partition function as a function defined on the space of exactly marginal couplings, i.e. the conformal manifold of the CFT. The marginal couplings, denoted as $\lambda^I$ with an index $I$ running over the multiple independent couplings, define a system of coordinates on this manifold. Conformal manifolds are equipped with the Zamolodchikov metric, computed by the two-point correlation function of the marginal operators $\langle O_I(x) O_J(y) \rangle = g_{IJ}(\lambda) |x-y|^{-8}$ \cite{Zamolodchikov:1986gt, Seiberg:1988pf, Kutasov:1988xb}. For $\mathcal{N}=2$ SCFTs in 4d these conformal manifolds are further constrained to be K\"ahler-Hodge \cite{Papadodimas:2009eu,Gomis:2015yaa}, a result that has recently been extended to the case with only 4 supercharges \cite{Niarchos:2021iax}. A K\"ahler-Hodge manifold is a K\"ahler manifold with the property that the flux of the K\"ahler form on any 2 cycle is quantized, or equivalently that the K\"ahler form is the first Chern class of a holomorphic line bundle on the manifold.

The K\"ahler potential $K(\lambda^I,\bar{\lambda}^{\bar{I}})$ on the conformal manifold is directly related to the supersymmetric partition function of the theory on $S^4$ \cite{Gerchkovitz:2014gta}
\begin{equation}\label{eq:KalZ4d}
	\log Z_{S^4} = \frac{K(\lambda^I,\bar{\lambda}^{\bar{I}})}{12} \,.
\end{equation}
Given this direct relation between the geometry of the conformal manifold and the sphere partition function, it is natural to ask whether a similar connection exists for the hemisphere partition function. 

In two dimensions a similar question was studied in \cite{Bachas:2013nxa} for superconformal interfaces $\mathcal I_{12}$ separating two $\mathcal N = (2,2)$ SCFTs, denoted as CFT$_1$ and CFT$_2$. This setup can be placed on $S^2$, with the interface on the equator and the two CFTs on the two hemispheres, and an interface partition function analogous to the boundary one \eqref{Eq: boundary F} can be defined as
\begin{equation}\label{eq:interfaceF}
F_{\mathcal{I}_{12}} = -\frac 12 \log \frac{Z_{S^2,\mathcal{I}_{12}}\overline{Z_{S^2,\mathcal{I}_{12}}}}{Z_{S^2,\text{CFT}_1}Z_{S^2,\text{CFT}_2}}~.
\end{equation}
A relation similar to \eqref{eq:KalZ4d} is also valid for the K\"ahler potential on the moduli space of 2d $\mathcal N = (2,2)$ SCFTs, namely
\begin{equation}\label{eq:KalZ2d}
	\log Z_{S^2} = - K(\lambda^I,\bar{\lambda}^{\bar{I}}) \,.
\end{equation}
Ref. \cite{Bachas:2013nxa} generalized this relation to the case of an interface that separates two CFTs belonging to the same conformal manifold and only differing by the values of the moduli $\lambda^I_1$ and $\lambda^I_2$, either K\"ahler or complex structure moduli. In this case the interface sphere partition function computes an analytic continuation of the K\"ahler potential to independent values of the two entries $\lambda^I$ and $\bar{\lambda}^I$, in such a way that the first entry is computed at $\lambda^I_1$ and the second at $\bar{\lambda}^I_2$.
This leads to the following identity
\begin{equation}
 -2 F_{\mathcal{I}_{12}}	 = K(\lambda^I_1 , \bar\lambda^{\bar I}_1) + K(\lambda^I_2 , \bar\lambda^{\bar I}_2)  - K(\lambda^I_1 , \bar\lambda^{\bar I}_2) - K(\lambda^I_2 , \bar\lambda^{\bar I}_1) \equiv \mathcal{D}_{12}\,,
\end{equation}
where $\lambda^I_a$ are the points in moduli space giving CFT$_a$, and $K$ is the K\"ahler potential. The function appearing on the right-hand side is the so called Calabi's diastasis function $\mathcal{D}_{12}$ on the moduli space. This function exists on any K\"ahler manifold, and it is independent of any K\"ahler transformation \cite{Calabi_Diastasis}. 
This function can be thought of as measuring the distance between the two CFTs separated by the interface. When the moduli become infinitesimally close it indeed agrees with the geodesic distance on the K\"ahler manifold determined by the Zamolodchikov metric, more precisely denoting with $d_{12}$ the geodesic distance the relation is $\mathcal{D}_{12} = d_{12}^2 +\mathcal{O}(d_{12}^4)$ \cite{Calabi_Diastasis}. As proposed in \cite{Bachas:2013nxa}, the fact that the diastasis can be expressed as the interface partition function invites a generalization to general pairs of CFTs: minimizing the partition function \eqref{eq:interfaceF} over the set of interfaces connecting the two CFTs, one gets the following notion of distance between the two CFTs
\begin{equation}\label{Eq: Distance between CFTs}
	d(\text{CFT}_1,\text{CFT}_2) = \min\limits_{{\mathcal I}_{12}} \sqrt{-2 F_{\mathcal{I}_{12}}}\,.
\end{equation}
As observed already in \cite{Bachas:2013nxa} this quantity however fails to define a distance in the  rigorous mathematical sense because it does not satisfy positivity and the triangular inequality. 

What we said about interface in 2d $\mathcal{N} = (2,2)$ CFTs can be readily imported to the setup with $\frac12$-BPS interfaces between $\mathcal{N}=2$ CFTs in 4d. In fact a similar interface partition function can be defined \cite{Gaiotto:2014gha}
\begin{equation}\label{eq:interfaceF4d}
F_{\mathcal{I}_{12}} = -\frac 12 \log \frac{Z_{S^4,\mathcal{I}_{12}}\overline{Z_{S^4,\mathcal{I}_{12}}}}{Z_{S^4,\text{CFT}_1}Z_{S^4,\text{CFT}_2}}~.
\end{equation}
Using the 4d relation \eqref{eq:KalZ4d}, one gets that for interfaces defined by changing the coupling within the same conformal manifolds this function computes the Calabi diastasis \cite{Goto:2018zrp,Goto:2020per}, like in 2d.  More generally, we get a similar notion of ``distance'' between 4d $\mathcal{N}=2$ CFTs 
\begin{equation}\label{Eq: Distance between 4d CFTs}
	d(\text{CFT}_1,\text{CFT}_2) = \min\limits_{{\mathcal I}_{12}} \sqrt{24\, F_{\mathcal{I}_{12}}}\,.
\end{equation}
The case of a boundary CFT can be seen as a special case of the interface in which one of the two sides is the empty theory. Restricting the definition \eqref{Eq: Distance between 4d CFTs} to this case, one defines a function that assigns a number to any CFT
\begin{equation}
	||\text{CFT}|| = d(\text{CFT},\text{empty}) = \min_{\mathcal B} \sqrt{24 F_{\partial,\mathcal{B}}}\,.
\end{equation}
where now the minimization is over the set of boundary conditions $\mathcal{B}$.

An notable property of the geometry of the $\mathcal{N}=2$ conformal manifold is the existence of the $tt^*$-equations \cite{Cecotti:1991me, Papadodimas:2009eu}, which strongly constrain the dependence of observables on the marginal couplings, see e.g. \cite{Baggio:2015vxa}. It would be interesting to explore the interplay of the hemisphere partition function with the $tt^*$-equations.

\section{Boundary conditions and Dualities in the Abelian theory}\label{Sec:Dualities}
We now specify the bulk theory to be the CFT of an $\mathcal{N}=2$ vector multiplet. This theory enjoys the electric-magnetic duality group SL(2,$\mathbb{Z}$) that acts on the exactly marginal coupling $\tau$ as
\begin{equation}
\tau \to g[\tau] \equiv \frac{a\tau + b}{c \tau + d}~,~~g =\begin{pmatrix} a & b \\
c & d
\end{pmatrix} \in \text{SL}(2,\mathbb{Z})~.
\end{equation}
In this section we study the interplay between the SL(2,$\mathbb{Z}$) duality group and the boundary conditions, in particular at the level of the $HS^4$ partition function and bulk one-point functions.

\subsection{Action of SL(2,$\mathbb{Z}$) on the boundary conditions}
Let us briefly review the action of SL(2,$\mathbb{Z}$) on the bondary conditions for the $\mathcal{N}=2$ vector multiplets. Very similar results apply to the 4d free vector CFT for any number of suspersymmetries and have been extensively discussed in the literature also for $\mathcal{N}=0$ and $\mathcal{N}=4$ \cite{Witten:2003ya, Gaiotto:2008ak, Kapustin:2009av, Dimofte:2011ju, DiPietro:2019hqe}.

Consider a modified Neumann boundary condition, in which the boundary component of the bulk gauge field gauges the U(1) symmetry of a certain 3d $\mathcal{N}=2$ SCFT $\mathcal{T}_\text{3d}$. Let us denote this boundary condition as $B\left(\tau, \mathcal{T}_\text{3d}\right)$. The following duality holds
\begin{equation}\label{eq:Bduality}
B\left(\tau, \mathcal{T}_\text{3d}\right) \simeq B\left(g[\tau], g\left[\mathcal{T}_\text{3d}\right]\right)~.
\end{equation}
Here $g\left[\mathcal{T}_\text{3d}\right]$ is 3d SCFT with a U(1) symmetry obtained acting on $\mathcal{T}_\text{3d}$ with Witten's SL(2,$\mathbb{Z}$) action \cite{Witten:2003ya}. To define this action, it is sufficient to say what is the action of the $S$ and $T$ transformations. The action of $S$ consists in gauging the U(1) symmetry with 3d gauge fields, and defining the new U(1) global symmetry to be the topological U(1). The action of $T$ only changes the theory by shifting the contact term in the two-point function of the conserved U(1) current. This shift amounts to adding to the partition function of the theory with a background gauge field $A_\mu$ the Chern-Simons counterterm $\frac{1}{4\pi}\int \left(A d A +\dots\right)$, where the dots stand for the supersymmetrization according to the 3d $\mathcal{N}=2$ symmetry preserved by the boundary condition.

An important consequence of the duality \eqref{eq:Bduality} is that the boundary condition admits several decoupling limits, whenever $g[\tau]\to +i\infty$ for any $g$, in which one is left with a pure Neumann boundary condition for the bulk field in the appropriate duality frame, and a decoupled 3d SCFT $g\left[\mathcal{T}_\text{3d}\right]$ on the boundary.

\subsection{$\SL(2,\mathbb{Z})$ and the hemisphere partition function}

The $\SL(2,\mathbb{Z})$ action we have just described translates into specific operations done on the partition functions computed through localization, that we reviewed in Section~\ref{SSec: supersymmetric localization on HS4}. In particular we will use the modified Neumann boundary condition with partition function given in \eqref{eq:Npf}. To describe the action on this function, we now consider the action of the generators $T$ and $S$.

For sake of notation, in this section we will leave the variables $z_\kappa$ implicit, as they do not play any role in the duality transformations.

\paragraph{$T$-transformation}

As we have explained above, the $T$ element acts on the 3d $\text{CFT}_3$ by adding a Chern-Simons term for the background vector multiplet coupled to U(1) current. At the level of 3d partition function this simply means
\begin{equation}\label{eq:Taction3dPF}
Z_{T\left[ \mathcal{T}_{3d}\right]}(a) =   e^{-i \pi a^2} Z_{\mathcal{T}_{3d}}(a) \:.
\end{equation}
Inverting the relation \eqref{eq:Taction3dPF} and plugging the result into \eqref{eq:Npf}, one obtains
\begin{align}\label{eq:4dTdualPF}
\begin{split}
\mathcal Z^{\text{N}}_{HS^4}( \tau, a' ,x') & = \int d a \,{e}^{-2\pi i a a'}\, e^{i \pi \tau a^2 }\, Z_{T\left[\mathcal{T}_{3d}\right]}(a-x')\, e^{i\pi(a-x')^2} \\
 & = e^{i\pi(x')^2}  \int d a \,{e}^{-2\pi i a (a'+x')}\, e^{i \pi (\tau+1) a^2 }\, Z_{T\left[\mathcal{T}_{3d}\right]}(a-x')\, .
\end{split}
\end{align}

In \eqref{eq:4dTdualPF} we have obtained the expression for a 4d supersymmetric Maxwell theory coupled to a boundary CFT$_3$ $T\left[\mathcal{T}_{3d}\right]$ with gauge coupling $T[\tau]=\tau+1$. In particular we obtain the relation:
\begin{equation}\label{eq:Tduality4dPF}
 T\left[{\mathcal  Z}_{HS^4}^{\text{N}}\right](\tau+1 ,a'+x',x')  =  e^{-i\pi(x')^2} \mathcal  Z_{HS^4}^{\text{N}}(\tau , a',x') 
\end{equation}
where by $T\left[{\mathcal  Z}_{HS^4}^{\text{N}}\right]$ we mean the 4d partition function of the 4d Maxwell theory coupled to $T\left[\mathcal{T}_{3d}\right]$.

\paragraph{$S$-transformation}

We now turn to the action of the $S$ element. 
As we said above, the S-dual CFT$_3$ is obtained by taking $\mathcal{T}_{3d}$ and coupling the U(1) symmetry to a dynamical 3d gauge boson. Localization then tells us that the partition functions of the dual 3d theories are connected by a Fourier transform:
\begin{align}\label{eq:3dDualPF}
\begin{split}
 Z_{S\left[\mathcal{T}_{3d}\right]}( x ) & = \int d a \,{e}^{-2\pi i a x}\,  Z_{\mathcal{T}_{3d}}(a)  ~.
\end{split}
\end{align}
We would expect that the 4d partition function \eqref{eq:Npf} can be rewritten in an analogous form, but in terms of $Z_{S\left[\mathcal{T}_{3d}\right]}( x )$. We now show that this is indeed the case. By inverting \eqref{eq:3dDualPF}, and inserting the result for $Z_{\mathcal{T}_{3d}}( a -x' )$ we obtain
\begin{equation}\label{eq:4dSdualPF}
	\begin{aligned}
		\mathcal Z_{HS^4}^{\text{N}}(\tau ,  a',x') =&\, \int \rmd a\,e^{-2\pi i a  a'} \, e^{i \pi \tau a^2} \, \int \rmd x \,e^{2\pi i (a-x') x}  Z_{S\left[\mathcal{T}_{3d}\right]}(x)\\
		=& \,\frac{1}{\sqrt{-i \tau}} \int \rmd x \, e^{-2\pi i x  x'}e^{-\frac{i \pi(x-a')^2}{\tau}}  Z_{S\left[\mathcal{T}_{3d}\right]}(x)\\ 
		=& \,\frac{e^{-2\pi i a'x'}}{\sqrt{-i \tau}}  \int \rmd x \,e^{-2\pi i x  x'}e^{- \frac{i \pi x^2}{\tau}}  Z_{S\left[\mathcal{T}_{3d}\right]}(x+a'),	
\end{aligned}
\end{equation}
where we did the following steps: in the second line we performed the Gaussian integral over the variable $a$; in the third line we shifted the integration variable $x\mapsto x+a'$.

In \eqref{eq:4dSdualPF} we have obtained the expression for a 4d supersymmetric Maxwell theory coupled to a boundary CFT$_3$ $S\left[\mathcal{T}_{3d}\right]$ with gauge coupling $S[\tau]=-1/\tau$. In particular we obtain the relation:
\begin{equation}\label{eq:Sduality4dPF}
 S\left[ {\mathcal  Z}_{HS^4}^{\text{N}} \right] (-1/\tau  , x', - a') = \sqrt{-i\tau}  e^{2\pi i a'x'} \mathcal  Z_{HS^4}^{\text{N}}(\tau  , a', x') 
\end{equation}
with the two fugacities $a'$ and $x'$ exchanged. Notice that the factor $\sqrt{-i\tau}\,  e^{2\pi i a'x'}$ cancels out of $F_\partial$, that then is left invariant under S-duality.

\

\noindent
{\bf Simple Example.}$\,\,\,$
We now consider a simple case, i.e. $\mathcal N=2$ supersymmetric Maxwell theory with Neumann boundary conditions and no 3d degrees of freedom. In this case $Z^{N}_{3\rmd}(a)=1$. 
Moreover there is no U(1) symmetry in the 3d theory (that is empty), hence we need to set $x'=0$ in the 4d partition function, that is:
\begin{align}\label{eq:NpfQED}
\begin{split}
\mathcal Z^{\text{N}}_{HS^4}( \tau,  a') & = \int d a \,{e}^{-2\pi i a a'}\, e^{i \pi \tau a^2 } \,=\,  \frac{1}{\sqrt{-i \tau}}e^{-i \pi (a')^2/\tau}  ~.
\end{split}
\end{align}
Using \eqref{eq:3dDualPF}, we can write the S-dual 3d partition function:
\begin{equation}
\begin{aligned}
 Z^{\text{D}}_{3\rmd}(x) =&\, \int \rmd a \,{e}^{-2\pi i a x} = \delta(x)\:.
\end{aligned}
\end{equation}
This induces Dirichlet boundary conditions. Inserting this expression in \eqref{eq:Npf} (and setting $a'=0$ because now there is no topological symmetry generated by the boundary value of $A^{4d}_\mu$), one obtains
\begin{align}\label{eq:DpfQED}
\begin{split}
\mathcal Z^{\text{D}}_{HS^4}( \tau,x') & = \int d a \, e^{i \pi \tau a^2 }\, \delta(a-x') \,=\, e^{i\pi\tau (x')^2}   ~,
\end{split}
\end{align}
that is in fact the 4d partition function for supersymmetric Maxwell theory with Dirichlet boundary conditions.

Notice that \eqref{eq:NpfQED} and \eqref{eq:DpfQED} satisfy the relation \eqref{eq:Sduality4dPF}, showing that Neumann and Dirichlet boundary conditions are exchanged by S-duality. 
For this simple case, we have reproduced the S-dual transformation laws for the partition function that were first written in \cite{Gaiotto:2014gha}.

\subsection{$\SL(2,\mathbb{Z})$ and one-point functions}

We now turn to how the one-point functions of Coulomb branch operators transform under $\SL(2,\mathbb{Z})$.  We remind their expression here, writing explicitly the dependence on the electric and magnetic fugacities
\begin{align}\label{eq:1ptFctForSLduality}
\begin{split}
		\langle A_n \rangle_{H\mathbb R^4}(\tau,a',x') &=\, \frac{1}{\mathcal Z_{HS^4}^{\text{N}}(\tau,a',x')} \cdot\\
		&\cdot \int\limits \rmd a\, \left( \frac{-i}{4\sqrt{2\pi \, \Im\,\tau}} \right)^n H_n(\sqrt{2\pi \, \Im\, \tau} a) e^{i  \pi \tau  a^2 } e^{-2\pi i a a'}   Z_{\mathcal{T}_{3d}}(a-x')\,.
\end{split}
\end{align}
where again we are omitting the dependence of the 3d partition function on the complexified mixing parameters $z_\kappa$.

\subsubsection*{$T$-transformation and one-point functions}

We follow the same steps done in \eqref{eq:4dTdualPF} in order to write the correlator \eqref{eq:1ptFctForSLduality} in terms of the correlator $ T\left[\langle A_n \rangle_{H\mathbb R^4}\right]$ in the 4d theory coupled to the 3d boundary theory $T\left[ \mathcal{T}_{3d}\right]$. We invert~\eqref{eq:Taction3dPF}, we plug it into \eqref{eq:Taction3dPF} and proceed as done above for the partition function: 
\begin{equation}\label{eq:1ptFctForSLduality2ndStep}
\begin{aligned}
		& \langle A_n \rangle_{H\mathbb R^4}(\tau,a',x') =\, \frac{e^{i\pi(x')^2}}{\mathcal Z_{HS^4}^{\text{N}}(\tau,a',x')}\cdot \\
		& \cdot \int\limits \rmd a\, \left( \frac{-i}{4\sqrt{2\pi \, \Im\,\tau}} \right)^n H_n(\sqrt{2\pi \, \Im\, \tau} a) \, e^{i  \pi (\tau+1)  a^2 } e^{-2\pi i a (a'+x')}  Z_{T\left[\mathcal{T}_{3d}\right]}(a-x')\\
		 &= \, \frac{1}{T\left[\mathcal Z_{HS^4}^{\text{N}}\right](\tau+1,a'+x',x')}\cdot \\ & \cdot \int\limits \rmd a\, \left( \frac{-i}{4\sqrt{2\pi \, \Im\,\tau}} \right)^n H_n(\sqrt{2\pi \, \Im\, \tau} a) \,  e^{i  \pi (\tau+1)  a^2 } e^{-2\pi i a (a'+x')}  Z_{T\left[\mathcal{T}_{3d}\right]}(a-x') \:,
\end{aligned}
\end{equation}
where in the last step we have used \eqref{eq:Tduality4dPF}.

The equation \eqref{eq:1ptFctForSLduality2ndStep} implies the following relation between the one-point functions of the  T-dual theories:
\begin{equation}\label{eq:Tduality4d1ptFnct}
\mathcal  \langle A_n \rangle_{H\mathbb R^4}(\tau  , a', x')  =  T\left[ \langle A_n \rangle_{H\mathbb R^4} \right] (\tau+1  , a'+x', x')\:. 
\end{equation}

\subsubsection*{$S$-transformation and one-point functions}

We again proceed as in \eqref{eq:4dSdualPF}, in order to write this correlator in terms of the correlator $ S\left[\langle A_n \rangle_{H\mathbb R^4}\right]$ in the 4d theory coupled to the 3d boundary theory $S\left[ \mathcal{T}_{3d}\right]$. Inverting \eqref{eq:3dDualPF},  plugging the result for $Z_{\mathcal{T}_{3d}}( a -x')$ into \eqref{eq:1ptFctForSLduality}  and rearranging the integrals, we obtain
 \begin{equation}\label{eq:1ptFctForSduality2ndStep}
\begin{aligned}
		\langle A_n \rangle_{H\mathbb R^4}(\tau,a',x') =&\, \frac{1}{\mathcal Z_{HS^4}^{\text{N}}(\tau,a',x')} \int\limits \rmd x\, \left( \frac{-i}{4\sqrt{2\pi \, \Im\,\tau}} \right)^n   e^{-2\pi i  x x'}   Z_{S\left[\mathcal{T}_{3d}\right]}(x)\,\cdot  \\
 & \,\,\,\cdot \int \rmd a \, e^{i  \pi \tau  a^2 } e^{2\pi i a (x-a')}  H_n(\sqrt{2\pi \, \Im\, \tau} a) \:.
\end{aligned}
\end{equation}
To perform the integration over $a$ on the second line, we use the following result for the Hermite polynomials:
\begin{equation}\label{eq:FTHermPol}
\int \rmd y  \, e^{i k y  }  \,  e^{-\frac{y^2 (1-i \gamma)}{2} }\,H_n(y)= 
\frac{i^n\sqrt{2\pi}}{\sqrt{1-i \gamma }}\frac{  (1+i \gamma )^{n/2}}{ (1-i \gamma )^{n/2}} e^{-\frac{k^2}{2 (1-i \gamma )}} H_n\left(\frac{k}{\sqrt{1+\gamma^2}}\right)\:.
\end{equation}
By redefining the integration variable as $a=\frac{y}{\sqrt{2\pi \, \Im\, \tau}}$ we can then write 
\begin{equation}\label{eq:1ptFctForSdualityIntermStep}
\begin{aligned}
 \int \rmd a \, e^{i  \pi \tau  a^2 } e^{2\pi i a (x-a')}  &H_n(\sqrt{2\pi \, \Im\, \tau} a)  =\, 
 \int \frac{\rmd y}{\sqrt{2\pi \, \Im\, \tau}}\, e^{\frac{2\pi (x-a')}{\sqrt{2\pi \, \Im\, \tau}}i y } \, e^{\frac{i  \pi \tau}{2\pi \, \Im\, \tau}  y^2 } \, H_n(y) \\
  &=\,  \frac{i^n}{\sqrt{-i \tau }}\frac{  (1+i \gamma )^{n/2}}{ (1-i \gamma )^{n/2}} \, e^{-\frac{i \pi (x-a')^2}{\tau}}
  H_n\left(\sqrt{2\pi \, \Im\, \left(-\frac{1}{\tau}\right)} (x-a')\right) \:,
\end{aligned}
\end{equation}
where in the last line we have applied \eqref{eq:FTHermPol} with  $k= \frac{\sqrt{2\pi}(x-a')}{\sqrt{\Im \tau}} $ and, as done before, $\gamma=\frac{\Re \tau}{\Im \tau}$. Moreover we have used the fact that $(1-i \gamma)\Im \tau= - i\tau$ and $\frac{1}{\Im \tau (1+\gamma^2)}=\frac{|\tau|^2}{\Im\tau} = \Im\left(-\frac{1}{\tau}\right)$. 

Plugging \eqref{eq:1ptFctForSdualityIntermStep} into \eqref{eq:1ptFctForSduality2ndStep} and shifing $x\mapsto x+a'$, we obtain
 \begin{equation}\label{eq:1ptFctForSduality3rdStep}
\begin{aligned}
		\langle A_n \rangle_{H\mathbb R^4}(\tau,a',x') =&\, \frac{e^{-2\pi i a'x'}}{\mathcal Z_{HS^4}^{\text{N}}(\tau,a',x')} \int\limits \rmd x\, \left( \frac{-i}{4\sqrt{2\pi \, \Im\,\tau}} \right)^n   e^{-2\pi i  x x'}   Z_{S\left[\mathcal{T}_{3d}\right]}(x+a')\,\cdot  \\
 & \,\,\,\cdot \frac{i^n}{\sqrt{-i \tau }}\frac{  (1+i \gamma )^{n/2}}{ (1-i \gamma )^{n/2}} e^{-\frac{i \pi x^2}{\tau}}
  H_n\left(\sqrt{2\pi \, \Im\, \left(-\frac{1}{\tau}\right)} x\right) \\
   =&\,     \frac{e^{-2\pi i a'x'}}{\sqrt{-i \tau }\mathcal Z_{HS^4}^{\text{N}}(\tau,a',x')}  \left(-\frac{1}{\tau}\right)^n \int\limits \rmd x\, \left( \frac{-i}{4\sqrt{2\pi \, \Im\,(-1/\tau)}} \right)^n  \,\cdot  \\
 & \,\,\,\cdot  H_n\left(\sqrt{2\pi \, \Im\, \left(-\frac{1}{\tau}\right)} x\right) \,e^{-\frac{i \pi x^2}{\tau}}    e^{-2\pi i  x x'}   Z_{S\left[\mathcal{T}_{3d}\right]}(x+a') \\
  =&\,     \frac{1}{S\left[\mathcal Z_{HS^4}^{\text{N}}\right](-1/\tau,x',-a')}  \left(-\frac{1}{\tau}\right)^n \int\limits \rmd x\, \left( \frac{-i}{4\sqrt{2\pi \, \Im\,(-1/\tau)}} \right)^n  \,\cdot  \\
 & \,\,\,\cdot  H_n\left(\sqrt{2\pi \, \Im\, \left(-\frac{1}{\tau}\right)} x\right) \,e^{-\frac{i \pi x^2}{\tau}}    e^{-2\pi i  x x'}   Z_{S\left[\mathcal{T}_{3d}\right]}(x+a')\:,
\end{aligned}
\end{equation}
where in the first step we have used $\Im\tau \left(\frac{1-i \gamma}{1+i \gamma}\right)=-\Im \tau\left( \frac{\tau}{\bar{\tau}}\right)=-\left(\frac{\Im\tau}{|\tau|^2}\right)\tau^2 = -\tau^2\,  \Im\left(-\frac{1}{\tau}\right)$ and in the second step the relation \eqref{eq:Sduality4dPF}.

We have then obtained the following expression that relates the one-point functions of the S-dual theories:
\begin{equation}\label{eq:Sduality4d1ptFnct}
\mathcal  \langle A_n \rangle_{H\mathbb R^4}(\tau  , a', x')  = \left(-\frac{1}{\tau}\right)^n  S\left[ \langle A_n \rangle_{H\mathbb R^4} \right] (-1/\tau  , x', - a') 
\end{equation}
The overall $\tau$-dependent factor is consistent with the action of the $S$ transformation on the vector multiplet, namely the self-dual components of the field strength under electric-magnetic duality transform as $S[F^+_{\mu\nu}] = - \tau F^+_{\mu\nu}$, and by supersymmetry the scalar $\phi$ must transform in the same way $S[\phi] = - \tau \phi$.

\section{Examples}\label{Sec:Examples}

In this section we will study different BCFTs to explicitly demonstrate our findings in the previous sections. In particular, we will study $\mathcal N=2$ supersymmetric $\text{U}(1)$ gauge theory in the bulk of $HS^4$, coupled to different CFTs on the boundary.\footnote{In this section, 
the partition functions and the one-point functions will 
be evaluated at zero fugacities for the global symmetries; they will
continue showing a dependence on the mixing parameters.}

Before proceeding with the examples, let us point out a general result about the mixing parameters $\tilde{\Delta}$ and $\hat{\Delta}$ for the magnetic and electric U(1)'s in \eqref{eq:NpfDeltatilde}: The values fixed by the maximization are both zero when 
$Z_{\mathcal{T}_{3d}}(a - i\hat{\Delta})$ is even under $a - i\hat{\Delta}\mapsto -(a - i\hat{\Delta})$. To see this, we use that when this condition is satisfied the boundary free energy \eqref{Eq: boundary F} is even under the simultaneous change of sign of both parameters i.e. $F_\partial(-\tilde{\Delta}, -\hat{\Delta})= F_\partial(\tilde{\Delta}, \hat{\Delta})$. Since the maximum is unique, it must then be at $(\tilde{\Delta}, \hat{\Delta})=(0,0)$. To prove this parity property of $F_\partial$ on can simply change the integration variable in \eqref{eq:NpfDeltatilde} by setting $a\mapsto -a$ to obtain that $\mathcal{Z}^{\text{N}}_{HS^4}(-\tilde{\Delta}, -\hat{\Delta})=\mathcal{Z}^{\text{N}}_{HS^4}(\tilde{\Delta}, \hat{\Delta})$.

\subsection{Chirals on the boundary}\label{sec:chirals}

Let us start by considering the 3d $\mathcal{N}=2$ SCFT of $N_{F}+N_{\bar{F}}$ free chiral fields. We need to specify the coupling to the bulk gauge field by picking a U(1) global symmetry, and we take it to be the symmetry that assigns charge $+1$ to $N_{F}$ chirals and $-1$ to the remaining $N_{\bar{F}}$ of them. We denote this theory as $\mathcal{T}_{3d} = (N_F, N_{\bar{F}})$. The hemisphere partition function and the values of the mixing parameters at the maximum were computed previously using a saddle-point analysis in \cite{Wu}, and also in \cite{KumarGupta:2019nay} which in addition studied the two-point functions of boundary currents, and performed an extensive numerical analysis. Our overlapping results are in agreement with those presented in these references.

The 3d partition function with fugacity $a$ for the selected U(1) global symmetry is \cite{Jafferis:2010un,Kapustin:2009kz}
\begin{align}\label{eq:NfNfbPF}
\begin{split}
	Z_{(N_F, N_{\bar{F}})}(a - i \hat{\Delta},\Delta) & = \left(Z^+_{\text{chiral}}(-a - i \hat{\Delta} - i\Delta)\right)^{N_F}\left(Z^-_{\text{chiral}}(a + i \hat{\Delta} - i \Delta)\right)^{N_{\bar{F}}}~,\\
	& Z^{\pm}_{\text{chiral}}(\xi) = e^{\pm\left[- \frac{i \pi}{12} + \frac{i \pi}{2} (i\xi-1)^2\right]}e^{\ell(1- i \xi)}~.
\end{split}
\end{align}
$\Delta$ and $\hat{\Delta}$ parametrize the most general choice of the $R$-charge for the chiral fields that is compatible with the global symmetry when we couple to the bulk gauge field with the above assignment of charges. The choice of $+$ or $-$ in $Z_{\text{chiral}}$ corresponds to the choice of background Chern-Simons contact terms that are needed to make the partition function coupled to the background gauge field invariant under background gauge transformations, see e.g. \cite{Closset:2019hyt}. The choice made here ensures that the theory is parity symmetric for $N_F=N_{\bar{F}}$.

Plugging the expressions for the 3d partition functions into \eqref{eq:NpfDeltatilde}, the integral takes the form
\begin{align}
\begin{split}\label{eq:chiralshifted}
	 &\mathcal Z_{HS^4}^{\text{N}} (\tau,-i\tilde \Delta,-i\hat\Delta,\Delta)\\& =  \int\limits \rmd a\,  e^{i \pi \tau a^2}  e^{-2\pi a \tilde\Delta} \left(Z^+_{\text{chiral}}(-a -i \hat\Delta- i\Delta)\right)^{N_F}\left(Z^-_{\text{chiral}}(a+i\hat\Delta - i \Delta)\right)^{N_{\bar{F}}} \\
	& = e^{i \pi \Phi} \int\limits \rmd a\, e^{ i \pi \tau_s a^2} e^{-2\pi a \tilde\Delta_s}e^{N_F \ell(1-\Delta -\hat\Delta+ i a) + N_{\bar F} \ell (1 - \Delta +\hat\Delta-i a)}~,
\end{split}	
\end{align}
where
\begin{align}\label{eshift}
\begin{split}
& \Phi = \left(\Delta ^2+\hat{\Delta}^2-2 \Delta +\frac{5}{6}\right) \frac{N_F-N_{\bar{F}}}{2}-(1-\Delta ) \hat{\Delta} (N_F+N_{\bar{F}})~,\\
&\tau_s = \tau-\frac{1}{2}  (N_F-N_{\bar{F}})~,\\
&\tilde\Delta_s = \tilde\Delta -\frac{N_F+N_{\bar{F}}}{2} (\Delta -1) -\frac{N_F-N_{\bar{F}}}{2} \hat{\Delta}~.
\end{split}	
\end{align}
The phase $\Phi$ does not affect the maximization, we can thus perform the maximization on the shifted variable $\tilde\Delta_s$, instead of $\tilde\Delta$. We will not consider the most general possibility but rather we restrict to two subclasses of examples: (1) $N_F = N_{\bar{F}}$, and (2) $N_{\bar{F}} = 0$, generic $N_F$. For the class (1) we can apply the parity condition mentioned above to the integral in terms of the shifted variables \eqref{eq:chiralshifted}, because the ``effective''  $Z_{\mathcal{T}_{3d}}$ given by $e^{N_F \ell(1-\Delta -\hat\Delta+ i a) + N_{\bar F} \ell (1 - \Delta +\hat\Delta-i a)}$ has the desired symmetry property. As a result in this case we obtain that the maximization gives $\tilde\Delta_{s,\text{max}} = \hat{\Delta}_{\text{max}} = 0$ and we only need to maximize with respect to $\Delta$. For the class (2) $\mathcal Z_{HS^4}^{\text{N}}$ depends on $\Delta$ and $\hat{\Delta}$ only through the combination $\Delta+\hat{\Delta}$, so we can reabsorb $\hat{\Delta}$ in $\Delta$. As a result, to cover both classes of examples it is enough to consider
\begin{align}
\begin{split}\label{Eq: U1 partition function}
	& \mathcal Z_{HS^4}^{\text{N}}  =  e^{i \pi \Phi} \int\limits \rmd a\,  e^{ i \pi \tau_s a^2} e^{-2\pi a \tilde\Delta_s}e^{N_F \ell(1-\Delta + i a) + N_{\bar F} \ell (1 - \Delta-i a)}~, 
\end{split}	
\end{align}
and perform extremization with respect to $\tilde{\Delta}_s$ and $\Delta$. We stress that in the most general case of arbitrary $N_F$ and $N_{\bar{F}}$ one would have instead to extremize with respect to three mixing parameters. 

For simplicity of presentation we define the variables
\begin{equation}
	\lambda^2 = {\frac{\Im\,\tau_s}{\tau_s \bar \tau_s}}\,,\quad\gamma =\frac{\mathrm{Re}\tau_s}{\mathrm{Im}\tau_s}\,,\quad \mu = \frac{N_F + N_{\bar F}}{2}\,,\quad \nu = \frac{N_F - N_{\bar F}}{2}~.
\end{equation}
The mixing parameters are then expanded in $\lambda$, with coefficients that can be an arbitrary function of the remaining parameters $\gamma, \mu, \nu$. Imposing the extremality constraints
\begin{equation}
	\partial_\Delta F_\partial = 0 \,,\quad \text{and} \quad \partial_{\tilde \Delta_s} F_\partial = 0\,,
\end{equation}
we can determine the coefficients of the expansion, order by order in the small $\lambda$ expansion. We have computed the expansions to order $\mathcal O(\lambda^{12})$. The full expression is, however, unwieldy to write down and instead we give up to $\mathcal{O}(\lambda^6)$
\begin{equation}\label{Eq: expansion of charges for (anti-)chirals}
	\begin{aligned}
		\Delta_{\text{max}} =&\, \frac12 - \frac{\lambda^2}{\pi} +  \left(\mu   + \left( \frac{2}{\pi} \right)^2 - \gamma^2 (\mu + 1) \right) \frac{\lambda^4}{2}  -\Bigg( \left(\frac{\pi  \mu(\mu+1)}{4} + \frac{\pi \nu^2}{6\mu} \right) (1-3\gamma^2)\\
		& - \left( \frac{2}{\pi^2} +  \frac{3}{4} + \mu \right)\pi\gamma^2 + \left( \frac{2}{\pi} \right)^3 +\frac{2}{3\pi} \Bigg) \lambda^6 + \mathcal O(\lambda^8)\,,\\
		\tilde \Delta_{s,\text{max}} =&\, \nu \gamma \Bigg( \frac{\lambda^2}{2} - \frac{\pi  (2\mu+1)}{4} \lambda^4  +\Bigg( \frac{\pi^2}{24} \left( \mu(\mu+14/3) - 2 {\nu^2}/{\mu}+1 \right)(1-3\gamma^2) \\
		& -\frac{\pi^2}{24} \left( 8 \mu( \mu + 8/3) + 5 \right) + 2 \mu - 1/2 \Bigg)\lambda^6  \Bigg) + \mathcal O(\lambda^8)\,.
	\end{aligned}
\end{equation}
We reiterate that $\Delta_{\text{max}}$ is the mixing parameter for an ordinary flavor symmetry when $N_F = N_{\bar{F}}$, while it should be interpreted as the mixing parameter $\hat{\Delta}_{\text{max}}$ of the electric U(1) for $N_{\bar{F}} =0$. The leading order terms in \eqref{Eq: expansion of charges for (anti-)chirals} are consistent with what one expects: when $\lambda\rightarrow 0$ the bulk and boundary decouple leaving a set of free chirals with charges $\Delta_{\text{max}}=1/2$, and $\tilde \Delta_{\text{max}} = 0$. Furthermore, for $\nu = 0$ we get $\tilde \Delta_{s,\text{max}} = 0$ as expected.\footnote{Also the vanishing of $\tilde \Delta_{s,\text{max}}$ for $\gamma = 0$ can be explained using simple symmetry arguments. Indeed, for $\gamma = 0$ we have $\bar{\tau}_s = - \tau_s$. Taking the complex conjugate of \eqref{eq:chiralshifted} and performing the change of variable $a\to-a$ we get $\bar{\mathcal Z}_{HS^4}^{\text{N}} (\tilde \Delta_s) = \mathcal Z_{HS^4}^{\text{N}} (-\tilde \Delta_s)$, where we used that $\widebar{e^{\ell(z)}} = e^{\ell(\bar{z})}$. As a result $F_\partial$ is even in $\tilde{\Delta}_s$ for $\gamma=0$ and the unique maximum can only be at $\tilde{\Delta}_s =0$.}

Using the expansion of $\Delta_{\text{max}}$ and $\tilde \Delta_{s,\text{max}}$ we can directly expand the partition function itself to the same order. Here we present it up to $\mathcal{O}(\lambda^4)$:
\begin{equation}\label{Eq: expansion Z for chiral-anti-chiral}
	\begin{aligned}
		&\mathcal Z^{\text{N}}_{HS^4} =\, \frac{2^{-\mu}e^{\frac{i\pi\nu}{12}}}{\sqrt{-i \tau}} \Bigg( 1 - \bigg(\frac{\pi\mu}{4}(1+i \gamma)-\frac{\nu}{\pi}i\bigg)\lambda^2 \\
		&\ \ \ \ \ \ \ \ \ \ \ \ \ \ +\bigg(-\frac{3\pi^2}{32}(-i+\gamma)^2\mu^2-\frac{\nu^2}{2\pi^2}+\frac{-i+\gamma+2i\pi(-1+\gamma^2)}{4}\mu\nu\\
		&\ \ \ \ \ \ \ \ \ \ \ \ \ \ \ \ \ \ \ \ -\frac{8-\pi^2+16i\gamma-2i\pi^2\gamma+\pi^2\gamma^2}{16}\mu+i\frac{2-4\pi^2+\pi^4\gamma^2}{2\pi^3}\nu\bigg)\lambda^4+\mathcal{O}(\lambda^6)\Bigg)\,.
	\end{aligned}
\end{equation}

We can then compute the coefficients of the one-point functions of the Coulomb branch operators $A_{2n} = \phi^{2n}$ using \eqref{Eq: 1-pt functions flat space}. %
We have obtained the result up to order $O(\lambda^4)$ for generic~$n$:
\begin{equation}\label{eq:chiralonept}
	\begin{aligned}
		\langle A_{2n}\rangle_{H\mathbb{R}^4} =&\,  \frac{(2n)!}{(n-1)!}\left(-\frac{(1+i\gamma)^2 \lambda^2 }{32 \pi ^2} \right)^n  \Bigg(\frac{1}{n} - \pi \mu \lambda^2\\
		 &\, +\frac{\mu}{6} \left(\pi^2(n-1)(2+ 3 \mu) - 24 + 3 \pi^2 (1+\mu)(1+i \gamma)\right)  \lambda^4 + \mathcal{O}(\lambda^6)\Bigg)\,.
	\end{aligned}
\end{equation}
As a consistency check we have performed a Feynman diagram analysis confirming the first two orders in this expression of the one point functions, which can be found in Section~\ref{SSec:perturbative}.

\subsection{XYZ on the boundary}\label{ref:XYZ}
As a second example, we couple the bulk field to the 3d $\mathcal{N}=2$ SCFT known as the XYZ model. Also this example was studied previously in \cite{KumarGupta:2019nay} and our overlapping results, namely the perturbative expansion for the mixing parameters, are in agreement. The XYZ model is defined as the IR fixed point of a theory of three chiral fields X, Y and Z deformed by the superpotential XYZ. We need to select a U(1) global symmetry to gauge with the bulk gauge field, and we take it to be the one that gives charge $+1$ to X, $-1$ to Y and $0$ to Z (as above, its mixing parameter is called $\hat{\Delta}$).  There is an additional U(1) global symmetry that can mix with the R symmetry, and assigns charge $1$ to X and Y, and $-2$ to Z. We call $\Delta$ the corresponding mixing parameter. The 3d partition function is\footnote{Note that this formula for $Z_{\text{XYZ}}(\xi,\Delta)$ differs from
\begin{equation}
Z^+_{\text{chiral}}(\xi - i(1-\Delta))Z^+_{\text{chiral}}(-\xi - i(1-\Delta))Z^+_{\text{chiral}}( -i 2\Delta)~,
\end{equation} 
by the prefactor 
\begin{equation}
e^{\frac{i \pi }{4}} e^{i \pi  \left(3 \Delta ^2-2 \Delta -\xi ^2\right)}~,
\end{equation}
which is the contribution of a properly quantized background Chern-Simons term. Therefore we make a definite choice of background contact terms in writing down this formula.}
\begin{equation}\label{eq:XYZPF}
	Z_{\text{XYZ}}(a -  i \hat{\Delta},\Delta) = \,e^{\ell(\Delta - \hat{\Delta} + i a) + \ell ( \Delta +\hat{\Delta} -i a) + \ell(1-2\Delta) }\,.
\end{equation}
This partition function satisfies the parity relation mentioned at the beginning of this section and therefore we can set $\tilde\Delta = \hat\Delta = 0$ in \eqref{eq:NpfDeltatilde} and maximize only with respect to $\Delta$. Similarly to the chiral case we use the following parameters for notational convenience
\begin{equation}
\lambda^2 = {\frac{\Im\,\tau}{\tau \bar \tau}}\,,\quad\gamma =\frac{\mathrm{Re}\tau}{\mathrm{Im}\tau}~.
\end{equation}
We have computed the maximizing value of $\Delta$ up to order $O(\lambda^{32})$, the first terms are
\begin{equation}
	\Delta_{\text{max,XYZ}} = \frac13 - \frac{2(\sqrt{3} \pi - 9)}{9(4\pi - 3 \sqrt{3})} \lambda^2 + \mathcal O(\lambda^4)\,.
\end{equation}
It turns out that the expansion of the partition function in this case is computationally more intense. Nevertheless, we have computed it to the order $\mathcal{O}(\lambda^{32})$ as well, albeit for the coefficients of the highest orders we have only the numerical result. The first orders are%
\begin{equation}\label{Eq: expansion Z for XYZ}
	\begin{aligned}
		\mathcal Z^{\text{N}}_{HS^4} =&\, \frac{e^{3\ell(1/3)}}{\sqrt{-i \tau}} \Bigg( 1 + i \frac{4 \sqrt{3} \pi - 9}{27(1+i \sqrt{3})}(1+i \gamma)\lambda^2 + \mathcal O(\lambda^4) \Bigg)\,.
	\end{aligned}
\end{equation}
Finally, for the one-point functions we have computed up to order $O(\lambda^{32})$, the first few terms are%
\begin{equation}\label{eq:exponepoint}
	\begin{aligned}
		\langle A_{2n}\rangle_{H\mathbb{R}^4} =&\,  \frac{2n!}{(n-1)!}\left(-\frac{(1+i\gamma)^2 \lambda^2 }{32 \pi ^2}\right)^n  \Bigg(\frac{1}{n} - \frac29 (3\sqrt{3} - 4 \pi) \lambda^2 + \mathcal O(\lambda^4) \Bigg)\,.
	\end{aligned}
\end{equation}
Note that the leading order in the one-point function is just fixed by the Neumann boundary condition and does not depend on the boundary degrees of freedom, and in fact it agrees with the result \eqref{eq:chiralonept} obtained for the boundary condition with chiral fields on the boundary.

\subsection{Dualities}
As described in Section~\ref{Sec:Dualities} the $\SL(2,\mathbb Z)$ duality group of the supersymmetric Maxwell theory relates different boundary conditions. In this section we consider examples of this action and use them to test the results for the partition function and for the one-point functions. We computed these observables in the previous sections perturbatively in the coupling, by expanding the localization integral. Since the $\SL(2,\mathbb Z)$ action maps weak coupling to strong coupling, the test requires extrapolating the perturbative answers. The tests we perform here therefore serve two complementary purposes: on one hand we use them to verify the localization formulas and their transformations under $\SL(2,\mathbb Z)$, and on the other hand we use them to check the validity of the extrapolations of the perturbative expansions. We consider two examples in which we can obtain predictions for the transformations of the partition functions using certain integral identities, that have already appeared in the literature in the context of 3d dualities, (see for example \cite{Dimofte:2011ju,Benvenuti:2016wet}).

\subsubsection{$(N_F,N_{\bar{F}}) = (1,1) \, \leftrightarrow \,$XYZ}\label{s11xyz}
The first explicit example we study is the the BCFT of chiral fields with $(N_F,N_{\bar{F}})=(1,1)$ in the notation of Section \ref{sec:chirals}. The consequences of the $\text{SL}(2,\mathbb{Z})$ duality in this setup were considered before in \cite{Wu,KumarGupta:2019nay}. Applying an $S$ transformation we find that the BCFT with this boundary condition and coupling $\tau$ is dual to the BCFT with coupling $-1/\tau$ and the boundary condition given by (the IR fixed point of) 3d SQED with the matter given by one chiral with charge +1 and one chiral with charge -1, coupled to the bulk via the gauging of the topological U(1) symmetry. Next, we use 3d mirror symmetry that tells us that this SQED theory is dual to the XYZ model, that we considered in Section \ref{ref:XYZ}. Denoting the theory of chirals as $(1,1)$, this 3d duality is verified at the level of the sphere partition function using the following integral identity, called the {\it pentagon identity} \cite{Benvenuti:2016wet},\footnote{This is an identity of the quantum dilogarithm \cite{Faddeev:2000if}. It was applied in this context in \cite{Dimofte:2011ju} with the following change of notation: $e^{\ell(-ix)}=s_{b=1}(x)$.}
\begin{equation}
e^{\ell(y+z)}e^{\ell(y-z)} = e^{\ell(2y-1)}\int d a \, e^{2\pi z a}\, e^{\ell(1-i a-y)}e^{\ell(1+i a-y)}~,
\end{equation}
which, upon setting $y= \Delta$ and $z=-i x$, implies 
\begin{equation}\label{eq:3ddualityXYZ}
Z_{S\left[(1, 1)\right]}( x ,\Delta) =\int d a \,{e}^{-2\pi i a x} Z_{(1,1)}(a,\Delta) = Z_{\text{XYZ}}(x,\Delta)~,
\end{equation} 
where $Z_{(1,1)}$ and $Z_{\text{XYZ}}$ were given in \eqref{eq:NfNfbPF} and \eqref{eq:XYZPF}, respectively.

\begin{figure}
\begin{center}
\begin{overpic}[scale=0.51,unit=1mm]{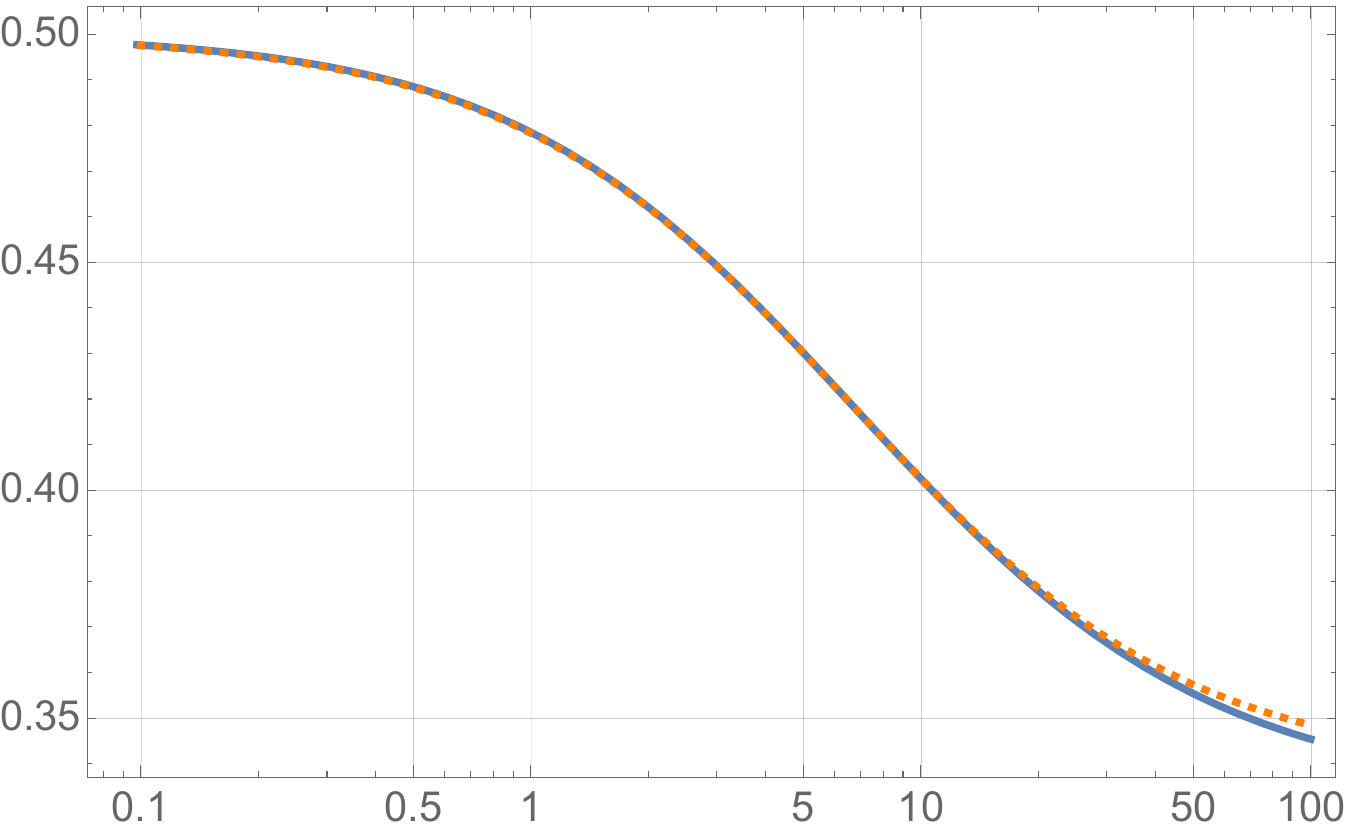}
\put(50,-4){\textcolor{gray}{$g^2$}}
\end{overpic}
\vspace{0.3cm}
\caption{The $\{8,8\}$ Pad\'e approximant of $\Delta^{(1,1)}_{\text{max}}(\tau)$, in blue, and of $\Delta^{\text{XYZ}}_{\text{max}}(-1/\tau)$, in dashed orange, at $\gamma=0$ as a function of $g^2$.}\label{fig:Delta}
\end{center}
\end{figure}

As a result we obtain that the BCFT with $(N_F,N_{\bar{F}})=(1,1)$ chiral fields on the boundary and coupling $\tau$ is dual to the BCFT with XYZ model on the boundary and coupling $-1/\tau$. Using \eqref{eq:Sduality4dPF} we then obtain
\begin{equation}
 {\mathcal  Z}_{HS^4}^{\text{N},\text{XYZ}} (-1/\tau  , -i\hat\Delta, +i\tilde\Delta, \Delta) = \sqrt{-i\tau}  e^{-2\pi i \hat\Delta\tilde\Delta} \mathcal  Z_{HS^4}^{\text{N},(1,1)}(\tau  , -i\tilde\Delta, -i\hat{\Delta}, \Delta)~. 
\end{equation}
This equation implies the following relations between the values of the mixing parameter that extremize the partition function (we write the argument simply as $\tau$ for convenience, but remember that these functions are real and not holomorphic)
\begin{align}\label{eq:DeltaS}
\begin{split}
\tilde\Delta_{\text{max}}^{\text{XYZ}}(-1/\tau) &= - \hat\Delta_{\text{max}}^{(1,1)}(\tau)~,\\
\hat\Delta_{\text{max}}^{\text{XYZ}}(-1/\tau) &=  \tilde\Delta_{\text{max}}^{(1,1)}(\tau)~,\\
\Delta_{\text{max}}^{\text{XYZ}}(-1/\tau) &= \Delta_{\text{max}}^{(1,1)}(\tau)~.
\end{split}
\end{align} 
The first two equations are trivially satisfied because in both theories $\tilde\Delta_{\text{max}} = \hat\Delta_{\text{max}} = 0$ by the symmetry argument explained above. The third line is a non-trivial constraint that relates the weak coupling behavior in one theory to the strong coupling one in the other theory. We test it by performing two independent Pad\'e extrapolations of the perturbative results obtained for $\Delta_{\text{max}}^{\text{XYZ}}$ and $\Delta_{\text{max}}^{(1,1)}$ in the previous section, and checking if the extrapolated functions satisfy the relation \eqref{eq:DeltaS}. We plot the result as a function of purely imaginary $\tau = \frac{4 \pi i}{g^2}$ in Figure \ref{fig:Delta}. We see a good agreement between the two sides of \eqref{eq:DeltaS} for all values of $g^2$, interpolating from the value of $1/2$ in the free $(1,1)$ theory to the value $1/3$ of the XYZ theory.

\begin{figure}
\begin{center}
\begin{overpic}[scale=0.5,unit=1mm]{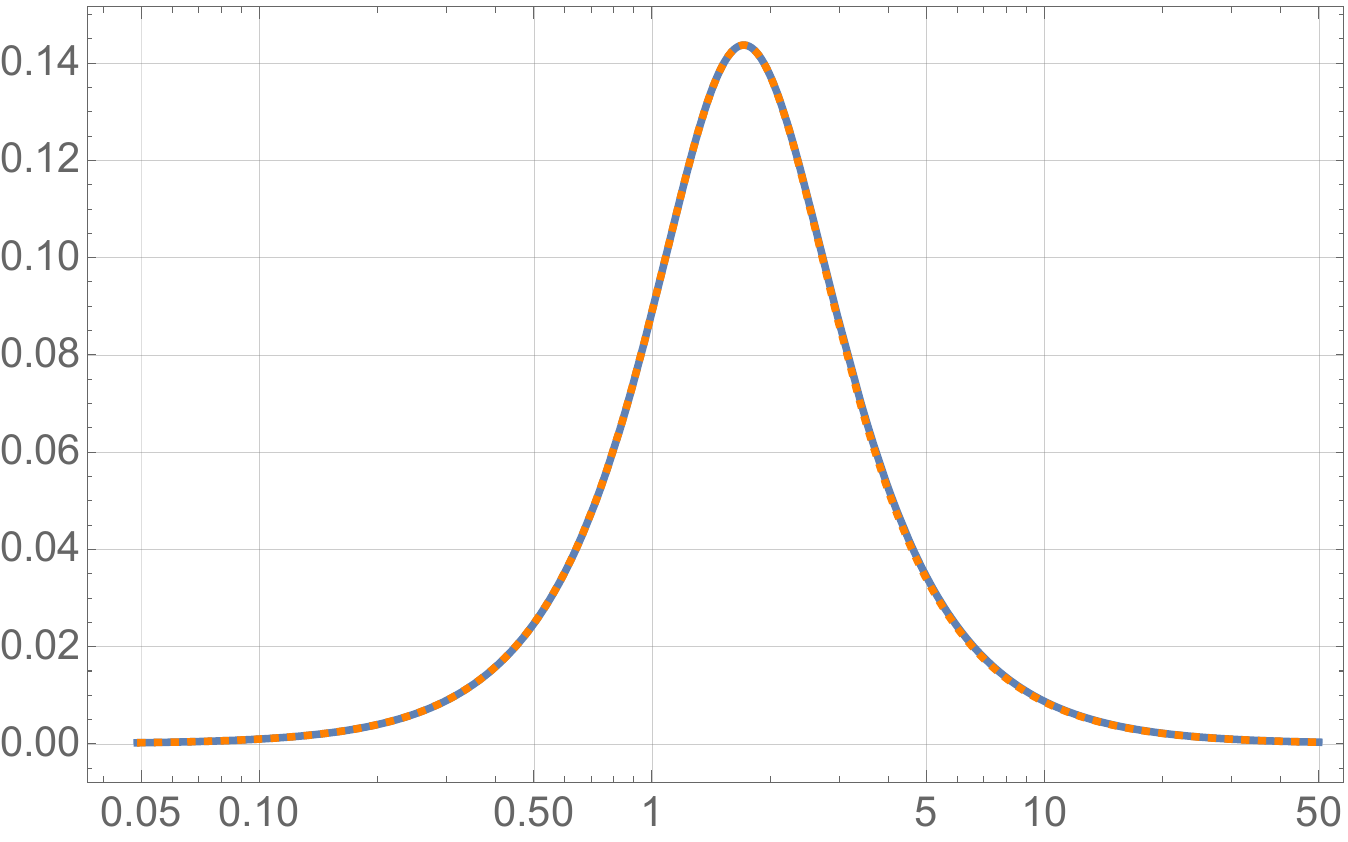}
\put(50,-4){\textcolor{gray}{$g^2$}}
\end{overpic}
\vspace{0.3 cm}
\caption{The $\{8,8\}$ Pad\'e approximant of $(e^{F_\partial})^{(1,1)}(\tau)$, in blue, and $(e^{F_\partial})^{\text{XYZ}}(-1/\tau)$, in dashed orange, at $\gamma=0$ as a function of $g^2$.}\label{fig:eF}
\end{center}
\end{figure}

We then plug these values for $\Delta_{\text{max}}$ into the partition functions and consider the two sides of Eq. \eqref{eq:3ddualityXYZ}. In order to get rid of prefactors, and to obtain a function which is analytic in the gauge coupling both in the weak and the strong coupling expansion, it turns out to be convenient to compare the combination
\begin{equation}
e^{F_\partial} =\left( \frac{\mathcal Z_{HS^4}^{\text{N}} \mathcal{\bar{Z}}_{HS^4}^N}{\mathcal Z_{S^4}}\right)^{-\frac 12}~,
\end{equation} 
which, by \eqref{eq:3ddualityXYZ}, satisfies
\begin{equation}
(e^{F_\partial})^{\text{XYZ}}(-1/\tau) = (e^{F_\partial})^{(1,1)}(\tau)~.
\end{equation}
Similar to $\Delta$, we perform independent Pad\'e extrapolations for this quantity, on the two sides of the duality, using the perturbative results of the previous section and compare them. The results are showed in~Figure~\ref{fig:eF}.

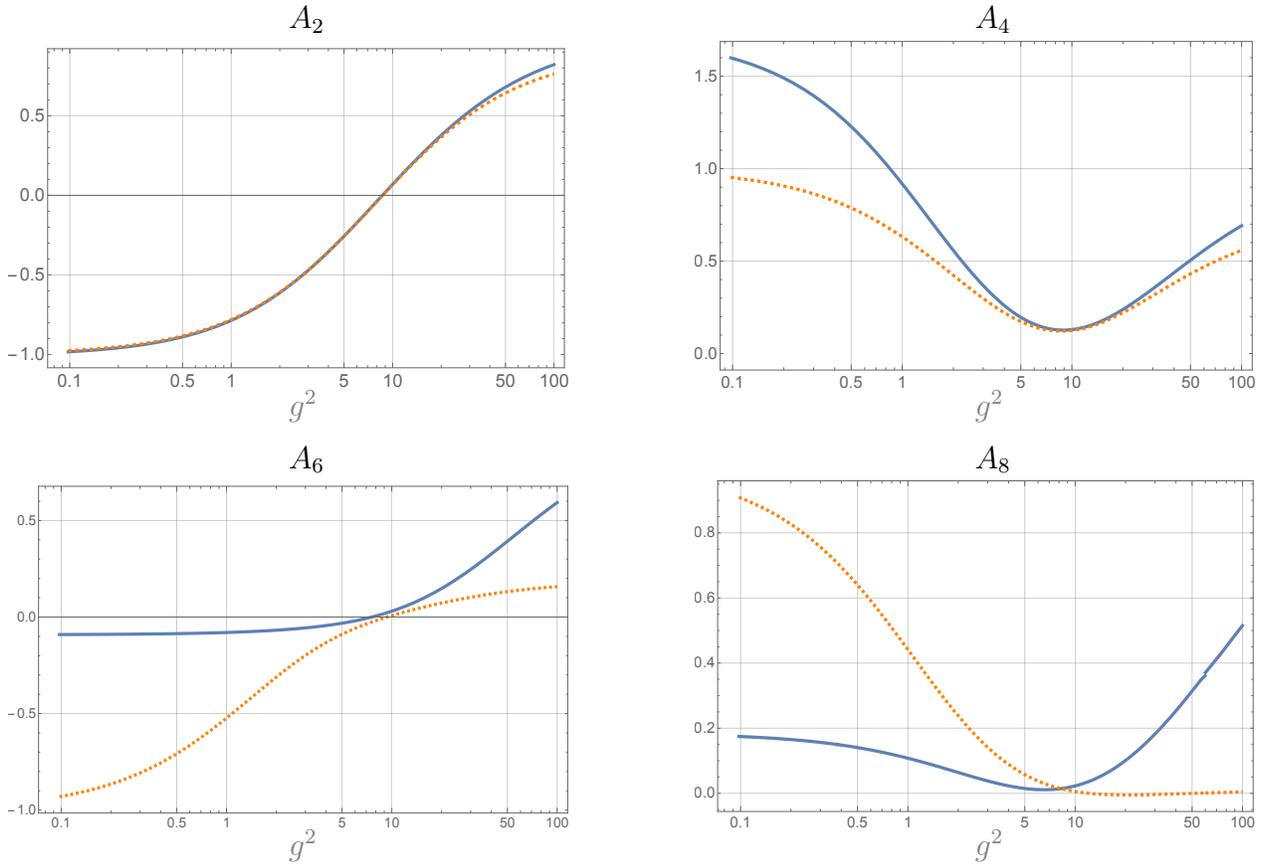
\begin{figure}
     \centering
     \begin{subfigure}{0.45\textwidth}
       \centering
       \begin{overpic}[width=\textwidth]{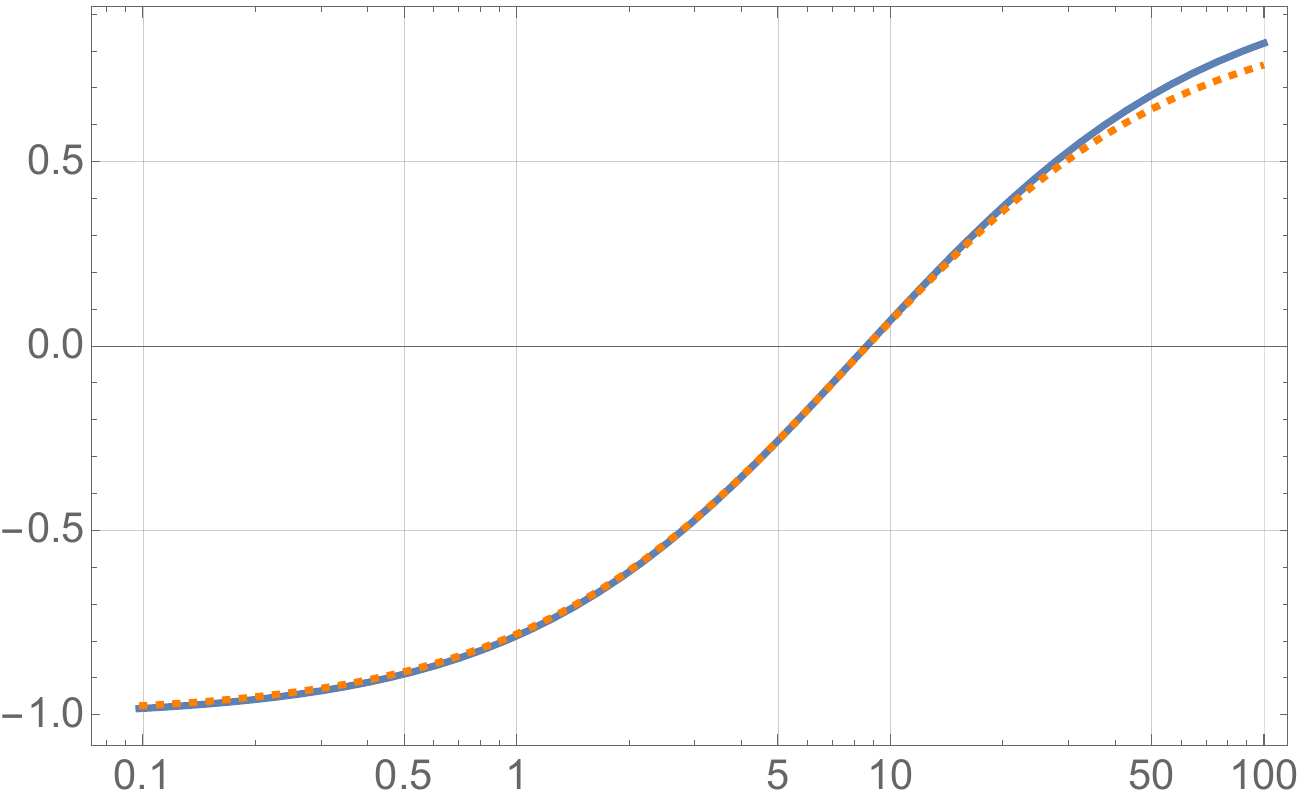}
        \put(50,-4){\textcolor{gray}{$g^2$}}
        \put(50,65){$A_2$}
       \end{overpic}
       \vspace{-0.2 cm}
     \end{subfigure}
     \hfill
     \begin{subfigure}{0.45\textwidth}
       \centering
       \begin{overpic}[width=\textwidth]{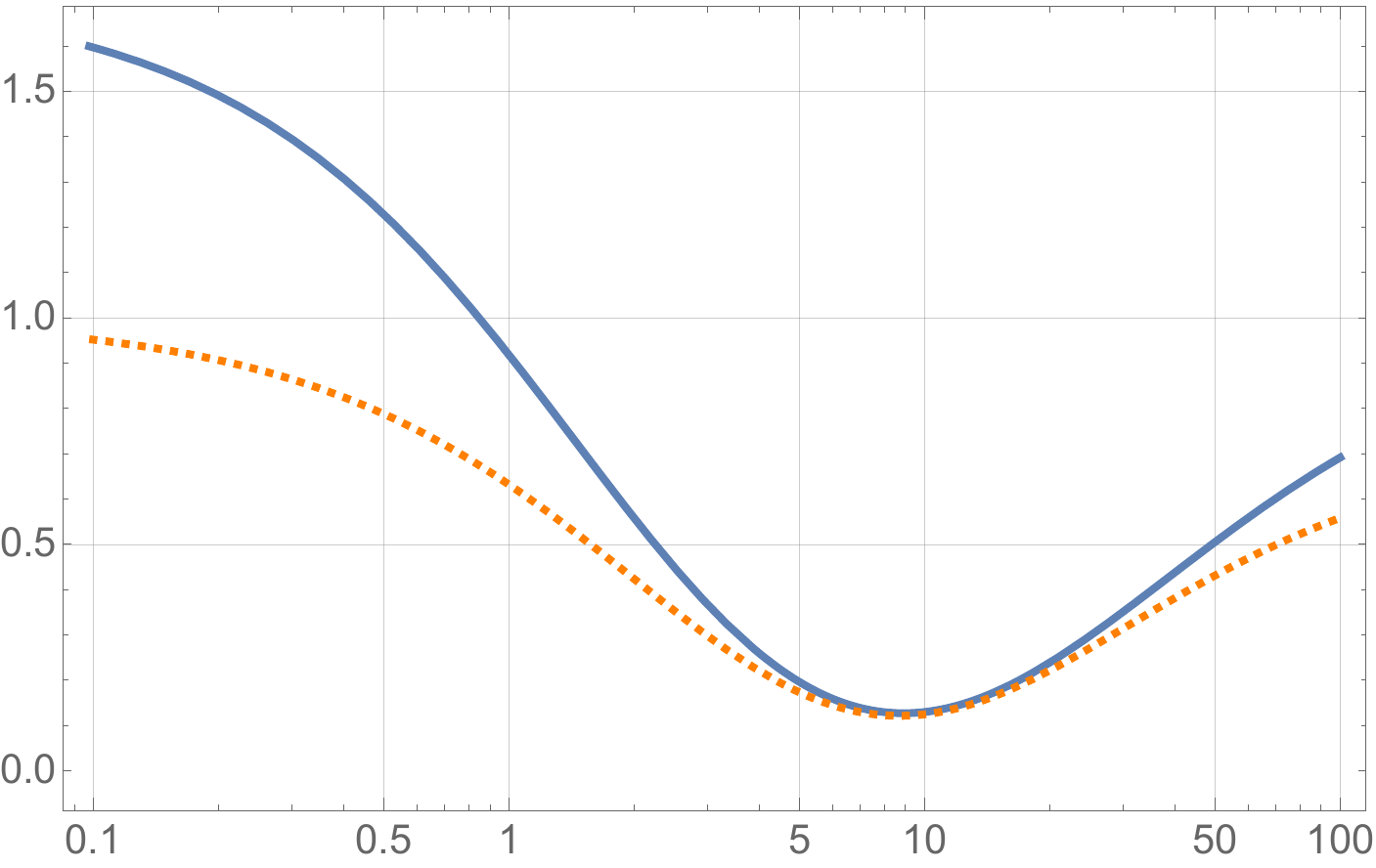}
        \put(50,-4){\textcolor{gray}{$g^2$}}
        \put(50,65){$A_4$}
       \end{overpic}
       \vspace{-0.2 cm}
     \end{subfigure}\\
     \vspace{0.8 cm}
     \begin{subfigure}{0.45\textwidth}
       \centering
       \begin{overpic}[width=\textwidth]{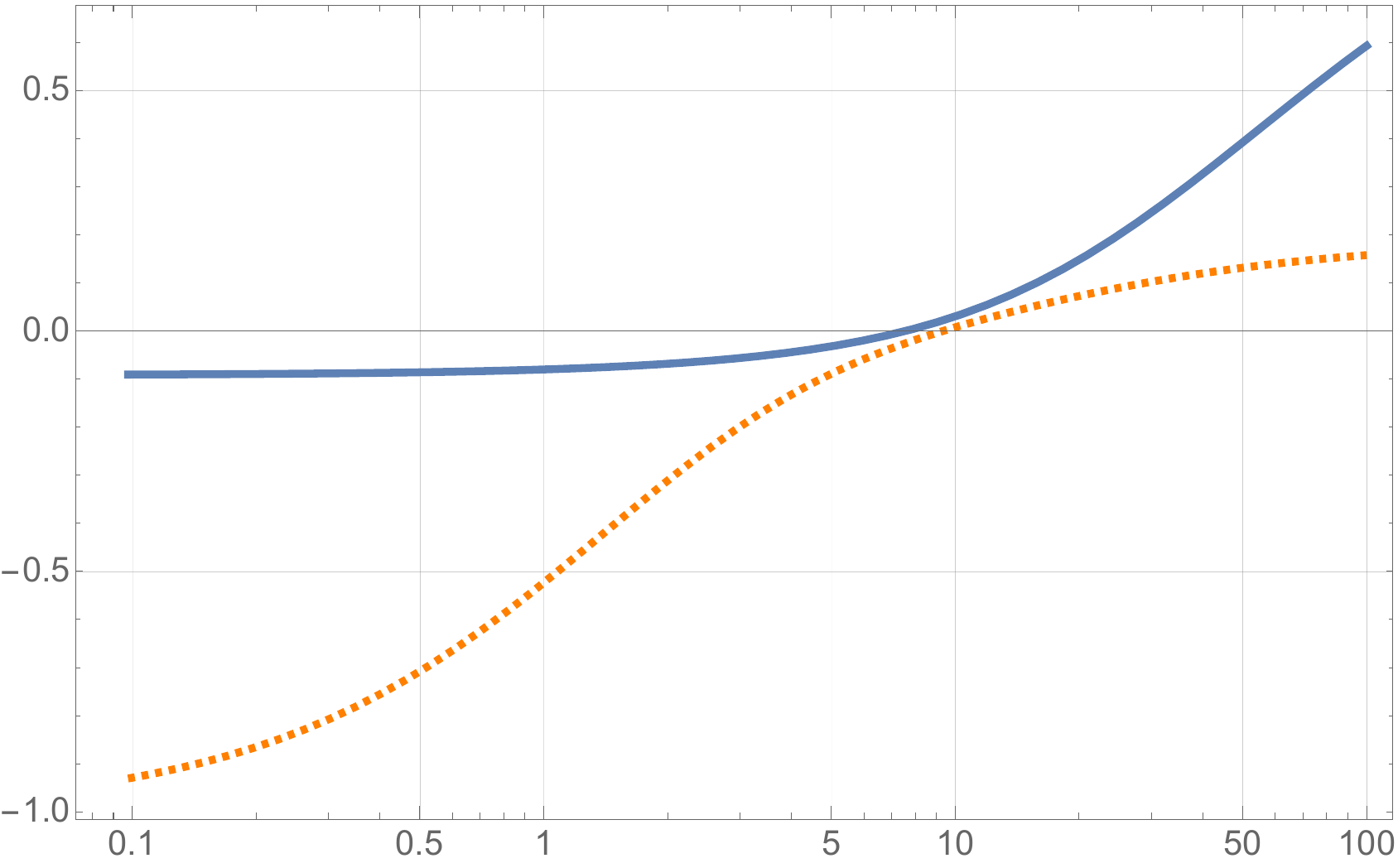}
        \put(50,-4){\textcolor{gray}{$g^2$}}
        \put(50,65){$A_6$}
       \end{overpic}
       \vspace{-0.2 cm}
     \end{subfigure}
       \hfill
     \begin{subfigure}{0.45\textwidth}
       \centering
       \begin{overpic}[width=\textwidth]{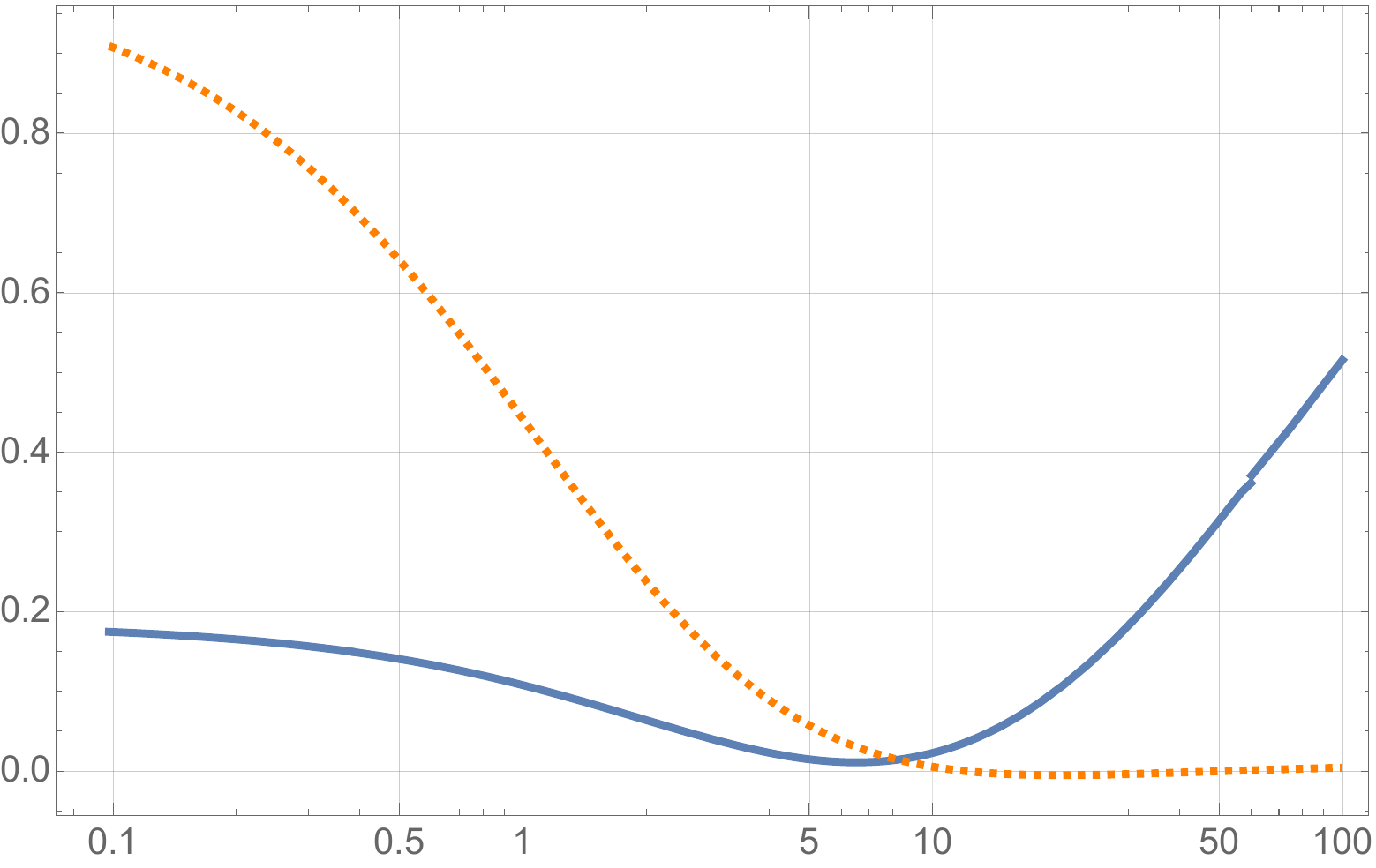}
        \put(50,-4){\textcolor{gray}{$g^2$}}
        \put(50,65){$A_8$}
       \end{overpic}
       \vspace{-0.2 cm}
     \end{subfigure}
        \caption{The plot labelled by $A_{2n}$ show the $\{8,8\}$ Pad\'e approximants for $(\tfrac{64\pi^2}{g^2})^n A_{2n}^{(1,1)}(\tau)$, in blue, and $(\tfrac{64\pi^2}{g^2\tau^2})^n A_{2n}^{XYZ}(-\tfrac 1\tau)$, in dashed orange, at $\gamma=0$ as a function of $g^2$. The normalization $(\tfrac{64\pi^2}{g^2})^n$ is chosen for graphical convenience.}
        \label{fig:AnS}
\end{figure}

Additionally, we test the $S$-duality action on the one-point functions. In the considered example, the $S$-duality relation~\eqref{eq:Sduality4d1ptFnct} becomes
\begin{equation}
\mathcal  \langle A_n \rangle^{(1,1)}_{H\mathbb R^4}(\tau)  = \left(-\frac{1}{\tau}\right)^n  \langle A_n \rangle^{\text{XYZ}}_{H\mathbb R^4}  (-1/\tau)~.
\end{equation}
We plot the results of the two independent Pad\'e extrapolations of the two sides in~Figure~\ref{fig:AnS}, for $n=2,4,6,8$. We see a good agreement for all values of $g^2$ for $n=2$. The agreement deteriorates as $n$ increase, the region where the two extrapolations agree progressively shrinks to values of $g^2$ of order 1. When they do not agree, this means that the number of available perturbative coefficients does not suffice to get a reliable Pad\'e extrapolation for these observables. The lesson we draw is that, perhaps expectedly, the question of how many orders are needed in order to get a reliable strong coupling extrapolation depends on the observable. The fact that the perturbative expansion becomes less and less reliable as we increase the charge $n$ of the Coulomb branch operators is also expected, because the perturbative coefficients grow with $n$, as exemplified by the first terms in \eqref{eq:chiralonept}. It should be possible to study a large charge limit of $n\to\infty$ with $g^2 n$ fixed, similarly to what was done in \cite{Grassi:2019txd} for the bulk two-point functions, in order to obtain exact results in $g^2 n$ in a $1/n$ expansion. We leave this as an interesting direction for the future.

\subsubsection{$(N_F , N_{\bar F}) = (1,0)$ triality}
Next, we consider the example of the $\text{U}(1)$ $\mathcal N=2$ Maxwell theory in the bulk coupled to one chiral field of charge $+1$ on the boundary, i.e. the theory $(1,0)$ in the notation of Section \ref{sec:chirals}. Applying the transformation $ST^{-1}$ we obtain that this boundary condition at bulk coupling $\tau$ is dual to the boundary condition at bulk coupling $-\tfrac{1}{\tau-1}$ with the following boundary degrees of freedom: (the IR fixed point of) 3d SQED with matter given by one chiral field of charge $+1$, and one unit of Chern-Simons term, coupled to the bulk via the gauging of the topological U(1) symmetry. The latter theory is typically referred to as $U(1)_{1/2}$ SQED with one chiral, where the half-integer level comes from taking the convention that a chiral field couples to the background gauge field with a $-1/2$ Chern-Simons contact term. This theory is also known as the ``tetrahedon theory'' because of the role it plays in the construction of 3d theories from compactifications of 6d the $(2,0)$ theory on 3 manifolds \cite{Dimofte:2011ju}.

We then use a 3d duality found in \cite{Dimofte:2011ju} that says that U(1)$_{1/2}$ SQED with one chiral is dual to the free SCFT of one chiral field. This is actually part of a triality, because both theories are also equivalent to U(1)$_{-1/2}$ SQED with one chiral. At the level of 3d partition functions, this duality is proven making use of the following integral identity (see e.g. \cite{Faddeev:2000if,Closset:2019hyt})
\begin{equation}
\int da \, e^{i\pi a^2}e^{2\pi a z}  e^{\frac{i \pi}{2} (1-r-i a)^2} e^{\ell(1-r-i a)} = e^{-\frac{7 \pi  i}{12} } e^{-i \pi(r^2 -2 r z -1)} \, e^{\frac{i\pi}{2} (r-z)^2}e^{\ell(r-z)}~,
\end{equation}
where, in order to make the integral convergent, the contour of the $da$ integration needs to be slightly tilted in the complex plane, i.e. ranging from $-\infty(1+ i \epsilon)$ to $+\infty(1+ i \epsilon)$.\footnote{To ensure convergence, also the variable $z$ needs to be restricted to a certain region in the complex plane, and outside of that region the identity should be interpreted in the sense of analytic continuation. We have not tried to fully determine the region of convergence, but we have checked numerically convergence and the validity of the identity in the region of small $|z|$ and $\mathrm{Arg}(z)$ in a right neighborhood of $-\frac \pi 2$.} Setting $r=1/2$ and $z=-i x$ this identity gives
\begin{equation}\label{eq:ST3d}
Z_{S T^{-1}[(1,0)]} (x) = \int da \, e^{-2\pi i a x}  e^{i \pi a^2}  \,Z_{(1,0)}(a-\tfrac{i}{2}) = e^{\frac{i\pi}{6}} e^{\pi x}\, Z_{(1,0)}(-x-\tfrac{i}{2})
\end{equation}
where $Z_{(1,0)}$ was given in \eqref{eq:NfNfbPF}, and the second argument of the function can be dropped in this case because $\Delta$ can be reabsorbed in $\hat\Delta$. The factors of $e^{\frac{i\pi}{6}}$ and $e^{\pi x}$ on the right-hand side correspond to background Chern-Simons terms that need to be added in order for the duality to work. Up to these contact terms, the transformation $ST^{-1}$ therefore leaves the theory $(1,0)$ invariant. The fact that $ST^{-1}$ is a transformation of order 3 implies that in fact this must be a triality, consistent with the claims that we reviewed above. 

This triality implies that the BCFT with boundary condition given by the single free chiral and coupling $\tau$ is left invariant under $\tau \to ST^{-1}[\tau] = -\frac{1}{\tau -1}$ and $\tau \to (ST^{-1})^2[\tau]=\frac{\tau-1}{\tau}$. More precisely, we need to take into account also the addition of background Chern-Simons terms on the left-hand side of \eqref{eq:ST3d}. Starting with \eqref{eq:ST3d} and performing manipulations analogous to those that lead to \eqref{eq:Tduality4dPF}-\eqref{eq:Sduality4dPF}, we obtain the following identity for the hemisphere partition function with $(1,0)$ boundary condition
\begin{align}
\begin{split}\label{eq:ST10}
&\mathcal{Z}_{HS^4}^{\text{N},(1,0)}(-\tfrac{1}{\tau-1}  , -i \hat\Delta , i(1+\tilde{\Delta}+\hat{\Delta})) \\
&\hspace{3cm} = \,\sqrt{\tau -1}\, e^{\frac{4 \pi  i}{3}} e^{i \pi  \hat\Delta \left(\hat\Delta-2 \tilde\Delta\right)} \mathcal{Z}_{HS^4}^{\text{N},(1,0)}(\tau  , -i\tilde\Delta, -i\hat{\Delta})~.
\end{split}
\end{align}
Some of the consequences of this triality on the BCFT setup were studied previously in \cite{KumarGupta:2019nay}. Note that the phase factor on the right hand side does not enter in the extremization. Performing extremization to determine $\tilde{\Delta}$ and $\hat{\Delta}$, this relation between the partition functions implies that
\begin{align}\label{eq:DeltaST}
\begin{split}
\tilde\Delta_{\text{max}}^{(1,0)}(\tau) &= -\hat\Delta_{\text{max}}^{(1,0)}(-\tfrac{1}{\tau-1})~,\\
\hat\Delta_{\text{max}}^{(1,0)}(\tau) &=  1+\tilde\Delta_{\text{max}}^{(1,0)}(-\tfrac{1}{\tau-1})- \hat\Delta_{\text{max}}^{(1,0)}(-\tfrac{1}{\tau-1})~.
\end{split}
\end{align} 
This transformation on the functions $\tilde\Delta$ and $\hat\Delta$ is easily verified to have order three, as it should. Moreover, it allows to solve for the values at the self-dual point $\tau=e^{i\frac{\pi}{3}}$, giving
\begin{equation}\label{edelta1chiral}
\tilde\Delta_{\text{max}}^{(1,0)}(e^{\frac{i\pi}{3}}) =-\frac 13~,~~\hat\Delta_{\text{max}}^{(1,0)}(e^{\frac{i\pi}{3}}) =\frac 13~.
\end{equation}
Consistently, plugging these values into \eqref{eq:ST10} one finds that the whole prefactor, including the $\sqrt{\tau -1}$, simplifies to 1. We can now proceed with the numerical tests.

Firstly, we check (\ref{edelta1chiral}) via a $\{7,7\}$ Pad\'e extrapolation from weak coupling along the complex curve $\tau=\tfrac 12 +i\frac{4\pi}{g^2}$
\begin{align}
\begin{split}
\tilde\Delta_{\text{max},\{7,7\}}^{(1,0)}(e^{i\frac{\pi}{3}})+\frac 13&\approx -0.001\,,\\
\hat\Delta_{\text{max},\{7,7\}}^{(1,0)}(e^{i\frac{\pi}{3}})-\frac 13&\approx -0.007\,,
\end{split}
\end{align}
and we find reasonable accordance.

\begin{figure}
     \centering
     \begin{subfigure}{0.45\textwidth}
       \centering
       \begin{overpic}[width=\textwidth]{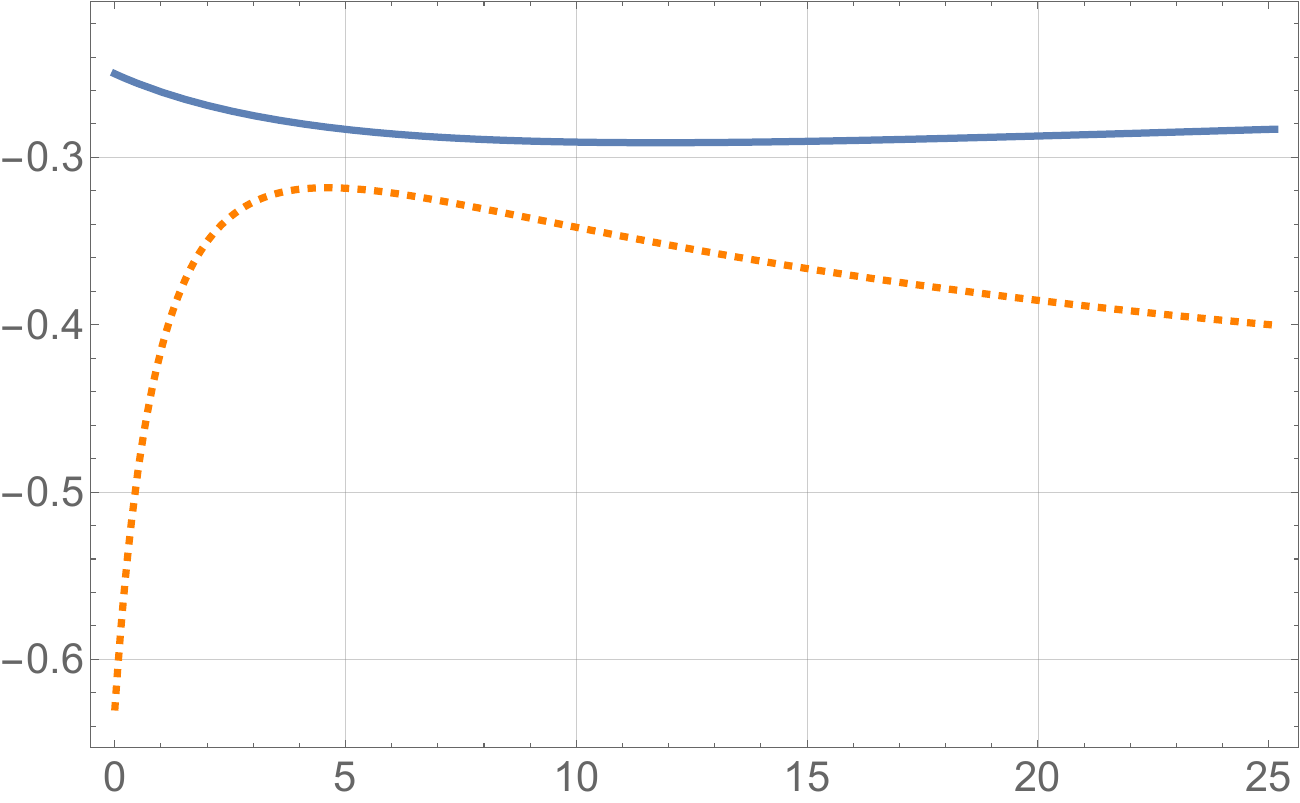}
        \put(50,-4){\textcolor{gray}{$g^2$}}
       \end{overpic}
     \end{subfigure}
     \hfill
     \begin{subfigure}{0.45\textwidth}
       \centering
       \begin{overpic}[width=\textwidth]{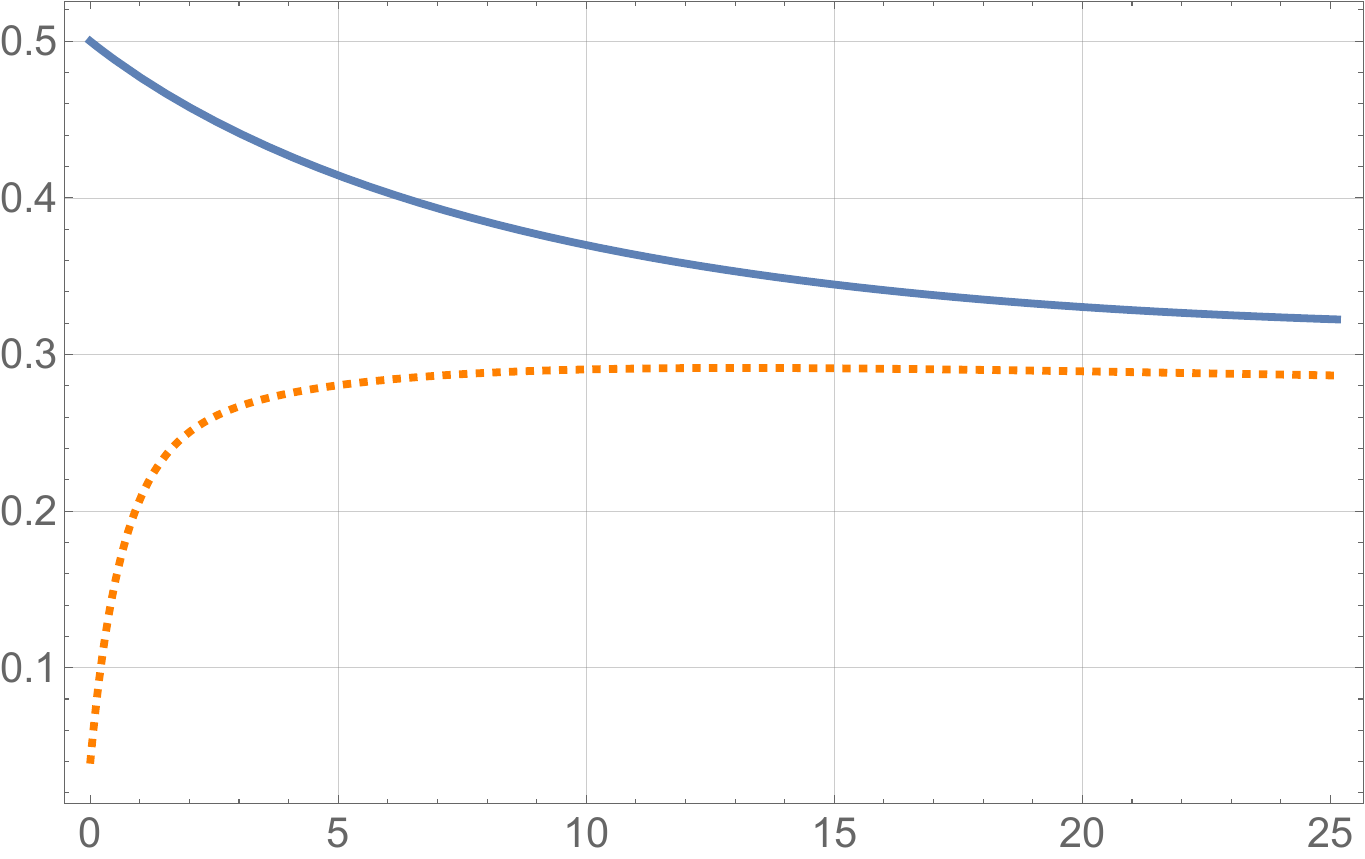}
        \put(50,-4){\textcolor{gray}{$g^2$}}
       \end{overpic}
     \end{subfigure}
     \vspace{0.1 cm}
        \caption{On the left, the $\{7,7\}$ Pad\'e approximant for $\tilde{\Delta}^{(1,0)}_{\text{max}}(\tau)$, in blue, and $-\hat{\Delta}^{(1,0)}_{\text{max}}(-\tfrac{1}{1-\tau})$, in dashed orange. On the right, the $\{7,7\}$ Pad\'e approximant for $\hat{\Delta}^{(1,0)}_{\text{max}}(\tau)$, in blue, and $1+\tilde{\Delta}^{(1,0)}_{\text{max}}(-\tfrac{1}{1-\tau})-\hat{\Delta}^{(1,0)}_{\text{max}}(-\tfrac{1}{1-\tau})$, in orange. The functions are evaluated at $\tau=1+i\tfrac{4\pi}{g^2}$, and plotted as a function of $g^2$.}\label{fig:deltas1chir}
\end{figure}

\begin{figure}
     \centering
     \begin{subfigure}{0.45\textwidth}
       \centering
       \begin{overpic}[width=\textwidth]{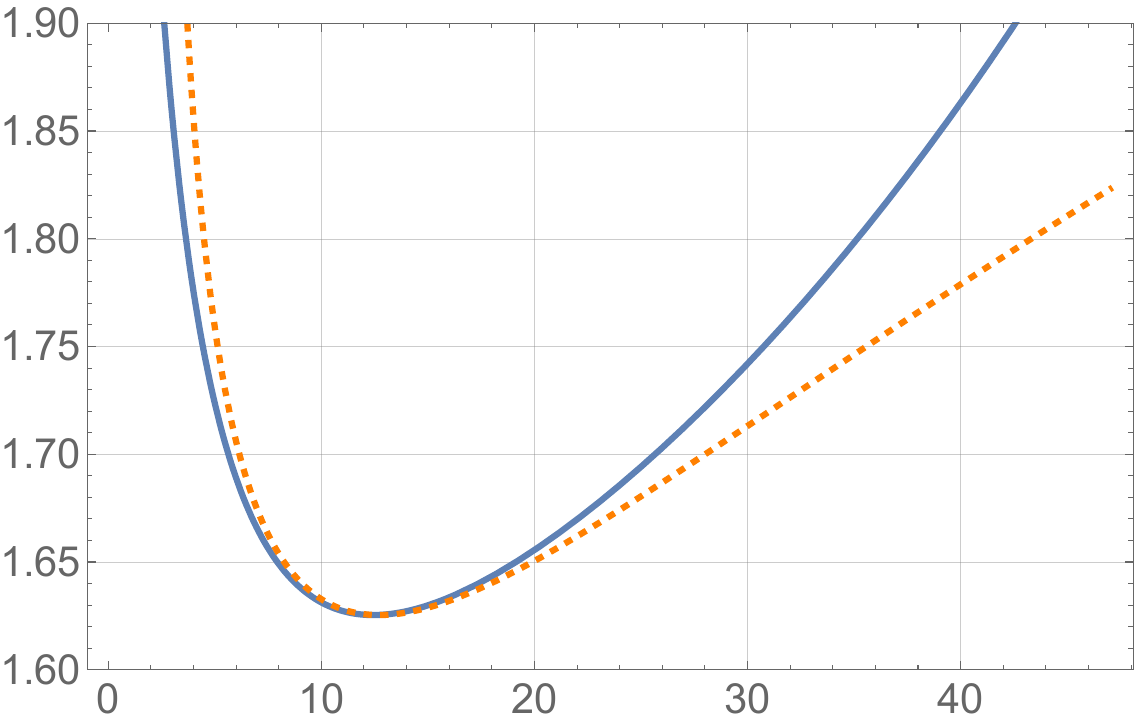}
        \put(50,-4){\textcolor{gray}{$g^2$}}
       \end{overpic}
     \end{subfigure}
     \hfill
     \begin{subfigure}{0.45\textwidth}
       \centering
       \begin{overpic}[width=\textwidth]{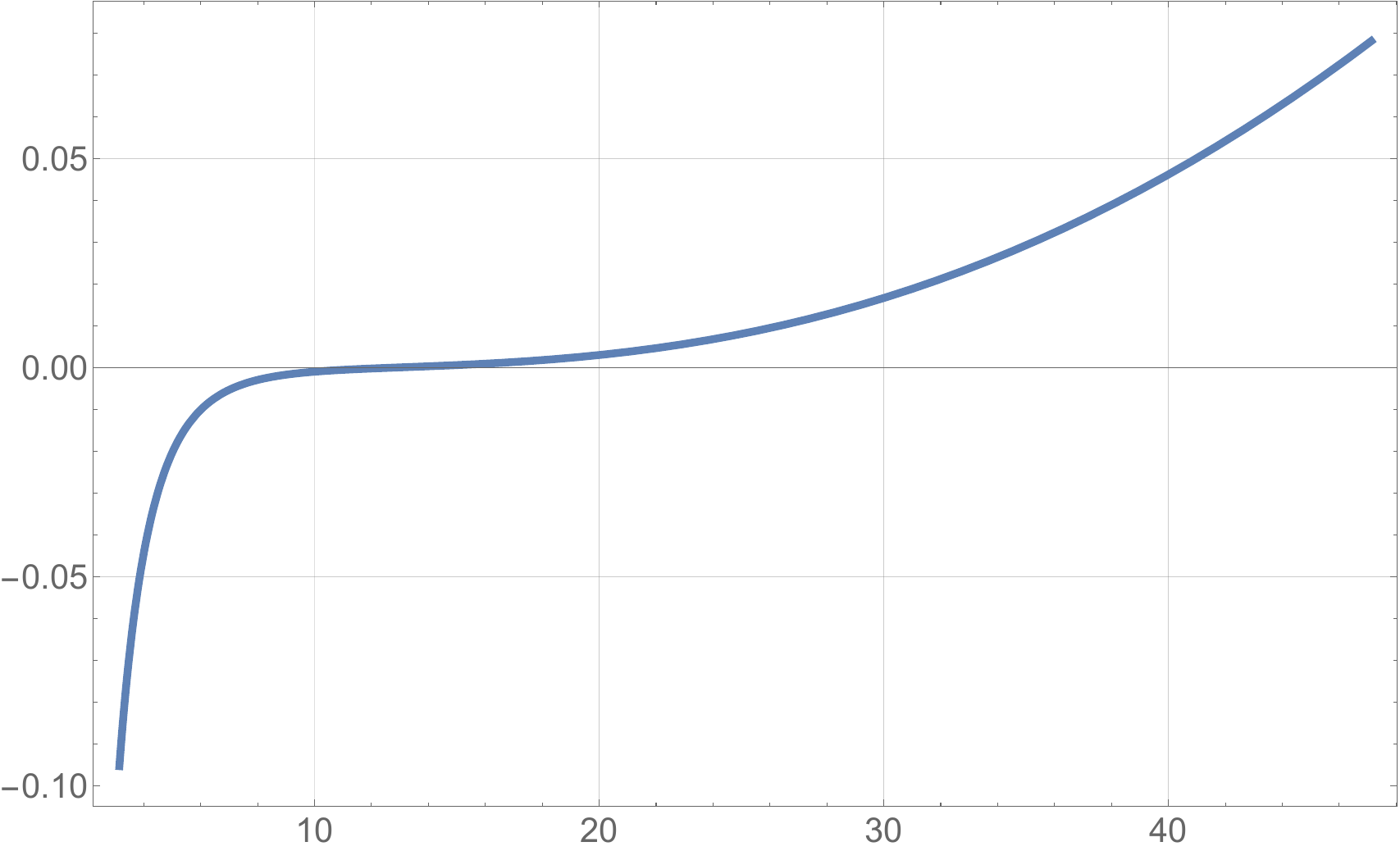}
        \put(50,-4){\textcolor{gray}{$g^2$}}
       \end{overpic}
     \end{subfigure}
     \vspace{0.1 cm}
        \caption{On the left, the $\{7,7\}$ Pad\'e approximant for $(e^{F_\partial})^{(1,0)}(\tau)$, in blue, and $(e^{F_\partial})^{(1,0)}(-\tfrac{1}{1-\tau})$, in dashed orange. On the right, the relative error (the difference between the two functions divided by the arithmetic mean). The functions are evaluated at $\tau=1+i\tfrac{4\pi}{g^2}$, and plotted as a function of $g^2$.}\label{fig:ef1chir}
\end{figure}

\begin{figure}
     \centering
     \begin{subfigure}{0.45\textwidth}
       \centering
       \begin{overpic}[width=\textwidth]{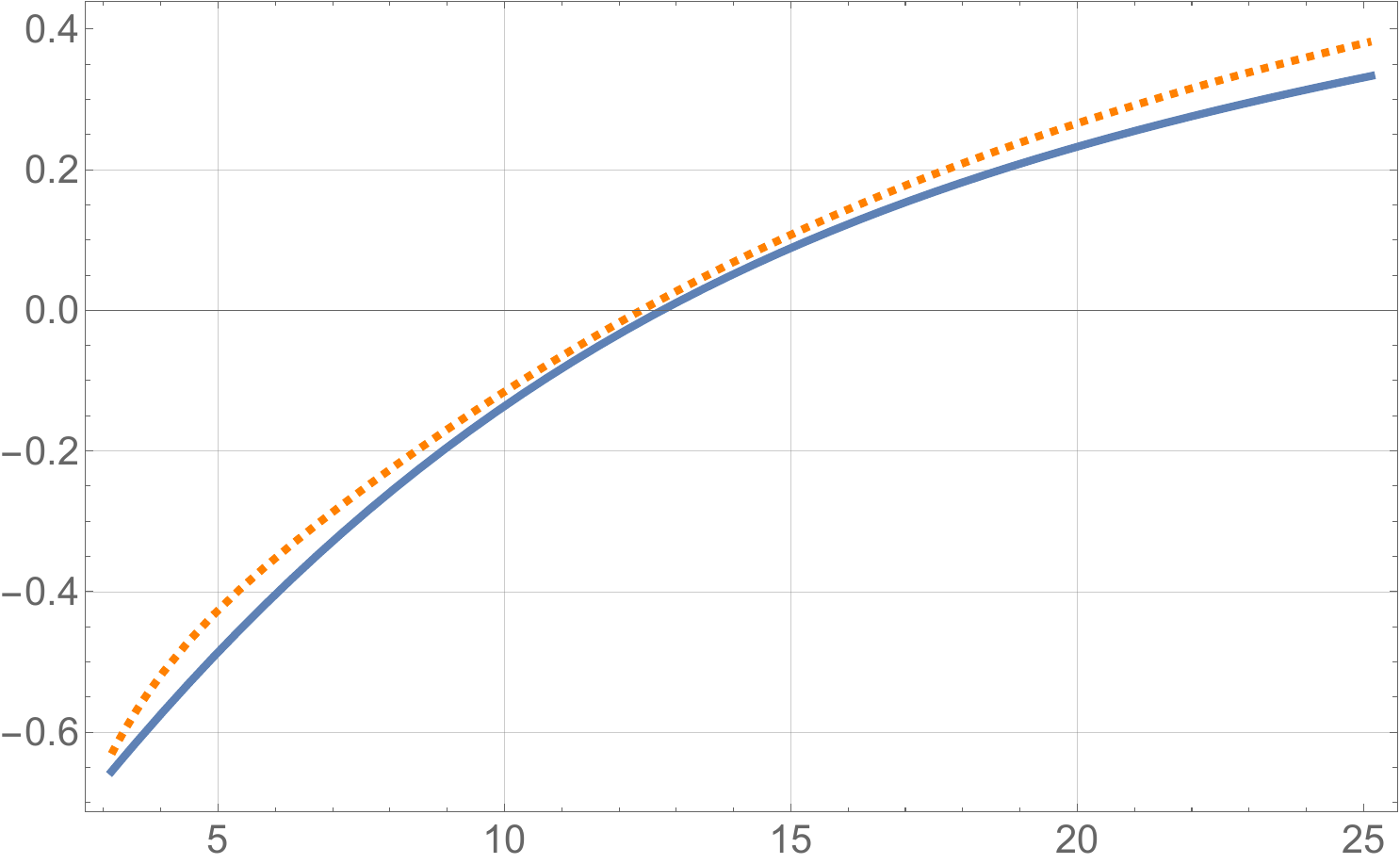}
        \put(50,-4){\textcolor{gray}{$g^2$}}
       \end{overpic}
     \end{subfigure}
     \hfill
     \begin{subfigure}{0.45\textwidth}
       \centering
       \begin{overpic}[width=\textwidth]{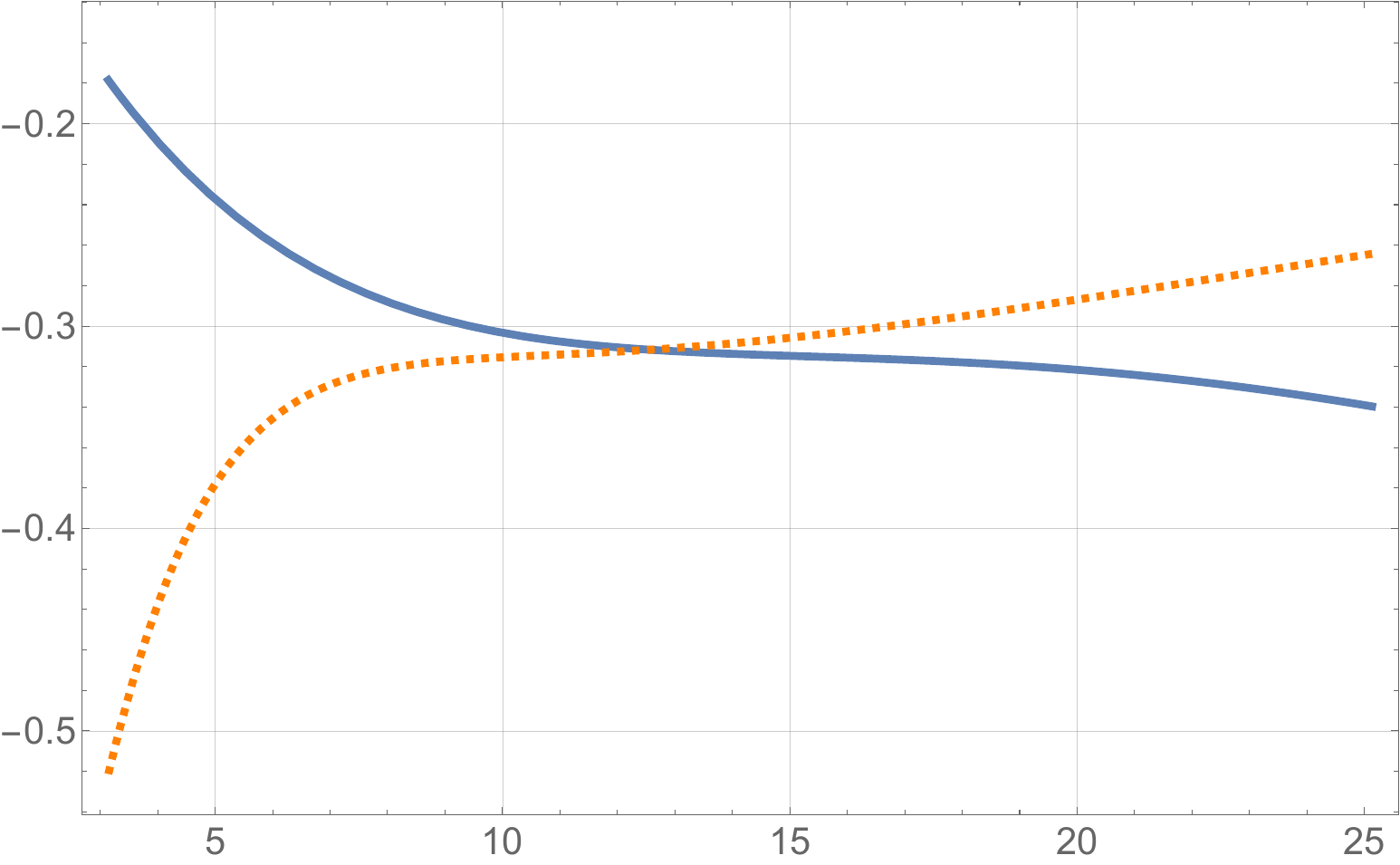}
        \put(50,-4){\textcolor{gray}{$g^2$}}
       \end{overpic}
     \end{subfigure}
     \vspace{0.1 cm}
        \caption{On the left, the real part of the $\{7,7\}$ Pad\'e approximant of $\tfrac{64\pi^2}{g^2}A_2(\tau)$, in blue, and of $\tfrac{64\pi^2}{g^2}(-\tfrac{1}{1-\tau})^2A_2(-\tfrac{1}{1-\tau})$, in dashed orange, evaluated at $\tau=1+i\tfrac{4\pi}{g^2}$, as a function of $g^2$. On the right the imaginary part. The normalization $\tfrac{64\pi^2}{g^2}$ is chosen for graphical convenience.}\label{fig:1pt1chir}
\end{figure}

The other checks we perform utilise a $\{7,7\}$ Pad\'e extrapolation along the complex curve $\tau=1 +i\frac{4\pi}{g^2}$ from both duality frames. In Figure \ref{fig:deltas1chir}, Figure \ref{fig:ef1chir} and Figure \ref{fig:1pt1chir} we show checks for (\ref{eq:DeltaST}), for $e^{F_\partial}$ and for the one-point function of $\phi^2$, which satisfies the relation
\begin{align}
\langle A_2^{(1,0)} \rangle(\tau)=\bigg(-\frac{1}{1-\tau}\bigg)^2 \langle A_2^{(1,0)} \rangle\big(-\frac{1}{1-\tau}\big)
\end{align}
that can be obtained via (\ref{eq:Tduality4d1ptFnct})-(\ref{eq:Sduality4d1ptFnct}). We find a good agreement with the predictions of the triality for the real part of the one-point function and for $e^{F_\partial}$, but generally the extrapolations are less accurate than the ones in the previous section. A possible interpretation is that perturbation theory is worse behaved for $\gamma\neq 0$. We leave some further comments for the conclusions.

\subsection{Feynman diagrams calculations}\label{SSec:perturbative}
In this section, we confirm the first two orders in \eqref{eq:exponepoint} via a Feynman diagram calculation. In order to compute the one-point function of $A_{2n} = \phi^{2n}$, we will repeatedly use the OPE in the bulk to bring $2n$ insertions of the field $\phi$ at the same point. 

\subsubsection{Tree-level}

At tree-level we have
\begin{align}\label{ewick}
	\bk{\phi(x_1)\hdots\phi(x_{2n})}=\bk{\phi(x_1)\phi(x_2)}\hdots\bk{\phi(x_{2n-1})\phi(x_{2n})}+\text{other Wick's contractions}~,
\end{align}
with $(2n-1)!!=(2n-1)(2n-3)...1$ addends. Sending $x_i\rightarrow x$ we remain with
\begin{align}
	\bk{\phi^{2n}(x)}=(2n-1)!!\bk{\phi^2(x)}^n
\end{align}
which leads to
\begin{align}\label{eaphi2n}
	a_{\phi^{2n}}=(2n-1)!!a_{\phi^2}^n~.
\end{align}
This matches the leading order in \eqref{eq:exponepoint} upon using the result in \eqref{ephi2N}. Note that the result of the perturbative calculation is naturally a function of $\tau_s$ rather than $\tau$, if we include a shift of the boundary Chern-Simons term compatible with the charge assignment of the boundary chiral fields.

\subsubsection{1-loop computations}\label{s1loop}

We now need to add boundary interaction vertices that will shift the above result by	$\d a^{N_F,N_{\bar{F}}}_{\phi^n}$.

\begin{figure}
	\begin{center}
		\includegraphics[scale=1]{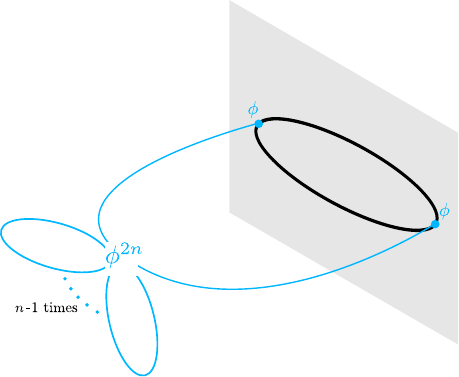}
	\end{center}
	\caption{Schematic representation of the NLO diagrams needed for the computation of $\bk{\phi^{2n}}$.}\label{fphin}
\end{figure}

A generic diagram will look like what is depicted in Figure \ref{fphin}.
Each possible contraction in the bulk can be seen as a correction to $a_{\phi^2}^\tt{tree}$ multiplied by $n-1$ closed loops. Each of these contributions has multiplicity $n(2n-1)!!$\footnote{There are $\frac{2n(2n-1)}{2}=n(2n-1)$ ways of choosing two $\phi$ operators between the $2n$ that will contract on the boundary. Then there are $(2n-3)!!$ ways of pairing the remaining $2n-2$ fields in the bulk.}. The contractions in the boundary simply produce the 1-loop diagrams computed in Appendix \ref{s1loopcomputations}. The NLO contribution can then be written as follows
\begin{align}
	\d a^{(N_F,N_{\bar{F}})}_{\phi^{2n}}=n(2n-1)!!(a^\tt{tree}_{\phi^2})^{n-1}\d a^{(N_F,N_{\bar{F}})}_{\phi^2}~.
\end{align}
It is easy to note that at the 1-loop order holds
\begin{align}
	\d a^{(N_F,N_{\bar{F}})}_{\phi^2}=(N_F+N_{\bar{F}})\d a^{(1,0)}_{\phi^2}\,,
\end{align}
thus we obtain
\begin{align}
	\d a^{(N_F,N_{\bar{F}})}_{\phi^{2n}}=(N_F+N_{\bar{F}})n(2n-1)!!(a^\tt{tree}_{\phi^2})^{n-1}\d a^{(1,0)}_{\phi^2}~.
\end{align}
Using (\ref{ephi2N}) and (\ref{ephi2l1}), again with $\tau\rightarrow\tau_s$, we find
\begin{align}
	\d a^{(N_F,N_{\bar{F}})}_{\phi^{2n}}=-(N_F+N_{\bar{F}})\frac{n(2n-1)!!}{32\pi}\bigg(\frac{i}{8\pi(\tau_s-\bar{\tau}_s)}\frac{\bar{\tau}_s}{\tau_s}\bigg)^{n-1}\tt{,}
\end{align}
in agreement with the NLO result shown in (\ref{eq:chiralonept}).

	\section{Discussion}
	In this paper we have described a method, based on supersymmetry, to obtain exact results for BCFT observables. The main application we have investigated so far is to the case of the $\mathcal{N}=2$ super Maxwell theory coupled to boundary degrees of freedom. In Section \ref{Sec:Examples} we have employed the localization formulas to compute many orders in perturbation theory for observables such as the hemisphere partition function, and the one-point functions, and compared the extrapolations to the predictions for the behavior at strong coupling coming from the SL(2,$\mathbb{Z}$) duality. A similar action of SL(2,$\mathbb{Z}$) on the boundary conditions exists also in the non-supersymmetric Maxwell theory, and also in that context the extrapolations of perturbative results to strong coupling can be used to estimate the observables of a large class of 3d CFTs, as discussed in \cite{DiPietro:2019hqe}. Our results in the supersymmetric context show encouraging agreements between the extrapolations and the expected behavior at strong coupling, at least for some observables, but on the other hand they also indicate that to obtain accurate results one needs many orders in perturbation, which at the moment seems beyond reach in the non-supersymmetric context. Given that here we have only used simple Pad\'e approximants, another possibility is to explore more sophisticated resummation techniques, and the supersymmetric examples can provide a valuable laboratory for this.
	
	An alternative nonperturbative method in BCFT, with or without supersymmetry, is the conformal bootstrap, which can be applied to the bulk two-point functions. The numerical implementation suffers from the lack of positivity of the coefficients in the bulk OPE expansion \cite{Liendo:2012hy}. Interesting results can still be obtained assuming that these coefficients are positive, or with non-rigorous approaches based on truncation \cite{Liendo:2012hy, Gliozzi:2015qsa, Padayasi:2021sik}. In the special case of a free CFT in the bulk, one can instead directly apply the numerical bootstrap to the boundary four-point functions of the boundary modes of the free field \cite{Behan:2020nsf, Behan:2021tcn}. Analytic bootstrap methods have been developed too, though they are mostly powerful in constraining deviations from mean-field theory \cite{Mazac:2018biw, Kaviraj:2018tfd}. It would be interesting to explore if the exact results presented in this paper can be used in conjunction with bootstrap techniques to constrain also non-protected data of the $\tfrac12$-BPS boundary conditions of 4d $\mathcal{N}=2$ SCFTs. Besides the one-point functions of chiral primaries, our method provides additional informations about the BCFT data that are potentially useful for the boostrap approach, such as the boundary OPE coefficients that can be extracted from the functions $\Delta_{\text{max}}(\tau)$ (see the discussion at the end of Section \ref{ssusyidentities}) and the bulk higher-point functions of the chiral primaries integrated over the hemisphere (see Section \ref{sec:mulder}).
	
	There are several directions in which the work presented here can be extended, and that we plan to explore in the future. One can consider the examples of non-abelian conformal gauge theories, studying the set of allowed boundary conditions and how they are mapped into each other by dualities, and performing detailed calculations using the localization formulas. The special case of $\mathcal{N}=4$ has already been studied in \cite{Wang:2020seq, Komatsu:2020sup, Dedushenko:2020vgd} using an alternative localization procedure available in that case. The hemisphere partition function for $\mathcal{N}=4$ has been also studied in \cite{Raamsdonk:2020tin} using the gravity dual. Particularly interesting $\mathcal{N}=2$ examples are given by SU$(N)$ SQCD with $2N$ flavors, and the two-node quiver theory containing the orbifold of $\mathcal{N}=4$ in its conformal manifold. For SQCD the duality wall has been derived in \cite{LeFloch:2015bto, Benini:2017dud} and it can be used to determine the action of the duality on the boundary conditions. These non-abelian theories and their dualities can often be realized with brane constructions, and one is naturally led to explore which boundary conditions can be engineered using branes ending on branes, similarly to what has been done for $\mathcal{N}=4$ \cite{Gaiotto:2008sa}. In the case of the orbifold of $\mathcal{N}=4$ it would be particularly interesting to study the BCFT data at large N, and reproduce their strong coupling behavior from a holographic calculation, similarly to what has been done in the maximally supersymmetric case in \cite{Aharony:2011yc}. Finally, it would be interesting to study the one-point functions in the limit of large charge, to see if there are simplifications that allow to obtain closed expressions for the coefficients at finite coupling.

	\bigskip
	\bigskip
	\leftline{\bf Acknowledgements}
	\smallskip
    \noindent We are grateful to C. Copetti, K. Zarembo, M. Bill\'o, A. Lerda and L. Bianchi for useful discussions. The authors also acknowledge support by INFN Iniziativa Specifica ST\&FI.
J.v.M. is supported by the ERC-COG grant NP-QFT No. 864583 ``Non-perturbative dynamics of quantum fields: from new deconfined phases of matter to quantum black holes''.
	
		\appendix

\section{Conventions}\label{app:conv}
In this appendix we write the conventions we use throughout the main text. We furthermore compare our convention to those used in relevant existing literature.

\subsection{Tensor conventions}

\begin{enumerate}

\item We work in Euclidean signature, and use the following index notations to denote the different representations:
\begin{equation}
\begin{aligned}[t]
&\mu,\ \nu,\ \hdots\ \text{4d curved-space indices}\tt{,}\notag\\
&i,\ j,\ \hdots\ \text{3d curved-space indices}\tt{,}\notag\\
&a,\ b,\ \hdots\ \text{4d tangent-space indices}\tt{,}\notag\\
&a',\ b',\ \hdots\ \text{3d tangent-space indices}\tt{,}\notag\\
&\a,\ \b,\ \hdots\ \tt{for }\psi_\a\in(\textbf{2},\textbf{1})_{Spin(4)}\tt{,}\\
&\da,\ \db,\ \hdots\ \tt{for }\bpsi^\da\in(\textbf{1},\textbf{2})_{Spin(4)}\tt{,}\\
&A,\ B,\ \hdots\ \tt{for }\l_A\in\textbf{2}_{SU(2)_R}\,.
\end{aligned}
\end{equation}

\item The Pauli-matrices are taken to be
\begin{align}
\tau^1=\begin{pmatrix}
0 & 1\\
1 & 0
\end{pmatrix}\,\ \ \ \tau^2=\begin{pmatrix}
0 & -i\\
i & 0
\end{pmatrix}\,\ \ \ \tau^3=\begin{pmatrix}
1 & 0\\
0 & -1
\end{pmatrix}\,.
\end{align}

\item All rank $n$ Levi-Civita tangent-space tensors are chosen such that $\varepsilon^{12\ldots n}= 1$. 

\item The Weyl/gamma matrices in four/three dimensions are taken to be
\begin{align}\label{e:gammamatrices}
\begin{split}
(\sigma^a)_{\a\db}=(-&i\vec{\tau},1)\,,\ \ \ \ (\bar{\sigma}^a)^{\da\b}=(i\vec{\tau},1)\,,\ \ \ \ (\g^{a'})_\a^{\ \b}=\tau^{a'}\,,\ \ \ \ \vec{\tau}_A^{\ B}=\vec{\tau}\,,\\
&\s^{ab}=\s^{[a}\bs^{b]}\,,\ \ \ \bs^{ab}=\bs^{[a}\s^{b]}\,,\ \ \ \g^{a'b'}=\g^{[a'}\g^{b']}\,.
\end{split}
\end{align}

\item All spinorial and $SU(2)_R$-symmetry indices are raised and lowered with the $\varepsilon$-tensors as in the Wess \& Bagger notation \cite{Wess:1992cp}. We employ the same notation for spinor bilinears as well as for the conjugation of Grassmann odd quantities $\theta_i$:
\begin{align}
(\t_1\t_2)^\dag=\t_2^\dag\t_1^\dag\,.
\end{align}

\item The symplectic indices are lowered with the symplectic tensors $\Omega_{IJ}$ and raised with $\Omega^{IJ}$ (defined such that $\Omega^{IJ}\Omega_{JK}=\delta^I_K$).
\end{enumerate}

\subsection{Matching conventions}
When considering supersymmetric theories on curved backgrounds, we have used the conventions of \cite{Hosomichi:2016flq}. For convenience of the reader we make connection with the conventions in \cite{Gerchkovitz:2016gxx,lauria2020supergravity}.

In \cite{Gerchkovitz:2016gxx}, objects in the fundamental representation of the $SU(2)_R$-symmetry group have indices $i,j,\hdots$ instead of our $A,B,...$. In \cite{Gerchkovitz:2016gxx} spinors are taken to have 4-components;  the $i$ index being up or down specifies the chirality of the spinor. We can map the spinors in \cite{Gerchkovitz:2016gxx} to ours through
\begin{equation}
	\ep^i_\text{\cite{Gerchkovitz:2016gxx}}\rightarrow\begin{pmatrix}
\frac{2}{cc_\s}\ep^A_\a\\
0
\end{pmatrix}\,,\quad \ep_i^\text{\cite{Gerchkovitz:2016gxx}}\rightarrow\begin{pmatrix}
0\\
2c\bar{\ep}_A^\da\,,
\end{pmatrix}
\end{equation}
where
\begin{align}
c^2= i\ \ \ \ \ \ \ \ \ c_\s^2=-1\,.
\end{align}
Correspondingly, the gamma matrices are mapped as follows
\begin{align}
\g^\mu_\text{\cite{Gerchkovitz:2016gxx}}&\rightarrow\begin{pmatrix}
&0 &c_\s \s^\mu_{\a\db}\\
&-c_\s \bs^{\mu,\ \da\b} &0
\end{pmatrix}\,.
\end{align}
Regarding the $\tau$-tensors and the $\ve$-tensors we have:
\begin{align}\label{emappingepsilon}
\tau_{\text{\cite{Gerchkovitz:2016gxx}}}\rightarrow i\tau\,,\ \ \ \ \ \ \varepsilon^{ij}_\text{\cite{Gerchkovitz:2016gxx}}\rightarrow \varepsilon^{AB}\,,\ \ \ \ \ \ \ve_{ij,\text{\cite{Gerchkovitz:2016gxx}}}\rightarrow -\ve_{AB}\,.
\end{align}
Since we are placing our theory on the sphere, the $SU(2)_R$-symmetry is broken and an explicit $\tau$-tensor appears (as in \ref{eksemain}). The breaking is chosen differently in \cite{Gerchkovitz:2016gxx}, and to compare to our result one has to implement an $SU(2)_R$-rotation such that
\begin{align}
\tau_{1,\text{\cite{Gerchkovitz:2016gxx}}}\rightarrow -i\tau_3\,.
\end{align}
Finally, the spinor denoted as $\bar{\Psi}$ in \cite{Gerchkovitz:2016gxx} is
\begin{align}
\bar{\Psi}=\Psi^T C
\end{align}
with $C$ the charge conjugation matrix 
\begin{align}
C=\begin{pmatrix}
&i\ve^{\a\b} &0\\
&0 &-i\ve_{\da\db}
\end{pmatrix}\,.
\end{align}

\section{Supersymmetry transformations}\label{App: Supersymmetry transformations}
In this appendix we will write down the supersymmetric transformations for the multiplets of interest, in rigid curved space.
In order to place a supersymmetric theory on curved manifold, one couples it to the corresponding off-shell supergravity multiplets, a similar procedure as done in \cite{Festuccia:2011ws}. The 3d $\cN=2$ gravity multiplet content can be found in Table \ref{tsugra3d} while the 4d one is in Table \ref{Tab: weyl multiplet}.
\begin{table}
\centering
\begin{tabular}{|c|c|c|}
\hline
\textbf{Bosons} & \textbf{Description} & \textbf{$\cR$-charge} \\ \hline
$g_{ij}$        & metric               & 0                     \\ \hline
$B_{ij}$    & 2-form gauge field   & 0                     \\ \hline
$C_i$         & photino              & 0                     \\ \hline
$A^{\cR}_i$   & $\cR$-gauge field    & 0                     \\ \hline
\end{tabular}
\caption{Content of the 3d $\mathcal{N}=2$ supergravity multiplet.}\label{tsugra3d}
\end{table}
In the 3d gravity multiplet it is customary to redefine the photino and the two-form as follows:
\begin{align}
V^i=-i\ve^{ijk}\pd_j C_k\,, \ \ \ \ \ H=\frac{i}{2}\ve^{ijk}\pd_i B_{jk}\,.
\end{align}
The off-shell supergravity background is then constructed by specifying $g_{\mu\nu}$ to be the metric of the manifold while the rest of the content is fixed via supersymmetry. On $S^4$ the gravity multiplet takes the following values
\begin{align}
	\tilde{M} = 0\:,\quad \mathcal A_\mu = 0 \:,\quad \mathcal V_\mu = 0\:,\quad  T_{\mu\nu} =0 \:,\quad  \bar T_{\mu\nu} = 0\,, \quad g_{\mu\nu} = g_{\mu\nu}^{S^4}\,.
\end{align}
On $S^3$ the gravity multiplet is fixed to be equal to
\begin{align}
H=\frac{i}{R}\,,\ \ \ \ \ \ V_i=A^\cR_i=0\,,\ \ \ \ \ \  g_{ij} = g_{ij}^{S^3}\,.
\end{align}

\subsection{$\cN=2$ 4d multiplets}\label{s4dmultiplets}

Here, we will list the curved space supersymmetry variations of the fields in the different multiplets. 

\paragraph{Vector multiplet}

The 4d $\mathcal N=2$ vector multiplet contains a vector, two complex scalars, two gaugini, and a triplet of auxilliary scalars. Their supersymmetric variations in curved space are given by
\begin{align}
\begin{split}\label{e4dvector}
&\d\phi=-i\ep^A\l_A\,,\\
&\d\bphi=i\bep^A\bl_A\,,\\
&\d\l_A=\frac{1}{2}\s^{\mu\nu}\ep_A(F_{\mu\nu}+8\bphi T_{\mu\nu})+2\s^\mu\bep_A D_\mu\phi+\s^\mu D_\mu\bep_A\phi+2i \ep_A[\phi,\bphi]+D_{AB}\ep^B\,,\\
&\d\bl_A=\frac{1}{2}\bs^{\mu\nu}\bep_A(F_{\mu\nu}+8\phi \bar{T}_{\mu\nu})+2\bs^\mu \ep_A D_\mu\bphi+\bs^\mu D_\mu \ep_A \bphi-2i\bep_A[\phi,\bphi]+D_{AB}\bep^B\,,\\
&\d A_\mu=i\ep^A\s_\mu \bl_A-i\bep^A\bs_\mu\l_A\,,\\
&\d D_{AB}=-2i\bep_{(A}\bs^\mu D_\mu\l_{B)}+2i\ep_{(A}\s^\mu D_\mu\bl_{B)}-4[\phi,\bep_{(A}\bl_{B)}]+4[\bphi,\ep_{(A}\l_{B)}]\,.
\end{split}
\end{align}

\paragraph{Hypermultiplet}

As explained in the main text, it is not possible to realize a 4d $\mathcal N=2$ hypermultiplet off shell without an infinite number of auxiliary fields. The three-dimensional $\mathcal N=2$ subalgebra, in which we are interested, is however realizable with a finite number of auxiliary fields. The realization of this subalgebra requires a constraint on the spinors, given in \eqref{eq:3dcondKS}. The supersymmetry variations then take the following form
\begin{align}
\begin{split}\label{ehyperssusy}
&\hspace{-3.5cm}\d q_{IA}=-i\ep_A\psi_{I}+i\bep_A\bpsi_{I}\,,\\
&\hspace{-3.5cm}\d \psi_{I}=2\s^\mu\bep_A D_\mu q^A_I+\s^\mu D_\mu \bep_A q^A_I-4i \ep_A\bphi q^A_I+2\hat{\ep}_{\hat{A}}F^{\hat{A}}_I\,,\\
&\hspace{-3.5cm}\d \bpsi_I=2\bs^\mu\ep_A D_\mu q^A_I+\bs^\mu D_\mu \ep_A q^A_I-4i \bep_A\phi q^A_I+2\bar{\hat{\ep}}_{\hat{A}}F^{\hat{A}}_I\,,\\
&\hspace{-3.5cm}\d F_{I\hat{A}}=i \hat{\ep}_{\hat{A}}\s^\mu D_\mu\bpsi_I-2\hat{\ep}_{\hat{A}}\phi\psi_I-2\hat{\ep}_{\hat{A}}\l_B q^B_I+2i\hat{\ep}_{\hat{A}}(\s^{\mu\nu}T_{\mu\nu})\psi_I\\
&\hspace{-3.5cm}\ \ \ \ \ \ \ \ \ \!-i \bar{\hat{\ep}}_{\hat{A}}\bs^\mu D_\mu\psi_I+2\bar{\hat{\ep}}_{\hat{A}}\bphi\bpsi_I+2\bar{\hat{\ep}}_{\hat{A}}\bl_B q^B_I-2i\bar{\hat{\ep}}_{\hat{A}}(\bs^{\mu\nu}\bar{T}_{\mu\nu})\bpsi_I\,,
\end{split}
\end{align}
with \cite{Hosomichi:2016flq}
\begin{align}
\hat{\ep}_{\hat{A}}=c^{\frac{1}{2}}\ep_A,\ \ \ \ \bar{\hat{\ep}}_{\hat{A}}=-c^{-\frac{1}{2}}\bep_A,\ \ \ \ c=-\frac{\bep^A\bep_A}{\ep^B\ep_B}\,.
\end{align}
\paragraph{(Anti-)Chiral multiplet} The supersymmetry variations of the fields inside the chiral multiplet and the anti-chiral multiplet take the following form
\begin{align}\label{eq:Chivar}
\begin{split}
&\d A=- i\ep^A \Psi_A~,\\
&\d \Psi_A=2 \sigma^\mu \bar{\epsilon}_A D_\mu A + B_{AB}\epsilon^B + \tfrac{1}{2}\s^{\mu\nu}\ep_A G^-_{\mu\nu} + w \,\sigma^\mu D_\mu \bar{\epsilon}_A A~,\\
& \d B_{AB} = -2 i\bar{\epsilon}_{(A}\bar{\sigma}^\mu D_\mu \Psi_{B)} + 2i \epsilon_{(A}\Lambda_{B)} + i (1-w) D_\mu \bar{\epsilon}_{(A}\bar{\sigma}^\mu \Psi_{B)}~,\\
& \d G_{\mu\nu}^-= -\tfrac{i}{2}\bar{\epsilon}^A \bar{\sigma}^\rho \sigma_{\mu\nu} D_\rho \Psi_A -\tfrac{i}{2}\epsilon^A \sigma_{\mu\nu}\Lambda_A -\tfrac{i}{4}(1+w)D_\rho \bar{\epsilon}^{A}\bar{\sigma}^\rho \sigma_{\mu\nu}\Psi_{A}~,\\
& \d \Lambda_A = -\tfrac{1}{2} D_\rho G_{\mu\nu}^-\sigma^{\mu\nu}\sigma^\rho \bar{\epsilon}_A + D_\mu B_{AB}\sigma^\mu \bar{\epsilon}^B - C \epsilon_A + 4 D_\rho A \,\bar{T}^{\mu\nu}\sigma^\rho\bar{\sigma}_{\mu\nu}\bar{\epsilon}_A \\
&~~~~~~~~~+\tfrac{1}{2}(1+w)B_{AB}\sigma^\mu D_\mu\bar{\epsilon}^B + \tfrac{1}{4}(1-w)G^-_{\mu\nu}\sigma^{\mu\nu}\sigma^\rho D_\rho \bar{\epsilon}_A+ 4 w A D_\rho \bar{T}^{\mu\nu}\sigma^\rho \bar{\sigma}_{\mu\nu}\bar{\epsilon}_A~,\\
& \d C = -2 i \bar{\epsilon}^A \bar{\sigma}^\mu D_\mu \Lambda_A + 4 i \bar{T}^{\mu\nu}\bar{\epsilon}^A  \bar{\sigma}_{\mu\nu}\bar{\sigma}^\rho D_\rho \Psi_A \\ 
& \,~~~~~~~ -i w D_\mu\bar{\epsilon}^A \bar{\sigma}^\mu \Lambda_A +4 i (w-1)D_\rho \bar{T}^{\mu\nu} \bar{\epsilon}^A \bar{\sigma}_{\mu\nu}\bar{\sigma}^\rho \Psi_A~.
\end{split}
\end{align}
and
\begin{align}\label{eq:AntiChivar}
\begin{split}
&\d \bar{A}= i\bar{\ep}^A \bar{\Psi}_A~,\\
&\d \bar{\Psi}_A=2 \bar{\sigma}^\mu \epsilon_A D_\mu \bar{A} + \bar{B}_{AB}\bar{\epsilon}^B +\tfrac{1}{2}\bar{\sigma}^{\mu\nu}\bar{\ep}_A G^+_{\mu\nu} + w \,\bar{\sigma}^\mu D_\mu \epsilon_A \bar{A}~,\\
& \d \bar{B}_{AB} = 2 i \epsilon_{(A}\sigma^\mu D_\mu \bar{\Psi}_{B)} + 2i \bar{\epsilon}_{(A}\bar{\Lambda}_{B)} - i (1-w) D_\mu \epsilon_{(A}\sigma^\mu \bar{\Psi}_{B)}~,\\
& \d G_{\mu\nu}^+= -\tfrac{i}{2}\epsilon^A \sigma^\rho \bar{\sigma}_{\mu\nu} D_\rho \bar{\Psi}_A -\tfrac{i}{2}\bar{\epsilon}^A \bar{\sigma}_{\mu\nu}\bar{\Lambda}_A +\tfrac{i}{4}(1+w)D_\rho \epsilon^{A}\sigma^\rho \bar{\sigma}_{\mu\nu}\bar{\Psi}_{A}~,\\
& \d \bar{\Lambda}_A = \tfrac{1}{2} D_\rho G_{\mu\nu}^+\bar{\sigma}^{\mu\nu}\bar{\sigma}^\rho \epsilon_A - D_\mu \bar{B}_{AB}\bar{\sigma}^\mu \epsilon^B + \bar{C}\bar{\epsilon}_A + 4 D_\rho \bar{A} \,T^{\mu\nu}\bar{\sigma}^\rho\sigma_{\mu\nu}\epsilon_A \\
&~~~~~~~~~-\tfrac{1}{2}(1+w)\bar{B}_{AB}\bar{\sigma}^\mu D_\mu \epsilon^B - \tfrac{1}{4}(1-w)G^+_{\mu\nu}\bar{\sigma}^{\mu\nu}\bar{\sigma}^\rho D_\rho \epsilon_A+ 4 w \bar{A} D_\rho T^{\mu\nu}\bar{\sigma}^\rho \sigma_{\mu\nu}\epsilon_A~,\\
& \d \bar{C} = -2 i \epsilon^A \sigma^\mu D_\mu \bar{\Lambda}_A +4 i T^{\mu\nu} \epsilon^A  \sigma_{\mu\nu}\sigma^\rho D_\rho \bar{\Psi}_A \\ 
& \,~~~~~~~ -i w D_\mu\epsilon^A \sigma^\mu \bar{\Lambda}_A +4 i (w-1) D_\rho T^{\mu\nu} \epsilon^A \sigma_{\mu\nu}\sigma^\rho \bar{\Psi}_A~.
\end{split}
\end{align}

\subsection{$\cN=2$ 3d multiplets}\label{s3d}
Due to the presence of a boundary, in this paper we will need $\cN=2$ 3d multiplets. Some of these multiplets can be found by constraining the four-dimensional multiplets to the boundary, as we will see below. We will largely follow the reference \cite{Closset:2012ru} for the 3d conventions. The $U(1)_R$-charge will be denoted with $r$ (anti-chiral fields will posses $U(1)_R$-charge equal to $-r$) while $z$ will be the central charge. The explicit supersymmetry transformations of the 3d chiral, vector, and linear multiplets are
\paragraph{Chiral multiplet}
\begin{align}
\begin{split}
&\hspace{-4cm}\d q=\sqrt{2}\zeta \psi\,,\\
&\hspace{-4cm}\d\psi=\sqrt{2}\z F-\sqrt{2}\rmi(z-r H)\tilde{\z} q-\sqrt{2}\rmi \g^i \tilde{\z} D_i q+\sqrt{2}\rmi \sigma \tilde{\z}q\,,\\
&\hspace{-4cm}\d F=\sqrt{2}\rmi(z-(r-2)H)\tilde{\z}\psi-\sqrt{2}\rmi D_i(\tilde{\z}\g^i\psi)-\sqrt{2}\rmi \sigma\tilde{\zeta}\psi+2\rmi\tilde{\zeta}\tilde{\l}q~.\label{e3dchiralsusy}
\end{split}
\end{align}

\paragraph{Anti-chiral multiplet} 
\begin{align}
\begin{split}
&\hspace{-4cm}\d \tilde{q}=-\sqrt{2}\tilde{\zeta}\tilde{\psi}\,,\\
&\hspace{-4cm}\d\tilde{\psi}=\sqrt{2}\tilde{\z}\tilde{F}+\sqrt{2}\rmi(z-r H)\z\tilde{q}+\sqrt{2}\rmi \g^i \z D_i \tilde{q}-\sqrt{2}\rmi \tilde{q}\sigma \z\,,\\
&\hspace{-4cm}\d\tilde{F}=\sqrt{2}\rmi(z-(r-2)H)\z\tilde{\psi}-\sqrt{2}\rmi D_i(\z\g^i\tilde{\psi})-\sqrt{2}\rmi \zeta\tilde{\psi}\s+2\rmi\tilde{q}\zeta\l~.\label{e3dantichiralsusy}
\end{split}
\end{align}

\paragraph{Vector multiplet}

\begin{align}
\hspace{-0.75 cm}\begin{split}\label{e3dvectorsusy}
&\delta\sigma=-\zeta\tilde{\lambda}+\tilde{\zeta}\lambda\,,\\
&\delta\lambda=\bigg(i(D+H\sigma)-\frac{i}{2}\varepsilon^{ijk}\gamma_k F_{ij}-i\gamma^i(D_i\sigma+iV_i\sigma)\bigg)\zeta\,,\\
&\delta\tilde{\lambda}=\bigg(-i(D+H\sigma)-\frac{i}{2}\varepsilon^{ijk}\gamma_k F_{ij}+i\gamma^i(D_i\sigma-iV_i\sigma)\bigg)\tilde{\zeta}\,,\\
&\delta A_i=-i(\zeta \gamma_i\tilde{\lambda}+\tilde{\zeta}\gamma_i\lambda)\,,\\
&\delta D=D_i(\zeta\gamma^i \tilde{\lambda}-\tilde{\zeta}\gamma^i \lambda)-i V_i(\zeta\gamma^i\tilde{\lambda}+\tilde{\zeta}\gamma^i\lambda)-H(\zeta\tilde{\lambda}-\tilde{\zeta}\lambda)-[\sigma,\zeta\tilde{\lambda}+\tilde{\zeta}\lambda]\,.
\end{split}
\end{align}

\paragraph{Linear multiplet}

\begin{align}
\hspace{-0.53 cm}\begin{split}\label{e3dlinear}
&\delta J=i\zeta j+i\tilde{\zeta}\tilde{j}\,,\\
&\delta j=\tilde{\zeta} K+i\gamma^i\tilde{\zeta}(j_i+i D_i J)+\tilde{\zeta}[\sigma,J]\,\\
&\delta\tilde{j}=\zeta K-i\gamma^i\zeta(j_i-i D_i J)-\zeta[\sigma,J]\,,\\
&\delta j_i=i\varepsilon_{ijk}D^j(\zeta\gamma^k j-\tilde{\zeta}\gamma^k\tilde{j})+[\sigma,\zeta\gamma_i j+\tilde{\zeta}\gamma_i\tilde{j}]+i[J,\tilde{\zeta}\gamma_i\lambda-\zeta\gamma_i\tilde{\lambda}]\,,\\
&\delta K=-i D_i(\zeta\gamma^i j+\tilde{\zeta}\gamma^i\tilde{j})+2iH(\zeta j+\tilde{\zeta}\tilde{j})-V_i(\zeta\gamma^i j+\tilde{\zeta}\gamma^i \tilde{j})+[\zeta\tilde{\lambda}+\tilde{\zeta}\lambda,J]~.
\end{split}
\end{align}
This definition of the linear multiplet matches the one of \cite{Closset:2012ru} in the abelian case.

\subsection{Reducing multiplets from four to three dimensions}\label{s4dto3d}

\paragraph{Vector multiplet}

When restricted to the boundary the fields in a 4d vector multiplet organize themselves into a 3d vector multiplet in the following way (on the left hand side there are the 3d fields, while one the right hand side the 4d fields):
\begin{align}\label{e3dmapvector}
\begin{split}
&A_i=A_i\,,\\
&\sigma=-i(\phi+\bar{\phi})=2\phi_1\,,\\
&\l=-\frac{1}{\sqrt{2}}(\l_1-i\s_4\bl_1)\,,\ \ \ \ \tilde{\l}=-\frac{1}{\sqrt{2}}(\l_2+i\s_4\bl_2)\,,\\
&D=-iD_{12}-D_\perp(\phi-\bar{\phi})=-iD_{12}-2D_\perp\phi_2\,.
\end{split}
\end{align}

\paragraph{4d hypermultiplet}
Using (\ref{ehyperssusy}) and the reduction of the Killing spinors on $S^3$ it is possible to find the relations between the 3d and 4d fields for the hypermultiplets:
\begin{align}\label{ehyperchiral3dmap}
\begin{split}
&q_I=q_{I1}\,,\\
&\chi_I=-\frac{i}{2}(\psi_I+i\sigma_\perp \bar{\psi}_I)\,,\\
&F_I=D_\perp q_{I2}+2\phi_2q_{I2}-i F_{I2}\,,
\end{split}
\ \ \ \ \ \ \ \ 
\begin{split}
&\tilde{q}_I=q_{I2}\,,\\
&\tilde{\chi}_I=\frac{i}{2}(\psi_I-i\sigma_\perp \bar{\psi}_I)\,,\\
&\tilde{F}_I=-D_\perp q_{I1}-2\phi_2 q_{I1}-i F_{I1}\,,
\end{split}
\end{align}
These fields organize themselves into three-dimensional off shell chiral multiplets $\Phi_{\tt{3d},I}=(q_I,\psi_I,F_I)$ and anti-chiral ones $\tilde{\Phi}_{\tt{3d},I}=(\tilde{q}_I,\tilde{\psi}_I,\tilde{F}_I)$ for each choice of $I\in 2N_{Sp(2N)}$. 

Due to the following conditions \cite{Hosomichi:2016flq} on $q_{IA}$
\begin{align}\label{erealcondhypers}
(q_{IA})^\dag=\ep^{AB}\Omega^{IJ}q_{JB}\doteq q^{AI}\,,
\end{align}
(and similar ones for the superpartners), not all of these multiplets are independent.

\section{One point functions in free BCFTs}\label{s1to2}
In this appendix we determine the perturbative one-point functions of scalar operators, via their two-point functions, in the case of Dirichlet and Neumann boundary conditions. We start with a generic free scalar and we will progressively move toward the free scalar of $\mathcal{N}=2$ $U(1)$ gauge theory on $H\mathbb{R}^4$ with the Neumann boundary conditions (\ref{ebc}).

\subsection{4d free massless scalar}\label{sfreescalar1to2}
Suppose we have a free 4d massless scalar on $\mathbb{R}^4$ with coordinates $x^\mu=(\vec{x},x_\perp)$. Suppose furthermore that the kinetic term in the Lagrangian is canonical, i.e. it takes the form of $\frac{1}{2}\pd_\mu\varphi\pd^\mu\varphi$. Then, through the Schwinger-Dyson equations, we can write
\begin{align}\label{eSDeq}
\bk{\Box\varphi(x)\varphi(y)}=\bk{\varphi(x)\Box\varphi(y)}=-\delta (x-y)
\end{align}
and find that
\begin{align}
\bk{\varphi(x)\varphi(y)}_{\text{no br.}}=\frac{1}{4\pi^2}\frac{1}{|\vec{x}-\vec{y}|^2+(x_\perp-y_\perp)^2}\,.
\end{align}
If we add a boundary at $x_\perp=0$ we will need to solve the same equation but in the region $x_\perp,y_\perp>0$ with some boundary conditions. For instance we can have
\begin{align}\label{efreepropwithbr}
\bk{\varphi(x)\varphi(y)}=\frac{1}{4\pi^2}\left(\frac{1}{|\vec{x}-\vec{y}|^2+(x_\perp-y_\perp)^2}\mp \frac{1}{|\vec{x}-\vec{y}|^2+(x_\perp+y_\perp)^2}\right)\,,\end{align}
where the sign is negative if $\varphi(\vec{x},0)=0$ and positive if $\pd_y\varphi(\vec{x},0)=0$. We then find that
\begin{align}
a_{\varphi^2}=\mp\frac{1}{16\pi^2}\ \ \ \begin{cases}
-\ \text{if }\varphi(\vec{x},0)=0\\
+\ \text{if }\pd_y\varphi(\vec{x},0)=0
\end{cases}\,.
\end{align}

\subsection{$U(1)\ \cN=2$ theory on $H\mathbb{R}^4$}

We will now move to the theory we are interested in. We will warm up first with the Dirichlet boundary conditions, then we will move to the Neumann ones (see \eqref{ebc}). We remind the reader that we use the following definitions \cite{Hama:2012bg, Gava:2016oep}:
\begin{align}\label{ephi1phi2}
\begin{split}
\phi&=\phi_2+i\phi_1\tt{,}\\
\bphi&=-\phi_2+i\phi_1\,.
\end{split}
\end{align}

\subsubsection*{Dirichlet boundary conditions}
In this case the boundary conditions for the scalars are
\begin{align}\label{eDscalars}
\begin{cases}
\phi_1=0\\
\pd_\perp\phi_2+\frac{i}{2}D_{12}=0
\end{cases}\,.
\end{align}
With these boundary conditions the $\theta$-term is zero and so $\bk{\phi_1(x)\phi_2(y)}=0$ due to a preserved $\bbZ\times\bbZ$ symmetry $\phi_i\longmapsto-\phi_i$. We can thus write:
\begin{align}\label{e1}
\bk{\phi(x)\phi(y)}=\bk{\phi_2(x)\phi_2(y)}-\bk{\phi_1(x)\phi_1(y)}\,.
\end{align}
Notice also that we have not considered the presence of the $D_{12}$ in the boundary conditions (\ref{eDscalars}). This is because at separated points the Schwinger-Dyson equations put that field to zero. We can use the result showed in (\ref{efreepropwithbr}) with the fields normalized in the action as $\frac{4}{g^2}\pd_\mu\phi_i\pd^\mu\phi_i=-\frac{i}{2\pi}(\tau-\btau)\pd_\mu\phi_i\pd^\mu\phi_i$ to find 
\begin{align}\label{ephiphiD}
\bk{\phi(x)\phi(y)}=\frac{i}{2\pi(\tau-\btau)}\frac{1}{|\vec{x}-\vec{y}|^2+(x_\perp+y_\perp)^2}\,.
\end{align}
In the absence of a boundary, thanks to what was explained around (\ref{enonsingularOPE}), we have a non-singular OPE for two chirals. Consistently then we have that (\ref{ephiphiD}) is non-singular in the limit $x\to y$. Taking the limit we find
\begin{align}\label{ephi2D}
a_{\phi^2}=\frac{i}{8\pi(\tau-\btau)}\,.
\end{align}
\subsubsection*{Neumann boundary conditions}

In this case there is no $\bbZ\times\bbZ$ symmetry. We still have the Schwinger-Dyson equations (\ref{eSDeq}) with $D_{12}=0$, with the different normalization dictated by the choice of the kinetic term, which we now solve for the boundary conditions
\begin{align}\label{egamma}
\begin{cases}
\g\phi_1+\phi_2=0\\
\pd_\perp\phi_1-\g\pd_\perp\phi_2=0
\end{cases}\,.
\end{align}

The solution can be found with the following ansatz
\begin{align}
\begin{cases}
\bk{\phi_1(x)\phi_1(y)}=\frac{g^2}{32\pi^2}\left(\frac{1}{|\vec{x}-\vec{y}|^2+(x_\perp-y_\perp)^2}+a\frac{1}{|\vec{x}-\vec{y}|^2+(x_\perp+y_\perp)^2}\right)\\
\bk{\phi_1(x)\phi_2(y)}=\frac{g^2}{32\pi^2}\left(b\frac{1}{|\vec{x}-\vec{y}|^2+(x_\perp-y_\perp)^2}+c\frac{1}{|\vec{x}-\vec{y}|^2+(x_\perp+y_\perp)^2}\right)\\
\bk{\phi_1(x)\phi_1(y)}=\frac{g^2}{32\pi^2}\left(\frac{1}{|\vec{x}-\vec{y}|^2+(x_\perp-y_\perp)^2}+d\frac{1}{|\vec{x}-\vec{y}|^2+(x_\perp+y_\perp)^2}\right)
\end{cases}\,.
\end{align}
Then we fix the constants via the boundary conditions
\begin{align}
\begin{cases}
\bk{\phi_1(x)\phi_1(y)}\Big|_{x\in\pd}=-\frac{1}{\g}\bk{\phi_2(x)\phi_1(y)}\Big|_{x\in\pd}\\
\bk{\phi_2(x)\phi_2(y)}\Big|_{x\in\pd}=-\g\bk{\phi_1(x)\phi_1(y)}\Big|_{x\in\pd}\\
\bk{\phi_1(x)\pd_\perp\phi_1(y)}\Big|_{y\in\pd}=\g\bk{\phi_2(x)\pd_\perp\phi_1(y)}\Big|_{y\in\pd}\\
\bk{\pd_\perp\phi_2(x)\phi_2(y)}\Big|_{x\in\pd}=\frac{1}{\g}\bk{\pd_\perp\phi_1(x)\phi_1(y)}\Big|_{x\in\pd}
\end{cases}\,.
\end{align}
Solving the above we find
\begin{align}\label{epropwithbr}
\begin{split}
\bk{\phi_1(x)\phi_1(y)}&=\frac{i}{4\pi(\tau-\btau)}\Bigg(\frac{1}{|\vec{x}-\vec{y}|^2+(x_\perp-y_\perp)^2}+\frac{1-\g^2}{1+\g^2}\frac{1}{|\vec{x}-\vec{y}|^2+(x_\perp+y_\perp)^2}\Bigg)\,,\\
\bk{\phi_2(x)\phi_2(y)}&=\frac{i}{4\pi(\tau-\btau)}\Bigg(\frac{1}{|\vec{x}-\vec{y}|^2+(x_\perp-y_\perp)^2}-\frac{1-\g^2}{1+\g^2}\frac{1}{|\vec{x}-\vec{y}|^2+(x_\perp+y_\perp)^2}\Bigg)\,,\\
\bk{\phi_1(x)\phi_2(y)}&=-\frac{i}{2\pi(\tau-\btau)}\frac{\g}{1+\g^2}\frac{1}{|\vec{x}-\vec{y}|^2+(x_\perp+y_\perp)^2}\,,
\end{split}
\end{align}
from which, using $\phi=\phi_2+i\phi_1$, we get:
\begin{align}\label{ephiphiN}
\bk{\phi(x)\phi(y)}=\frac{i}{2\pi(\tau-\btau)}\frac{\btau}{\tau}\frac{1}{|\vec{x}-\vec{y}|^2+(x_\perp+y_\perp)^2}\,.
\end{align}
For the same reasons explained below (\ref{e1}), we have a non-divergent result and in the limit $x\rightarrow y$ we find
\begin{align}\label{ephi2N}
a_{\phi^2}=\frac{i}{8\pi(\tau-\bar{\tau})}\frac{\bar{\tau}}{\tau}\,.
\end{align}

\section{Diagrammatics}\label{sdiagrammatics}
In this appendix we compute the 1-loop contributions to $\bk{A_2}=\bk{\phi^2(x)}$ in $U(1)\ \cN=2$ Maxwell theory on $H\mathbb{R}^4$ conformally coupled to a single chiral in the boundary. The complete diagrammatics needed for this calculation, using dimensional regularization 
\begin{align}\label{edimreg}
d=4-2\ep
\end{align}
is given below

\begin{align}\label{ebulkbrprop}
\eqfig{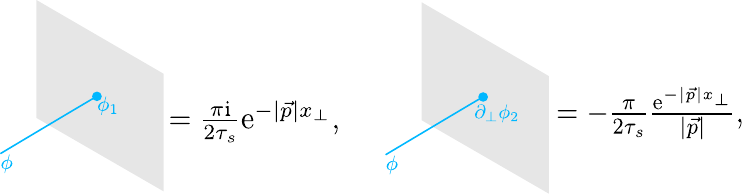}{0.70}
\end{align}

\begin{align}\label{ebrbrprop}
\eqfig{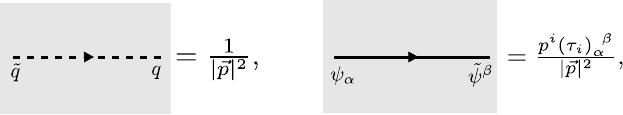}{0.65}
\end{align}

\begin{align}\label{evertices}
\eqfig{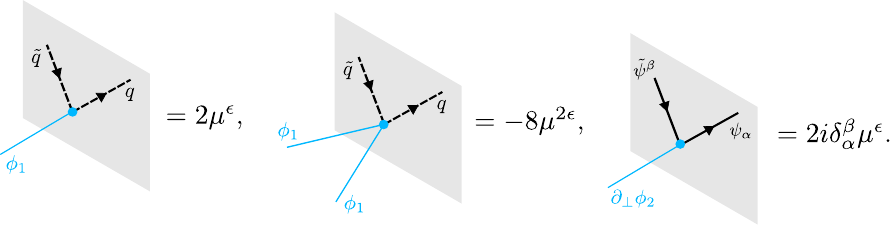}{0.85}
\end{align}

Here we already employed the shift $\tau\rightarrow \tau_s$ showed in (\ref{eshift}) due to the Chern-Simons contact terms we need to add for the chirals (see Section \ref{sec:chirals}).

\subsection{Bulk-boundary propagators in phase space}\label{sbulkbrprop}

Following what found in (\ref{epropwithbr}) and placing one operator on the boundary, we can perform the Fourier transformations over $\vec{x}-\vec{y}$:
\begin{align}
\begin{split}
\bk{\phi_1(x)\phi_1(\vec{y})}&=\frac{i}{4\pi(\tau_s-\btau_s)}\bigg(1+\frac{1-\g^2}{1+\g^2}\bigg)\frac{1}{(\vec{x}-\vec{y})^2+x_\perp^2}=\frac{i}{2\pi(\tau_s-\btau_s)}\frac{1}{1+\g^2}\frac{1}{(\vec{x}-\vec{y})^2+x_\perp^2}\\
&=-\frac{i}{8\pi}\frac{\tau_s-\btau_s}{\tau_s\btau_s}\frac{1}{(\vec{x}-\vec{y})^2+x_\perp^2}=\int\frac{d^{d-1}\vec{p}}{(2\pi)^{2-1}}e^{+i\vec{p}\cdot(\vec{x}-\vec{y})}\bigg(-\frac{\pi i}{4}\frac{\tau_s-\btau_s}{\tau_s\btau_s}\frac{e^{-|\vec{p}|x_\perp}}{|\vec{p}|}\bigg)\,,\\
\bk{\phi_2(x)\phi_1(\vec{y})}&=\bk{\phi_1(x)\phi_2(\vec{y})}=-\g\bk{\phi_1(x)\phi_1(\vec{y})}=-\frac{1}{8\pi}\frac{\tau_s+\btau_s}{\tau_s\btau_s}\frac{1}{(\vec{x}-\vec{y})^2+x_\perp^2}\\
&=\int\frac{d^{d-1}\vec{p}}{(2\pi)^{2-1}}e^{+i\vec{p}\cdot(\vec{x}-\vec{y})}\bigg(-\frac{\pi}{4}\frac{\tau_s+\btau_s}{\tau_s\btau_s}\frac{e^{-|\vec{p}|x_\perp}}{|\vec{p}|}\bigg)\,,\\
\bk{\phi_1(x)\pd_\perp\phi_2(\vec{y})}&=\pd_{y_\perp}\bk{\phi_1(x)\phi_2(y)}\bigg|_{y_\perp=0}=\frac{1}{8\pi}\frac{\tau_s+\btau_s}{\tau_s\btau_s}\frac{2x_\perp}{\big((\vec{x}-\vec{y})^2+x_\perp^2\big)^2}\\
&=\int\frac{d^{d-1}\vec{p}}{(2\pi)^{2-1}}e^{+i\vec{p}\cdot(\vec{x}-\vec{y})}\bigg(\frac{\pi}{4}\frac{\tau_s+\btau_s}{\tau_s\btau_s}e^{-|\vec{p}|x_\perp}\bigg)\,,\\
\bk{\phi_2(x)\pd_\perp\phi_2(\vec{y})}&=\pd_{y_\perp}\bk{\phi_2(x)\phi_2(y)}\bigg|_{y_\perp=0}=\frac{i}{4\pi(\tau_s-\btau_s)}\bigg(1+\frac{1-\g^2}{1+\g^2}\bigg)\frac{2x_\perp}{\big((\vec{x}-\vec{y})^2+x_\perp^2\big)^2}\\
&=\int\frac{d^{d-1}\vec{p}}{(2\pi)^{2-1}}e^{+i\vec{p}\cdot(\vec{x}-\vec{y})}\bigg(-\frac{i\pi}{4}\frac{\tau_s-\btau_s}{\tau_s\btau_s}e^{-|\vec{p}|x_\perp}\bigg)\,.
\end{split}
\end{align}
Then, using (\ref{ephi1phi2}), we get what is shown in (\ref{ebulkbrprop}).

\subsection{Boundary-boundary propagators and vertices}

We shall write the action for one $\cN=2$ 3d chiral multiplet interacting with a vector following the conventions in \cite{Willett:2016adv} (whose 3d supersymmetry transformations are equal to the ones in \cite{Closset:2012ru} that we decided to follow). Since we have a conformal coupling with the bulk we have to choose $r=\frac{1}{2}$ and $z=0$ in (\ref{e3dchiralsusy}) and in (\ref{e3dantichiralsusy}). We shall write everything in function of the 3d gauge-content and then we will use  (\ref{e3dmapvector}) to put everything in function of the 4d field content:
\begin{align}\label{echirallagr}
\cL_{\text{chiral}}&=D_i \tilde{q} D^i q-i\tilde{\psi}\cancel{D}\psi +\tilde{q}\bigg(D+\s^2\bigg)q-i\tilde{\psi}\s\psi+\sqrt{2}i(\tilde{q}\l\psi+\tilde{\psi}\tilde{\l}q)-\tilde{F}F=\\
&=D_i \tilde{q} D^i q-i\tilde{\psi}\cancel{D}\psi +\tilde{q}\bigg(-iD_{12}-2\pd_\perp\phi_2+4\phi_1^2\bigg)q+\\
&\ \ \ -i2\tilde{\psi}\phi_1\psi+\sqrt{2}i(\tilde{q}(\l_1-i\s_4\bl)\psi+\tilde{\psi}(\l_2+i\s_4\bl_2)q)-\tilde{F}F\,.
\end{align}

With this we arrive to the boundary-boundary propagators showed in (\ref{ebrbrprop}) and the vertices in (\ref{evertices}). 

\subsection{1-loop correction to $A_2$, 1 chiral multiplet}\label{s1loopcomputations}
We now compute the 1-loop contribution to $\bk{A_2}$.

There are only three diagrams contributing to the 1-loop correction. Employing dimensional regularization (\ref{edimreg}) we can write the first contribution
\begin{align}
\eqfig{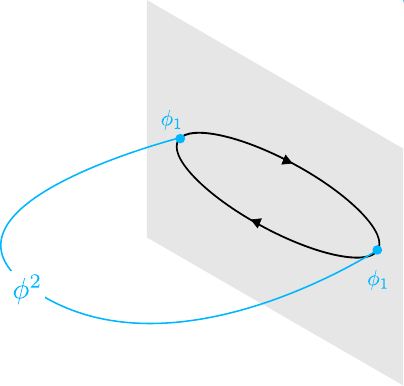}{0.3}=\frac{\pi^2}{\tau_s^2}\mu^{2\epsilon}\int\! \frac{d^{d-1}\vec{q}}{(2\pi)^{d-1}}\int\! \frac{d^{d-1}\vec{p}}{(2\pi)^{d-1}} \frac{\tr(\cancel{\vec{q}}(\cancel{\vec{q}}-\cancel{\vec{p}}))}{\vec{q}^2(\vec{q}-\vec{p})^2\vec{p}^2}e^{-2|\vec{p}|x_\perp}=\frac{\Y_1}{x_\perp^{2-4\epsilon}}+\frac{\Y_2}{x_\perp^{1-2\epsilon}}
\end{align}

with

\begin{align}
\Y_1&=-\frac{\pi^2}{2\tau_s^2}\frac{(4\mu)^{2\epsilon}\G(2-2\epsilon)\G\big(\frac{5-d}{2}\big)}{(4\pi)^{d-1}\G\big(\frac{d-1}{2}\big)}\mintt{0}{1}dx\ \big[x(1-x)\big]^{-\frac{1}{2}(1+2\epsilon)}\text{,}\\
\Y_2&=\frac{4\pi^2}{\tau_s^2}\frac{(2\mu)^{2\epsilon}\G(1-2\epsilon)}{(4\pi)^{d-1}\G^2\big(\frac{d-1}{2}\big)}\mintt{0}{+\infty}dq\ q^{d-4}\,.
\end{align}
Note that the $\Y_2$ contribution must cancel in the end result. Indeed, not only does it add an IR-divergence (which is a problem for the theory at hand if we do not have other diagrams cancelling it), it also adds a term scaling like $\sim\frac{1}{x_\perp}$ not allowed by conformal invariance which fixes $\bk{\phi^2(x)}\propto\frac{1}{x_\perp^2}$.

Indeed the following further two corrections are given by:
\begin{align}
\eqfig{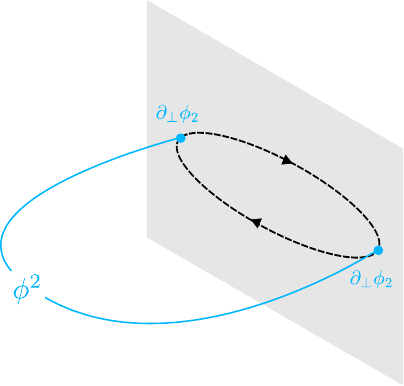}{0.3}=-\frac{2\pi^2}{\tau_s^2}\mu^{2\epsilon}\int\! \frac{d^{d-1}\vec{q}}{(2\pi)^{d-1}}\int\! \frac{d^{d-1}\vec{p}}{(2\pi)^{d-1}} \frac{1}{\vec{q}^2}\frac{e^{-2|\vec{p}|x_\perp}}{\vec{p}^2}=-\frac{\Y_2}{x_\perp^{1-2\epsilon}}
\end{align}
\begin{align}
\eqfig{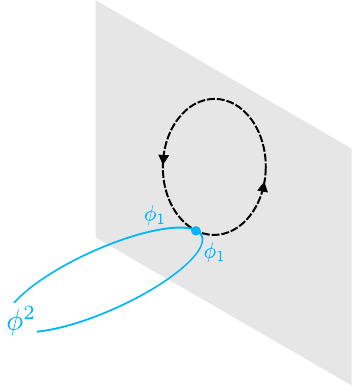}{0.3}=-\frac{\pi^2}{\tau_s^2}\mu^{2\epsilon}\int\! \frac{d^{d-1}\vec{q}}{(2\pi)^{d-1}}\int\! \frac{d^{d-1}\vec{p}}{(2\pi)^{d-1}} \frac{e^{-2|\vec{p}|x_\perp}}{\vec{q}^2(\vec{q}-\vec{p})^2}=\frac{\Y_1}{x_\perp^{2-4\epsilon}}
\end{align}
The total contribution goes precisely as $\sim\frac{1}{x_\perp^2}$ as it should in the limit for $\ep\ar 0$. We thus obtain
\begin{align}\label{ephi2l1}
\d a^{(1,0)}_{\phi^2}=\underset{\ep\ar 0}{\lim}\ 2\Y_1=-\frac{1}{32\tau_s^2}\,.
\end{align}

\newpage	
	
	\bibliography{HS4bib}
	\bibliographystyle{utphys}
	
\end{document}